\documentclass[a4paper,12pt]{article}
\pdfoutput=1
\usepackage[a4paper,text={16.8cm,22.8cm}]{geometry} 
\usepackage{amsmath,amssymb,bm,graphicx}
\usepackage{mathtools}
\usepackage{mathrsfs} 
\usepackage[dvipsnames]{xcolor}
\usepackage{extarrows}
\usepackage[utf8]{inputenc}
\usepackage{amsthm}
\usepackage{slashed}
\usepackage{bm}
\usepackage{rotating}
\usepackage{hyperref}
\usepackage{array}
\usepackage{multirow}
\usepackage{bbold}
\usepackage{cite}
\usepackage{changepage}
\usepackage{pifont}
\usepackage[]{relsize}
\usepackage{tabularx}
\usepackage{stackengine}
\usepackage{gensymb}
\usepackage[]{cancel} 
\usepackage{caption}
\usepackage{subcaption}
\usepackage{tikz}
\usepackage{tikz-feynman}
\tikzfeynmanset{compat=1.1.0}
\usepackage{soul} 

\usetikzlibrary{external}
\allowdisplaybreaks 
\renewcommand{\arraystretch}{1.2}

\newcommand{\pkg}[1]{\texttt{#1}} 

\newcommand\FN{\ensuremath{\text{U(1)}_{\text{FN}}}}
\newcommand\LUV{\ensuremath{\Lambda_{\text{UV}}}}
\newcommand\LFN{\ensuremath{\Lambda_{\text{FN}}}}
\newcommand\LSM{\ensuremath{\Lambda_{\text{SM}}}}

\newcommand\DFN{\ensuremath{d_{\text{FN}}}}
\newcommand\DSM{\ensuremath{d_{\text{SM}}}}
\newcommand\SFN{\ensuremath{\mathcal{S}_{\text{FN}}}} 
\newcommand\ZFN{\ensuremath{\mathbb{Z}_{\text{FN}}}}

\newcommand{\p}{\partial} 
\newcommand{\hc}{^{\dagger}} 
\newcommand{\hdh}{H^\dagger H} 
\newcommand\ptwiddle[1]{\mathord{\mathop{#1}\limits^{\scriptscriptstyle(\smaller{\sim})}}}

\newcommand{\Darr}{\mbox{${i \, \raisebox{2mm}{\boldmath ${}^\leftrightarrow$}\hspace{-4mm} D_\mu}$}} 
\newcommand{\DarrI}{\mbox{${i \, \raisebox{2mm}{\boldmath ${}^\leftrightarrow$}\hspace{-4mm} D_\mu^{\,I}}$}}
\newcommand{\parr}{\mbox{${i \, \raisebox{2mm}{\boldmath ${}^\leftrightarrow$}\hspace{-4mm} \partial_\mu}$}}

\newcommand{\tabletitle}[1]{
\begin{tabularx}{0.99\textwidth}{>{\centering\arraybackslash}X}
#1 \\[-0.2cm]
\end{tabularx}
}

\let\originalleft\left
\let\originalright\right
\renewcommand{\left}{\mathopen{}\mathclose\bgroup\originalleft}
\renewcommand{\right}{\aftergroup\egroup\originalright}

\newcommand\be{\begin{equation}}
\newcommand\ee{\end{equation}}

\begin{document}

\begin{titlepage}

\begin{flushright} 
LA-UR-24-21703
\end{flushright}

\vspace{1.2cm}
\begin{center}
\Large\bf
\boldmath
Froggatt-Nielsen Meets the SMEFT
\unboldmath
\end{center}
\vspace{0.2cm}
\begin{center}
{\large{Eetu Loisa$^{a}$ and Jim Talbert$^{a,b}$}}\\
\vspace{1.0cm}
{\small\sl 
${}^a$\,DAMTP, University of Cambridge, Wilberforce Rd., Cambridge, CB3 0WA, United Kingdom\\[0.1cm]
${}^b$\,Theoretical Division, Group T-2, MS B283, Los Alamos National Laboratory, P.O. Box 1663, \\Los Alamos, NM  87545, USA}\\[0.5cm]
{\bf{E-mail}}: eal47@cam.ac.uk, rjt89@cantab.ac.uk\\[1.0cm]
\end{center}

\vspace{0.5cm}
\begin{abstract}
\vspace{0.2cm}
\noindent 
We study the matching of Froggatt-Nielsen theories of flavour onto the Standard Model Effective Field Theory (SMEFT), upon integrating out a heavy Beyond-the-Standard-Model (BSM) scalar `flavon' whose vacuum expectation value breaks an Abelian flavour symmetry at energies $ \LFN$ well above the electroweak scale, $\LFN > \LSM$.  We include matching contributions to the infrared $d_{\text{SM}}=6$ (Warsaw basis) SMEFT sourced from ultraviolet contact terms suppressed up to order $1/\LUV^2$ in the Froggatt-Nielsen Lagrangian, where $\LUV > \LFN$ is an arbitrary ultraviolet scale where further unspecified BSM particles are dynamical. This includes tree-level (one-loop) ultraviolet diagrams with $ d_{\text{FN}}=6$ $(5)$ effective vertices.  We first do so with a toy model, but then generalize our findings to arbitrary Frogatt-Nielsen charges. 
Our results indicate a rich and non-trivial signature of Froggatt-Nielsen theories on the (otherwise) model-independent operators of the SMEFT, and we briefly speculate on extending our analysis to broader classes of BSM flavour models, e.g. non-Abelian and/or gauged theories.  We thus take an important step towards determining how to use rapidly developing theoretical and experimental SMEFT technologies to gain unambiguous insight into the SM's longstanding fermion flavour puzzle.
\end{abstract}
\vfil

\end{titlepage}


\tableofcontents
\noindent \makebox[\linewidth]{\rule{16.8cm}{.4pt}}


\section{Introduction}
\label{sec:INTRO}

If not the oldest, the Froggatt-Nielsen (FN) mechanism \cite{Froggatt:1978nt} is arguably the most famous attempt to dynamically resolve the Standard Model's (SM) \emph{flavour puzzle}, i.e. to explain the otherwise arbitrary (and often dramatic) hierarchies of fermionic mass, mixing, and CP-violation inferred from decades of experiment.  Allowing for simple generalizations of the original FN idea, the ingredients necessary for solving such a longstanding puzzle of the SM are strikingly simple: a new symmetry (gauged or global, Abelian or non-Abelian, continuous or discrete), a new scalar sector to break it, and a sufficiently high mass scale to suppress entries in the SM's Yukawa couplings.

Indeed, the FN mechanism\footnote{...or, rather, what we actually mean when we refer to said mechanism!} is most transparent at intermediate energy scales, where we can discuss the Beyond-the-SM (BSM) dynamics most relevant to the flavour puzzle without reference to fully ultraviolet (UV)-complete Lagrangians.  Namely, when the Higgs boson is non-trivially charged under an Abelian (continuous or discrete) BSM flavour symmetry \SFN{}, the SM's renormalizable Yukawa couplings are, a priori, disallowed for arbitrary flavours.  Non-renormalizable corrections to the Yukawa sector can appear, however, given successive insertions of a novel scalar flavon $\theta$, itself non-trivially charged under \SFN{} :
\begin{equation}
\label{eq:basicidea}
\mathcal{L} \supset y_{ij} \,\overline{\psi}_i \, H\, \psi_j \longrightarrow \mathcal{L} \supset c_{ij} \, \overline{\psi}_i \, H \, \psi_j \left(\frac{\theta}{\Lambda_{\text{UV}}}\right)^{x_{ij}} \,\,\,\,\,\text{with}\,\,\,\,\,\,x_{ij} = -\frac{(q_H + q_{\overline{\psi}_i}+q_{\psi_j})}{q_\theta}\,,
\end{equation}
where $\psi$ are SM fermions, $q$ is the respective \SFN{} charge of a given field, $x_{ij}$ represents the power suppression necessary to achieve \SFN{} invariance, $c_{ij}$ are generally considered to be $\mathcal{O}(1)$ parameters, and $\Lambda_{\text{UV}}$ represents an arbitrary UV mass scale where heavier degrees of freedom propagate.  For example, heavy vector-like fermions can `complete' the effective interaction on the right-hand-side (RHS) of \eqref{eq:basicidea}, yielding a fully renormalizable Lagrangian, although we wish to remain agnostic about such (likely unfalsiable) UV-completions. Regardless, well below $\Lambda_{\text{UV}}$, the scalar flavon $\theta$ can obtain a vacuum expectation value (vev) which breaks the \SFN{} symmetry of the effective Lagrangian, such that $\langle \theta \rangle / \Lambda_{UV} \equiv \lambda \lesssim 0.1 $.  Then, upon the Higgs itself acquiring a vev in the electroweak phase transition, $\langle H\hc H \rangle \equiv v^2_{H}/2$, the SM fermions obtain masses,
\begin{equation}
\label{eq:basicideamass}
    M^\psi_{ij} \simeq c_{ij}\cdot\frac{v_{H}}{\sqrt{2}}\cdot\lambda^{x_{ij}}\,,
\end{equation}
where we use a capitalized $M$ to indicate that the mass matrix is in an interaction basis, and not the physical mass-eigenstate basis, and where $c_{ij} \rightarrow y_{ij}$ for matrix elements allowed at the renormalizable level ($x_{ij} = 0$).  In this way, hierarchical patterns of mass and mixing can be built up via hierarchies in powers of $\lambda$, in accord with observation, as opposed to unnatural hierarchies in Lagrangian parameters.

Unfortunately, however, FN-type models of flavour have not convincingly predicted the complete structure of the deeply hierarchical SM quark mass spectrum, including the well-constrained \cite{ParticleDataGroup:2022pth} Cabibbo-Kobayashi-Maskawa (CKM) matrix, much less the mass scales and mixing patterns that analogously appear in the lepton sector given non-zero neutrino masses (see e.g. \cite{Esteban:2020cvm}).\footnote{Although certain predictions for and relations among the fermion masses and mixing angles from FN-type constructions (see e.g. \cite{Leurer:1992wg,Leurer:1993gy}) and other constructions (see e.g. \cite{Gatto:1968ss,Georgi:1979df}) have been enticing for decades.}  Indeed, given that only the leptonic Dirac CP-violating phase, further phases associated to the potential Majorana nature of neutrinos, and the exact values of the neutrino-mass eigenvalues (and hence also their ordering) remain largely unresolved in the SM's pure mass and mixing sector, most predictions coming from new FN-type models should actually be considered \emph{retrodictions}.  
Furthermore, the mass scale associated with $\theta$ is often assumed to be very heavy for simplicity, such that any additional phenomenology associated to $\theta$-dependent vertices at (e.g.) high-energy colliders and/or precision flavour factories is often limited (purposely overlooked) in the literature --- for recent analyses of FN-type phenomenology that extend beyond mass and mixing predictions, see e.g.  \cite{Bonnefoy:2019lsn,Fedele:2020fvh,Allanach:2022blr,Cornella:2023zme,Asadi:2023ucx, Greljo:2024evt}.  

In short, the falsifiability of FN-type models (as typically studied) is becoming increasingly challenged.  Given this somewhat uninspiring context, a natural question arises:
\medskip
\begin{adjustwidth}{20pt}{20pt}
\emph{Is it possible to probe the UV dynamics of FN-type theories in a largely model-independent manner, thereby exposing any universal infrared (IR) predictions that can be efficiently falsified by experiment?} ($\bigtriangleup$)
\end{adjustwidth}
\medskip
One is immediately led to effective field theory (EFT) considerations, and in particular to the SMEFT \cite{Buchmuller:1985jz,Grzadkowski:2010es} (see \cite{Brivio:2017vri,Isidori:2023pyp} for reviews), for an answer. The SMEFT is the theory including all renormalizable and non-renormalizable interactions of the SM field content, invariant under spacetime and linearly-realized SM gauge symmetries $\mathcal{G}_{\text{SM}} = \text{SU(3)}_\text{c} \times \text{SU(2)}_\text{L} \times \text{U(1)}_\text{Y}$, assuming new BSM physics is parametrically heavier than characteristic SM mass scales:\footnote{We have suppressed flavour labels in \eqref{eq:SMEFT}.}
\begin{equation}
\label{eq:SMEFT}
    \mathcal{L}_{\text{SMEFT}} \equiv \mathcal{L}_{\text{SM}} + \sum_i \frac{C_i^{(d)}}{\Lambda^{d-4}} \mathcal{O}_i^{(d)} \,.
\end{equation}
Here $C_i$ are arbitrary Wilson coefficients that encode the impact of UV dynamics on local, IR contact interactions $\mathcal{O}_i$ of mass-dimension $d$, suppressed by appropriate powers of a BSM mass scale $\Lambda > \LSM$, above which the unspecified UV dynamics can propagate. 
The counting of operators at arbitrary mass dimensions $d$ is by now greatly simplified via Hilbert Series methods \cite{Lehman:2015via}, while complete and non-redundant bases up to $d \le 6$ are well-utilized in the literature --- the lone operator appearing at $d=5$ is the Weinberg operator \cite{Weinberg:1979sa} giving a non-zero Majorana neutrino mass term, while a plethora of operators already appear at $d=6$ \cite{Grzadkowski:2010es}.  
Critically, \eqref{eq:SMEFT} parameterizes model-dependent BSM physics into a single, largely model-independent IR Lagrangian; constraining $C_i$ experimentally amounts to constraining every BSM model that `matches' to $\mathcal{L}_{\text{SMEFT}}$ (with specific predictions for $C_i$).  
Whilst technically challenging, global fits to $C_i^{(6)}$ are now becoming available in the combined top, Higgs, and electroweak sector \cite{Ellis:2020unq,Ethier:2021bye,Iranipour:2022iak,Brivio:2022hrb}, while fits addressing embedded flavour assumptions in particular observables are also appearing \cite{Falkowski:2017pss,Falkowski:2019hvp,Cirigliano:2019vfc,Aoude:2020dwv,Bissmann:2020mfi,Cirigliano:2021img,Bellafronte:2023amz,Grunwald:2023nli,Allwicher:2022gkm,Fajfer:2023gie,Allwicher:2023shc}.

In an attempt to address ($\bigtriangleup$), we aim to `match' the FN mechanism distilled in \eqref{eq:basicidea} to the SMEFT in \eqref{eq:SMEFT}.  That is, we will determine which operators $\mathcal{O}_i$ have generically non-zero Wilson coefficients $C_i$ upon integrating the flavon field $\theta$ out of the UV Lagrangian.  
We can therefore identify $\Lambda \sim \LFN$ in \eqref{eq:SMEFT}, and hence our job is to calculate $C_i$ explicitly. 
We will do so up to $d \equiv \DSM = 6$ in \eqref{eq:SMEFT} (assuming the Warsaw basis of \cite{Grzadkowski:2010es}), including contributions from operators up to $\DFN \equiv 4 + x_{ij} = 6$ in the intermediate-energy FN EFT $\mathcal{L}_{\text{FN}}$ introduced in \eqref{eq:LFNgeneral} below.  Pursuing matching at $\DFN = 6$ has also required enumerating a minimal, non-redundant basis for the FN EFT at this order, including SM and flavon degrees of freedom.

We will use analytic functional techniques (cf. the recent presentations in (e.g.) \cite{Fuentes-Martin:2016uol, Cohen:2020fcu}) when matching tree-level FN interactions to the SMEFT, and the automated functional techniques embedded in {\tt{Matchete}} \cite{Fuentes-Martin:2022jrf} when considering one-loop UV contributions (many of which we will also check/intuit diagramatically by hand). 
In general, we find that a host of SMEFT operators appear already at $\DSM = 6$, including Higgs-scalar ($H^6$), Higgs-derivative ($H^4 D^2$), Higgs-gauge ($X^2 H^2$), four-fermion ($\psi^4$), Higgs-kinetic ($\psi^2 H^2 D $), and Higgs-enhanced Yukawa ($H^3 \psi^2$) operators.  
These effects are driven both by the \SFN{} charge structure of the FN theory, giving distinct IR SMEFT flavour phenomenology, but also by flavour-\emph{independent} effects, including Higgs-flavon mixing. 
{We often stop to compare the exact matching results with the expectations from an analysis where the Froggatt-Nielsen parameter $\lambda$ is treated as a spurion of the \SFN{} symmetry breaking.}
Our results are given explicitly in Tables~\ref{tab:treeleveldim4and5}--\ref{tab:4fermions}.

The IR signatures of FN-type theories are therefore quite distinct, and hence we address how one might use the SMEFT to unambiguously probe theories aiming to dynamically resolve the flavour puzzle.  Simply put, phenomenological analyses that constrain the SMEFT operators we have identified simultaneously constrain the FN-type theories that ubiquitously turn them on.   
Our work therefore adds to a growing list of theoretical studies addressing what one can learn about UV BSM flavour physics from the flavour structure of the IR SMEFT \cite{Descotes-Genon:2018foz,Hurth:2019ula,Helset:2019eyc,Aebischer:2020lsx,Faroughy:2020ina,Bruggisser:2021duo,Kobayashi:2021pav,Talbert:2021iqn,Bonnefoy:2021tbt,Isidori:2021gqe,Greljo:2022cah,Dawson:2022bxd,Bruggisser:2022rhb,Machado:2022ozb,Greljo:2023adz,Bonnefoy:2023bzx,Antusch:2023shi}. 

The paper develops as follows: in Section \ref{sec:MODEL} we introduce a toy model of \eqref{eq:basicidea}, review the functional matching techniques/tools we will use, and then perform a tree- and loop-level matching to the SMEFT.  As mentioned above, we also present a complete and minimal $\DFN = 6$ EFT basis for the FN theory in this Section, before also generalizing our matching to arbitrary toy models of this type when presenting final results in Tables~\ref{tab:treeleveldim4and5}--\ref{tab:4fermions}.   
Then, in Section \ref{sec:GENERALIZE}, we briefly discuss a number of generalizations to the naive toy setup of Section \ref{sec:MODEL}, including considerations of non-Abelian flavour models, as well as continuous (global or gauged) constructions.
Finally, we provide a summary and outlook in Section \ref{sec:CONCLUDE}.

\section{Matching the Froggatt-Nielsen Mechanism}
\label{sec:MODEL}

We begin with a generic FN effective Lagrangian valid at scales $\LFN > \LSM$, which is composed of both renormalizable and non-renormalizable interactions furnished by the SM field content and the novel flavon field $\theta$, which we take to be a complex scalar charged under \SFN{} but transforming as a singlet of the SM gauge group.  We can decompose the resulting $\mathcal{G}_{\text{SM}} \times \SFN{}$ symmetric Lagrangian as:
\begin{equation}
\label{eq:LFNgeneral}
\mathcal{L}_{\text{FN}} \equiv \mathcal{L}_{\text{SM}^\prime} + \mathcal{L}^4_{\theta} + \sum_{d>4}^\infty \mathcal{L}^d_{\text{FN}}\,,
\end{equation}
where $\mathcal{L}^4_\theta$ is the set of $\theta$-dependent renormalizable operators,\footnote{Here we have anticipated working in the broken \SFN{} phase.} 
\begin{equation}
\label{eq:Ltheta}
\mathcal{L}^4_\theta = 
	- \theta^* \Box \theta + \mu_\theta^2 \theta^* \theta - \lambda_{02} \left( \theta^* \theta \right)^2 - \lambda_{11} \left( H\hc H \right) \left( \theta^* \theta \right),
\end{equation}
$\mathcal{L}_{\text{SM}^\prime}$ is the \SFN{}-symmetric SM, and $\mathcal{L}^d_{\text{FN}}$ is the set of effective, non-renormalizable operators that parameterize further BSM effects appearing in the UV.  
At this stage, we wish to remain agnostic about any specific UV completions of the effective FN Lagrangian, and thus allow all operators invariant under $\mathcal{G}_{\text{SM}} \times \SFN{}$ to be turned on in \eqref{eq:LFNgeneral}.  
Hence the non-renormalizable terms in the equation can be further decomposed into \SFN{}-symmetric SMEFT operators (that is, terms independent of $\theta$) and operators with $\theta$-dependence,
\begin{equation}
    \mathcal{L}^{d>4}_{\text{FN}} \equiv \mathcal{L}^{d>4}_{\text{SMEFT}^\prime} + \mathcal{L}^{d>4}_{\theta}\,.
\end{equation}
Hence one critical task for our study is to enumerate the minimal non-redundant operator basis for $\mathcal{L}^{d>4}_{\text{FN}}$ order-by-order, i.e. in a way that is consistent with the output of a Hilbert series calculation.  
Thankfully this task is somewhat simplified given our assumptions. 
For example, we assume that both $H$ and $\theta$ are non-trivially charged under \SFN{}\, and that $\theta$ is a $\mathcal{G}_{\text{SM}}$-singlet.  In this case it is easy to deduce that there are no odd-mass-dimension terms in the effective scalar potential, for instance.  We will return to the issue of operator bases in Section \ref{sec:dim6matching} below.

Given \eqref{eq:LFNgeneral}, we must adopt a prescription for integrating $\theta$ out of the dynamical action.  We will do so with functional matching techniques, where the one-light-particle-irreducible (1LPI) effective action of the UV theory is matched with that of the IR EFT. 
At tree-level:
\begin{equation}
\mathcal{L}_\text{EFT}^\text{tree} [\phi] = \mathcal{L}_\text{UV}^\text{tree}[\Phi,\phi]\vert_{\Phi = \Phi_c[\phi]}   \,,
\end{equation} 
where $\Phi$ collectively labels the heavy fields, $\phi$ are the light SM fields and $\Phi_c[\phi]$ means the solutions to the equations of motion (EOM) of the heavy fields in terms of the light fields (see \cite{Cohen:2020fcu} for a review). 
At tree-level, functional matching thus reduces to a simple application of the Euler-Lagrange equation,
\begin{equation}
\label{eq:EulerLagrange}
    \frac{\delta \mathcal{L}}{\delta \theta^{(\star)}} = \partial_\mu \frac{\delta \mathcal{L}}{\delta (\partial_\mu \theta^{(\star)})}\,,
\end{equation}
yielding the classical EOM for the heavy dynamical fields (the flavon $\theta$ in our case).  
To obtain an IR EFT Lagrangian up to a fixed operator dimension, one solves the EOM~\eqref{eq:EulerLagrange} iteratively by expanding the solution $\theta_c [\phi]$ as a series organized in inverse powers of $m_\theta^2$, with each term written in terms of the SM fields $\phi$ assumed to be light compared to $\theta$. 
Substituting the classical power series solution back into the UV Lagrangian and truncating at the desired operator order yields the tree-level-matched EFT.

We apply \eqref{eq:EulerLagrange} to \eqref{eq:LFNgeneral} in the broken \SFN{} phase, in order to match to the SMEFT where \SFN{} is no longer a linearly-realized symmetry.  That is, at scales far above the electroweak scale, but below the \SFN{} breaking scale, we may expand the theory about its classical vacuum, 
 \begin{equation}
 \label{eq:thetavev}
	\langle H\hc H \rangle = 0, \quad \langle \theta^* \theta \rangle = \frac{v_\theta^2}{2} = \frac{\mu_\theta^2}{2\lambda_{02}} \,,
\end{equation}
by writing 
\begin{equation}
\label{eq:thetavevexpand}
	H \rightarrow H, \quad \theta \rightarrow \frac{v_\theta + \vartheta + i\pi_\theta}{\sqrt{2} }\,,
\end{equation}
where $ \vartheta $ is the new, massive CP-even scalar state and $ \pi_\theta $ is a potential CP-odd Nambu-Goldstone boson associated with the \SFN{} symmetry breaking.

At this point we must further discuss the group properties of \SFN{}.  The classic scenario is to take \SFN{} $\sim$ \FN, i.e. a continuous Abelian symmetry which, a priori, can either be global or gauged. 
While the former choice implies that $\pi_\theta$ is classically massless, one might expect it to acquire a small mass through radiative effects, e.g. mixed \FN-gauge-gauge anomalies.\footnote{We refer the reader to e.g. \cite{Calibbi:2016hwq, Ema:2016ops}, where $\pi_\theta$ is identified as the QCD axion, for related studies.}  As such, matching to the SMEFT may be inappropriate when considering a global symmetry-breaking scenario, given the expectation that $\pi_\theta$ will be light (cf. Section \ref{sec:BEYONDABELIAN}).
In the latter case, where \FN{} is gauged, $\pi_\theta$ can be absorbed into the longitudinal component of the family non-universal \FN{} gauge boson, whose mass will be of the order of the symmetry breaking scale $v_\theta$. 
This scenario necessitates the fermionic charges to be fixed such that all gauge anomalies cancel (see e.g. \cite{Bonnefoy:2019lsn,Smolkovic:2019jow}), unless anomaly cancellation is assumed to occur through the Green-Schwartz mechanism \cite{Green:1984sg,Binetruy:1994ru,Ibanez:1994ig,Nir:1995bu},  
and also warrants further study regarding the SMEFT matching effects sourced from integrating out the presumably heavy BSM gauge boson alongside the flavon. 

On the other hand, the dominant flavour mechanism of interest in \eqref{eq:basicidea} (ultimately responsible for solving the flavour puzzle) can just as easily be realized in a \emph{discrete} Abelian scenario.\footnote{Anomaly constraints can also be important in models employing (Abelian or non-Abelian) discrete symmetries \cite{Ibanez:1991hv,Ibanez:1991wt,Banks:1991xj,Araki:2006sqx,Araki:2008ek,Ishimori:2010au,Chen:2015aba,Talbert:2018nkq,Gripaios:2022vvc,Davighi:2022icj}, which arguably must be gauged in the deep UV to avoid quantum gravity (wormhole) constraints \cite{Krauss:1988zc}.}  In this case we can neglect the subtleties surrounding $\pi_\theta$ while preserving the SMEFT matching results associated with the physics of the vev $v_\theta$ and the real component $\vartheta$ of the flavon field, thereby exposing the most pertinent IR signatures of the FN mechanism.  Hence, for the remainder of the section, we will effectively take
\begin{equation}
\mathcal{S}_{\text{FN}} \sim \mathbb{Z}_{\text{FN}}
\end{equation}
and thus content ourselves with the assumption that either {\bf{(1)}} a discrete symmetry may truly be the ultimate UV source of flavour (as can easily be motivated in stringy scenarios --- see e.g. \cite{Baur:2019kwi,Baur:2019iai,Baur:2021bly} for recent progress),  {\bf{(2)}} that the \ZFN{} represents an IR residual symmetry of a (yet further) UV continuous symmetry, perhaps realized at or above $\LUV$,  or that  {\bf{(3)}} the IR physics arising from the flavon Goldstone or the flavourful gauge boson plays a phenomenological `second fiddle' to that of $\vartheta$.  Regardless, in Section \ref{sec:BEYONDABELIAN} we further discuss the generalization of \ZFN{} to both non-Abelian and/or continuous symmetry constructions.

No matter what the symmetry structure is, the matching procedure becomes more complex at the one-loop level despite still remaining methodical. 
Matching the 1LPI effective actions of the IR EFT and the full UV theory at one-loop, one finds, using the method of regions, that only the hard momentum region of the UV effective action contributes to the Wilson coefficients of the EFT
\begin{equation}
\label{eq:loopfunctional1}
    \int d^d \mathcal{L}_\text{EFT}^\text{1-loop} [\phi] = \Gamma_{\text{L,UV}}^\text{1-loop}[\phi] \big\vert_\text{hard}.
\end{equation}
The background field method (see \cite{Abbott:1981ke} for an introduction) is used to write this in a more practical form,
\begin{equation}
\label{eq:loopfunctional2}
    \int d^d x \, \mathcal{L}_\text{EFT}^\text{1-loop} = \frac{i}{2} \log \text{Sdet} \left(- \frac{\partial^2 S_\text{UV}}{\partial \varphi^2} \bigg\vert_{\Phi = \Phi_c[\phi]} \right) \Bigg\vert_\text{hard}\,,
\end{equation}
where $S_\text{UV}$ is the action of the full theory, $\varphi$ stands for both the heavy and light fields and Sdet is a generalization of the functional determinant capturing integration over both bosonic and fermionic fields.
Following on from \cite{Fuentes-Martin:2020udw}, the {\pkg{Matchete}} package \cite{Fuentes-Martin:2022jrf} version 0.1.5 is able to match a UV Lagrangian to the SMEFT at the one-loop level in an automated way using this functional method.
The only manual task is to manipulate the output Lagrangian into the desired physical basis.  Note of course that diagrammatic techniques for one-loop matching (e.g. those embedded in {\tt{Matchmakereft}} \cite{Carmona:2021xtq}) are also applicable, and in what follows we will often intuit the loop-level functional output {\pkg{Matchete}} with UV diagrams that we also use to check many of {\pkg{Matchete}}'s results, particularly those unique to FN models.

Furthermore, besides these normal complications regarding functional matching, there are also additional subtleties in our calculation stemming from the fact that we are matching an EFT to an EFT, as opposed to a renormalizable theory to an EFT.  
For example, our principle expansion parameter is phenomenologically bounded by fermionic mass and mixing textures to $\lambda \sim \mathcal{O}(10^{-1})$, which necessitates considering the IR matching effects of higher-order operators in the UV (see the discussion below).
Additionally, care has to be taken in the treatment of evanescent operators when matching UV models to EFTs at the one-loop-level or higher, and when transforming the results into a physical basis (see \cite{Fuentes-Martin:2022vvu} for details).
The subtleties arise because the complete operator basis in $d= 4-2\varepsilon$ dimensions is infinite dimensional whereas the $d=4$ basis of dimension-6 operators is finite dimensional.
Whether this will be a problem at one-loop hinges on which EFT operators are generated when the heavy new physics is integrated out at the tree-level.
For the FN model in question, the flavon couplings to the SM are such that no evanescent operators are generated up to the operator dimensions reached in this paper.
The absence of such operators means that we can safely relate the redundant set of operators in the \pkg{Matchete} output into the physical Warsaw basis using the Fierz identities.  Indeed, all of our upcoming matching results at mass-dimension 6 are presented in the non-redundant, physical Warsaw basis \cite{Grzadkowski:2010es}.

\subsection{A Toy Model for the Down Quarks}
\label{sec:DOWNTOY}
\begin{table*}[t]
\centering
{\renewcommand{\arraystretch}{1.7}
\begin{tabular}{|c|c|c|c|c|c|c|c|c|}
\hline
\SFN{} Charge & $\overline{Q}_1$ & $\overline{Q}_2$ & $\overline{Q}_3$ & $d_1$ & $d_2$ & $d_3$ & $H$ & $\theta$ \\
\hline \hline
$q$ & 6 & 4 & 0 & 5 & 3 & 3 & $-3$ & $-2$ \\
\hline
\end{tabular}}
\caption{The charge assignments for a toy three-generation FN model in the down-quark sector. 
The subscripts denote the fermion generation. 
}
\label{tab:DownToy}
\end{table*}

We begin by setting up a simple benchmark model of flavour, concentrating only on the down-type quarks and aiming to achieve somewhat realistic Yukawa textures. 
This model, whose charge assignments are displayed in Table~\ref{tab:DownToy}, allows us to illustrate the key features of our results in a concrete way and paves the way for the more abstract discussion of Section \ref{sec:GENERALIZE}.\footnote{We are aware of this toy model thanks to a publicly available talk given by C. Luhn at FLASY 2014. To the best of our knowledge, the earliest use of these textures traces back to \cite{Binetruy:1996xk, Chun:1996xv}, where one can also find more details about the implied mass and mixing phenomenology they generate.}
All SM fields not listed in Table~\ref{tab:DownToy} are assumed to carry unspecified \SFN{} charges such that some renormalizable Yukawa elements are allowed for the charged leptons and up quarks.  
In other words, while we allow FN suppression patterns to arise in these family sectors, we do not assume that FN dynamics \emph{successfully} resolve/predict their flavour structure.

Imposing this set of \SFN{} charges, the down-quark mass matrix and the corresponding mass dimension of the operators required to populate its individual matrix elements are given by
\begin{equation}
\label{eq:toymD}
M^d \simeq \frac{v_H}{\sqrt{2}}
\left(
\begin{array}{ccc}
c^{d_8}_{11}\,\lambda^4 & c^{d_7}_{12}\,\lambda^3 & c^{d_7}_{13}\,\lambda^3 \\
c^{d_7}_{21}\,\lambda^3 & c^{d_6}_{22}\,\lambda^2 & c^{d_6}_{23}\,\lambda^2\\
c^{d_5}_{31}\,\lambda & y^d_{32}\,\lambda^0 & y^d_{33}\,\lambda^0 
\end{array}
\right) \, \Longrightarrow 
\left[\mathcal{O}^d\right] = \left(
\begin{array}{ccc}
8 & 7 & 7 \\
7 & 6 & 6 \\
5 & 4 & 4
\end{array}
\right)\,,
\end{equation}
where it is clear that the FN charges allow for a reduced renormalizable Yukawa sector given by
\begin{equation} \label{eq:toyLdim4}
\mathcal{L}_{\text{SM}^\prime} \supset y^d_{32}\, \overline{Q}_3 H d_2 +  y^d_{33}\, \overline{Q}_3 H d_3 + \text{h.c.}\,,
\end{equation}
in addition to the aforementioned unspecified flavour(s) $\lbrace i, j \rbrace$ of up quark and charged lepton Yukawas, $y^{u}_{ij} \overline{Q}_i \tilde{H} u_j$ and $y^{e}_{ij} \overline{L}_i H e_j$, that turn on at mass-dimension 4.
The superscript $d_k$ in the coefficients $c^{d_k}_{ij}$ above denotes a down-type Yukawa Wilson coefficient which enters at operator dimension $k$ --- the Yukawa element is thus suppressed by $\lambda^{k-4}$.
The exception to this notation is the dimension-5 coefficient $c^{d_5}_{ij}$, which will appear so frequently in our analysis that we leave the dimension label implicit; the coefficient $c^d_{ij}$ will everywhere stand for $c^{d_5}_{ij}$ in what follows. 

The matrix \eqref{eq:toymD} indicates that, in order to be fully consistent, one would need to consider the effective FN Lagrangian corresponding to this toy model up to $\DFN=8$. However, the phenomenological implications of the FN theory beyond \eqref{eq:toymD} are of course most relevant at \emph{lower} orders in $1/\LUV$.  Indeed, we will find in upcoming sections that the IR SMEFT effects sourced from FN operators as low as $\DFN \le 6$ are already extremely rich, and sufficiently challenging to enumerate on a technical level.  This comment is true upon considering both tree-level \emph{and} one-loop UV contributions with FN vertices.  The latter loop contributions are especially important in the context of our study due to the fact that, unlike most matching analyses, we have a relatively fixed hierarchy of new physics scales built-in, thanks to the constraint $\lambda \lesssim \mathcal{O}(10^{-1})$ driven by mass and mixing considerations.  Since $(4\pi)^{-1} \sim 10^{-(1-2)}$, naive power counting tells us that $\lambda/(4\pi) \sim \lambda^2$.  
Assuming that dimensionless Lagrangian parameters are $\mathcal{O}(1)$, this indicates that a tree-level UV contribution sourced from a dimension-$\DFN$ vertex can become competitively large with a one-loop UV contribution sourced from a $\DFN-2$ dimension vertex. 

In what follows we will therefore systematically match our toy model implied in \eqref{eq:toymD} to the $\DSM = 6$ SMEFT, considering up to $\DFN \le 6$ UV operators, and including loop contributions from $\DFN\le5$ operators. When presenting our results, we adopt the notational conventions of \cite{Grzadkowski:2010es} unless otherwise stated.

\subsubsection{Dimension Four ($\normalfont{\DFN=4}$)}
\label{sec:dim4matching}
We first consider our toy model's renormalizable FN Lagrangian, which is given by \eqref{eq:Ltheta} and the \SFN{}-invariant SM.  Working in the broken \SFN{} phase, we expand the flavon field about its vacuum and, ignoring operators without $\vartheta$ or $v_\theta$ dependence, obtain\footnote{The slashed notation here and below, $\slashed{\text{FN}}$, simply indicates interactions in the broken \SFN{} phase.}
\begin{equation}
\label{eq:L4thetaexpanded}
	\mathcal{L}_{\slashed{\text{FN}}}^4 \supset 
	- \frac{1}{2} \vartheta \Box \vartheta 
	- \mu_\theta^2 \vartheta^2 
	- \frac{\mu_\theta^2}{v_\theta} \vartheta^3 
	- \frac{\mu_\theta^2}{4 v_\theta^2} \vartheta^4 
	- \frac{\lambda_{11}}{2} v^2_\theta \left( H\hc H \right) -\lambda_{11} v_\theta \left( H \hc H \right) \vartheta
	- \frac{\lambda_{11}}{2} \left( H\hc H \right) \vartheta^2 \,,
\end{equation}
 which of course includes `static' terms proportional to the flavon vev $v_\theta$ and terms dynamical in $\vartheta$, which we will integrate out with \eqref{eq:EulerLagrange}.  In doing so, we will find familiar results, given that $\mathcal{L}_{\text{FN}}^4$ effectively reduces to the well-studied case of the SM extended by a singlet (complex) scalar, with the only difference being that the Yukawa matrices controlled by FN-type dynamics are mostly populated by zeros at the renormalizable level thanks to \SFN{}.  Therefore we can and will compare our findings with those in (e.g.) \cite{Jiang:2018pbd,Haisch:2020ahr}.
\subsubsection*{Tree-Level Matching}
From \eqref{eq:L4thetaexpanded} we can immediately derive the following EOM for $\vartheta$:
\begin{equation}
\label{eq:ThetaEOMDim4}
	 \left(\Box + 2 \mu_\theta^2 \right) \vartheta =
	 - \lambda_{11} v_{\theta} \left( H\hc H \right) 
	 -\lambda_{11} \left( H\hc H \right) \vartheta
	 - \frac{3 \mu_\theta^2}{v_\theta} \vartheta^2
	 - \frac{\mu_\theta^2}{v_\theta^2} \vartheta^3.
\end{equation}
Rearranging and identifying $ 2 \mu_\theta^2 \equiv m_\theta^2 $, one obtains the (recursive) expression for $ \vartheta $:
\begin{equation}
\label{eq:ThetaRecursiveDim4}
	\vartheta = \frac{-\lambda_{11} v_\theta}{\Box + m_\theta^2} \left( H\hc H \right)
	-\frac{\lambda_{11}}{\Box + m_\theta^2} \left( H\hc H \right) \vartheta 
	- \frac{3 m_\theta^2}{2 v_\theta} \frac{1}{\Box + m_\theta^2}  \vartheta^2
	-\frac{m_\theta^2}{2 v_\theta^2} \frac{1}{\Box + m_\theta^2}  \vartheta^3.
\end{equation}
Following (e.g.) \cite{Ellis:2017jns}, we will solve this expression iteratively, and proceed 
by seeking a power series solution with $ 1 / m_\theta^2 $ the power-counting parameter. We thus write
\begin{equation}
\label{eq:ThetaRecursiveGeneralExpand}
	\vartheta = \vartheta^{\left( 2 \right) } + \vartheta^{(4)} + \vartheta^{(6)} + \ldots 
\end{equation}
which, upon collecting terms and solving, yields
\begin{align}
\nonumber
	\vartheta^{(2)} &= 
    - \frac{\lambda_{11} v_\theta \left( H\hc H \right) }{m_\theta^2} \,,\\
\nonumber
	\vartheta^{(4)} &= 
	- \frac{\lambda_{11}^2 v_\theta}{2 m_\theta^4} \left( H\hc H \right)^2 
	+ \frac{\lambda_{11} v_\theta}{m_\theta^4} \Box \left( H\hc H \right)\,, \\
\label{eq:ThetaIterativeDim4}
    \vartheta^{(6)}	&= 
	- \frac{\lambda_{11}^3 v_\theta}{2 m_\theta^6} \left( H\hc H \right)^3
	+ \frac{\lambda_{11}^2 v_\theta}{2 m_\theta^6} \Box \left( H\hc H \right)^2 + \frac{2 \lambda_{11}^2 v_\theta}{m_\theta^6} \left( H\hc H \right) \Box \left( H\hc H \right) 
	- \frac{\lambda_{11} v_\theta}{m_\theta^6} \Box^2 \left( H\hc H \right) \,.
\end{align}
Substituting the flavon EOM back into \eqref{eq:L4thetaexpanded}, keeping terms up to $\DSM = 6$, and appreciating a number of nice cancellations, one finally obtains 
\begin{equation} 
\label{eq:dim4treematchfinal}
	\mathcal{L}_{\text{SMEFT}} \supset  
     - \frac{\lambda_{11} v_\theta^2}{2} \left( H\hc H \right)
	+ \frac{\lambda_{11}^2 v_\theta^2}{2 m_\theta^2} \left( H\hc H \right)^2
 - \frac{\lambda_{11}^2 v_\theta^2}{2 m_\theta^4} \left( H\hc H \right) \Box \left( H\hc H \right).
\end{equation}
We can make a few important observations from this simple result.
	To begin with, two renormalizable terms, $ H\hc H $ and $ (H\hc H)^2 $, receive contributions from integrating out the flavon.
    This changes the quadratic and quartic Higgs potential terms from
 \begin{equation}
     V(H) \supset - \mu_H^2 \left(H\hc H \right) + \lambda_{20} \left( H\hc H \right)^2
 \end{equation}
 to 
 \begin{equation}
 \label{eq:higgscorrections}
     V(H) \supset - \underbrace{\left(\mu_H^2 - \frac{\lambda_{11} v_\theta^2}{2}  \right)}_{\mu_H'^2} \left(H\hc H \right) + \underbrace{\left(\lambda_{20} - \frac{\lambda_{11}^2 v_\theta^2}{2 m_\theta^2}  \right)}_{\lambda_{20}'} \left( H\hc H \right)^2 \,,
 \end{equation}
	where the shifts may be absorbed into redefinitions of the quadratic and quartic Higgs couplings as shown.

As it stands, the flavon contributes to the Higgs hierarchy problem by shifting the Higgs mass term, as expected for a heavy BSM scalar. 
We do not address the hierarchy problem in this work, although we note that approaches to solving it by suppressing the Higgs-flavon coupling $\lambda_{11}$ simultaneously suppress a number of the IR FN signatures we present in terms of matched SMEFT Wilson coefficients.\footnote{As a concrete example, it may be possible to devise a supersymmetric version of the model where both the Higgs superfields and the flavon superfield are charged under a gauged FN symmetry. 
A small but Dirac natural FN gauge coupling, say $g_\text{F} \sim 10^{-2}$, would result in a $D$-term contribution to the quartic Higgs-flavon coupling of order $\lambda_{11} \sim g_\text{F}^2 \sim 10^{-4}$, producing Higgs mass corrections small enough that flavon masses up to tens of TeV would not imply a significantly fine-tuned Higgs mass. 
The same principle naturally applies to quartic scalar terms derived from $F$-terms. 
Such supersymmetric models would of course have to satisfy a number of phenomenological and model building constraints, but developing them in detail is beyond the scope of this work.
} 
We thank an anonymous referee for this latter comment, and further highlight that such theories would result in a more suppressed SMEFT phenomenology than the generic FN setup.
This is because there are no $\lambda_{11}$-independent Wilson coefficients observed in our tree-level matching results up to $\DFN < 6$ --- cf.\ \eqref{eq:dim4treematchfinal} and results below.  
On the other hand, the fine-tuning implied in \eqref{eq:higgscorrections} is less than 1 in $\mathcal{O}\left(\lambda_{11}\cdot3200\right)$ (i.e.\ not too troubling), assuming a 125 GeV Higgs and a low-scale FN model with $v_\theta~\lesssim~10$~TeV.

	In addition to the renormalizable terms, one dimension-6 operator in the Warsaw basis, the bosonic $ \mathcal{O}_{H \Box} $, is turned on in \eqref{eq:dim4treematchfinal}.
    This operator universally modifies the couplings between the physical Higgs field $h$ and the other SM fields in the broken EW phase and can be interpreted as capturing the mixing between the Higgs and the flavon field.\footnote{
    As shown and explained in \cite{Alonso:2013hga}, after renormalizing the Higgs kinetic term, all \textit{renormalizable} operators containing the Higgs field will see their couplings $\{g_i\}$ modified by a shift $ \delta g_i \propto C_{H,\text{kin}}$ where 
    \begin{equation*}
        C_{H,\text{kin}} \equiv \left( C_{H \Box} - \frac{1}{4} C_{HD}\right) v_H^2 = - \frac{\lambda_{11}^2 v_\theta^2 v_H^2}{2 m_\theta^4}.
    \end{equation*}
    For instance, the trilinear Higgs self-coupling will shift as
    \begin{equation*}
        - \lambda_{20} h^3 \rightarrow - \lambda_{20} \left( 1 + 3 C_{H,\text{kin}} \right) h^3.
    \end{equation*}
    Had we not integrated out the flavon $\vartheta$, this phenomenon would be attributed to Higgs-flavon mixing which makes the scalar mass matrix non-diagonal.
    Diagonalizing the mass matrix requires a redefinition of the physical scalar fields $h$ and $\vartheta$ which modifies the couplings between $h$ and the SM fields by an amount shown above. 
    When accounting for the modified couplings in the case of dimension-6 SMEFT operators, the shift arising from the rescaling of the kinetic term is formally of dimension-8 and can thus be neglected when working at dimension-6 level in the EFT. 
    }
	Another ``Higgs-like operator'' which one may have expected, $ \mathcal{O}_H $, vanishes when we expand about the minimum of the scalar potential in the broken phase. 
	
	Solving for $ \vartheta  $ up to mass-dimension 4 terms and substituting back into the UV Lagrangian of \eqref{eq:L4thetaexpanded}, we find that dependence on $ \vartheta^{(4)} $ cancels out when the series is truncated at dimension-6 terms. The higher dimension term $ \vartheta^{(4)} $ first appears at dimension-8 in the SMEFT.
 
	Hence, before even considering the FN-type operators driving the flavour mechanism of core interest, we see that the presence of the additional \SFN{} symmetry (and its breaking via a scalar flavon) generates non-trivial consequences in the SMEFT.  Indeed, the Wilson coefficient of $\mathcal{O}_{H \Box}$ is already experimentally bounded by available global SMEFT fits \cite{Ellis:2020unq} which, for $\lambda_{11} = \mathcal{O}(1) $ and $m_\theta \sim v_\theta$, imply a lower bound $m_\theta \gtrsim 1 $ TeV on the flavon mass, although we expect bounds on other SMEFT operators to push this limit higher.

 As a final comment, we note that we have checked our analytic result in \eqref{eq:dim4treematchfinal} against the output of {\tt{Matchete}}, which gives 
\begin{align} 
\nonumber
	\mathcal{L}_{\text{SMEFT}} \big\vert_{{\tt{Matchete}}} \supset
	&- \frac{\lambda_{11} v_\theta^2}{2}  H\hc H 
	+ \frac{\lambda_{11}^2 v_\theta^2}{2 m_\theta^2} \left( H\hc H \right) ^2 
	-\frac{\lambda_{11}^2 v_\theta^2 }{m_\theta^4} \left( H\hc H \right) \left( D_\mu H \right)\hc \left( D^\mu H \right) \\
\label{eq:dim4treematchMatchete}
	&-\frac{\lambda_{11}^2 v_\theta^2 }{2 m_\theta^4} \left( H\hc H \right) \left( H\hc D_\mu D^\mu H + \text{h.c.} \right)
\end{align}
upon asking that \eqref{eq:L4thetaexpanded} be matched to the $\DSM=6$ SMEFT.
The reason this looks different from our analytic result calculated in \eqref{eq:dim4treematchfinal} is that both of the two-derivative bosonic operators listed here are not in the Warsaw basis, whereas our result from above is.
However, using the standard IBP identity
\begin{equation} \label{eq:Higgs_IBP}
 \bigl( H\hc H \bigr) \left( D_\mu H \right)\hc \bigl( D^\mu H \bigr) 
 = \frac{1}{2} \bigl( H\hc H \bigr) \left( \Box \bigl( H\hc H \bigr) - H\hc D_\mu D^\mu H - \left( D_\mu D^\mu H \right)\hc H \right) \,,
\end{equation}
we may exchange the bosonic two-derivative operators in \eqref{eq:dim4treematchMatchete} for $ (H\hc H) \Box ( H\hc H ) $ to arrive at the same result, thus completing our crosscheck and giving us further confidence in {\tt{Matchete}}'s functionality, which we will now use for loop-level matching.\footnote{In what follows we will perform this identity and others (e.g. Fierz identities) by hand, in order to translate {\tt{Matchete}} output to the Warsaw basis for all operators that appear.}

\subsubsection*{One-Loop Matching}
Following the functional algorithm briefly reviewed around \eqref{eq:loopfunctional1}--\eqref{eq:loopfunctional2}, we have allowed {\tt{Matchete}} to calculate the one-loop contributions sourced by matching \eqref{eq:L4thetaexpanded} to the $\DSM \le 6$ SMEFT.  
In so doing we have uncovered a host of further non-trivial operator structures in the Yukawa, Higgs-scalar, Higgs-gauge, Higgs-derivative, Higgs-kinetic, and four-fermion sectors.  
For the remainder of this work, we use $\hbar$ as shorthand for the loop suppression factor $(4 \pi)^{-2}$, a choice that follows the convention in \pkg{Matchete}.

We first consider contributions to the (Higgs-enhanced) \emph{Yukawa} sector ($\psi^2 H^{1+\DSM-4}$), including both modifications to the renormalizable SM terms and higher-dimensional SMEFT operators:
\begin{align}
\nonumber
&\mathcal{L}_{\text{SMEFT}} \supset \frac{\hbar  \lambda_{11}^2  v_\theta^2 y_{ij}^e}{12 m_\theta^4}\left[3 m_\theta^2 - m_H^2\left(17 + 6 \mathbb{L}\right)\right]\overline{L}_i H e_j +\frac{\hbar \lambda_{11}^2 v_\theta^2 y^d_{ij}  }{12 m_\theta^4} \left[ 3 m_\theta^2 - m_H^2 (17 + 6 \mathbb{L} ) \right] \overline{Q}_i H d_j \\
\nonumber
&  +\frac{\hbar y^u_{ij}\lambda_{11}^2 v_\theta^2 }{12 m_\theta^4} \bigl[ 3 m_\theta^2 - m_H^2 (17 + 6 \mathbb{L} ) \bigr] \overline{Q}_i \tilde{H} u_j + \frac{\hbar \lambda_{11}^2 v_\theta^2 }{36 m_\theta^4} \Bigl[ 9 (y^e y^{e\dagger} y^e)_{ij}  (5 + 6 \mathbb{L}) + 6 y^e_{ij} \lambda_{20}' (29 + 6 \mathbb{L})\\
\nonumber
& - y^e_{ij} g^2  (31 + 30 \mathbb{L}) \Bigr] \left(H^\dagger H\right)\overline{L}_i H e_j + \frac{\hbar \lambda_{11}^2 v_\theta^2 y^d_{ij}}{36 m_\theta^4} \left[9 \left( \lvert y^d_{32} \rvert^2 + \lvert y^d_{33} \rvert^2 \right) (5 + 6 \mathbb{L}) \right. \\
\nonumber
&\left. - g^2 ( 31 + 30 \mathbb{L}) + 6 \lambda_{20}' (29 + 6 \mathbb{L} )  \right] \left(H^\dagger H\right)\overline{Q}_i H d_j 
+ \frac{\hbar \lambda_{11}^2 v_\theta^2 }{36 m_\theta^4} \left[ 9 (y^u y^{u \dagger} y^u)_{ij}   (5 + 6 \mathbb{L} ) \right. \\
&\left. + 6 y^u_{ij} \lambda_{20}' (29 + 6 \mathbb{L} ) - y^u_{ij} g^2  (31 + 30 \mathbb{L} ) \right]\left(H^\dagger H\right)\overline{Q}_i \tilde{H} u_j \,, 
\label{eq:dim4loopmatchYukawa}
\end{align}
where $m_H$ is the Higgs mass parameter in the UV Lagrangian and $\mathbb{L} \equiv \ln (\mu^2/m_\theta^2)$ with $\mu$ the $\overline{\text{MS}}$ renormalization/matching scale). We also see in \eqref{eq:dim4loopmatchYukawa} the first manifestation of our \SFN{} charges from Table \ref{tab:DownToy}, in that generic flavour indices appear on up-quark and charged-lepton coefficients, whereas specific indices ((32) and (33)) appear on the down quarks.

There are also contributions to the purely bosonic sector of the $\DSM\le6$ SMEFT, including augmented \emph{Higgs-scalar potential} and \emph{Higgs-derivative} $ (H^4 D^2) $ terms given by
\begin{align}
    \label{eq:dim4loopmatchHiggsDerivative}
	\nonumber
	& \mathcal{L}_{\text{SMEFT}}  \supset \frac{\hbar \lambda_{11}^2}{9 m_\theta^8} \biggl[ 6 \lambda_{11} m_\theta^6 -6 \lambda_{11}^4 v_\theta^6   + 18\lambda_{11}^2 v_\theta^4 m_\theta^2   ( \lambda_{11} - 6 \lambda_{20}' )  \\
		\nonumber
	& + v_\theta^2 m_\theta^4 \Bigl( - 18 \lambda_{11}^2 + 108 \lambda_{20}' \lambda_{11}+  \lambda_{20}' \bigl( 12 \lambda_{20}' (41 + 30 \mathbb{L} ) - g^2 (31 + 30 \mathbb{L} )   \bigr) \Bigr) \biggr]\left(\hdh \right)^3 +\\ 
	\nonumber
	& \left\{ - \frac{\lambda_{11}^2 v_\theta^2}{2 m_\theta^4} + \frac{\hbar \lambda_{11}^2}{12 m_\theta^2} \bigg[ 16 - 18 \mathbb{L}  - \frac{2 \lambda_{11} v_\theta^2 }{ m_\theta^2} \left( 7  - 12 \mathbb{L}  \right) + \frac{\lambda_{11}^2 v_\theta^4 }{m_\theta^4} \left( 19 + 12 \mathbb{L}  \right) - \frac{g_Y^2 v_\theta^2}{6 m_\theta^2} \left( 31 + 30 \mathbb{L}  \right) \right.\\
	&   \left. - \frac{g^2 v_\theta^2}{2 m_\theta^2} \left( 31 + 30 \mathbb{L} \right)  \bigg] \right\} \left( \hdh \right) \Box \left( \hdh \right) - \frac{\hbar g_Y^2 v_\theta^2 \lambda_{11}^2}{18 m_\theta^4 } \left[ 31 + 30 \mathbb{L}  \right]( H\hc D^\mu H )\hc ( H\hc D_\mu H )\,,
\end{align}
as well as mixed \emph{Higgs-gauge} ($H^2 X^2$) effective operators turning on 
\begin{equation}
\label{eq:dim4loopmatchHiggsGauge}
\mathcal{L}_{\text{SMEFT}} \supset \frac{\hbar \lambda_{11}^2 v_\theta^2}{12 m_\theta^4} \left[\left(H^\dagger H\right) \left(g_Y^2 B^{\mu\nu}B_{\mu\nu} + g^2 W^{I\mu\nu}W^I_{\mu\nu}\right) + 2 g g_Y\left(H^\dagger \tau^I H\right)B^{\mu\nu}W^I_{\mu\nu} \right]
\end{equation}
with hypercharge ($B$), weak isospin ($W$), and mixed ($BW$) field-strength tensors appearing.   

Additional \emph{Higgs-kinetic terms} with non-trivial fermion structure ($ \psi^2 H^2 D $) also appear,
\begin{align}
\nonumber
    &\mathcal{L}_{\text{SMEFT}} \supset \frac{\hbar \lambda_{11}^2 v_\theta^2 }{144 m_\theta^4}\Big[ 9 \left( \lvert y^d_{32} \rvert^2 + \lvert y^d_{33} \rvert^2 \right) \delta_{3i} \delta_{3j} \left(5+ 2\mathbb{L}\right) - 9 \left( y^u y^{u\dagger} \right)_{ij} \left(5+ 2\mathbb{L}\right) \\
\nonumber
    & - \frac{g_Y^2}{3} \left(17 + 6 \mathbb{L}\right)\delta_{ij}\Big] \left(H^\dagger \Darr H\right) \overline{Q}_i \gamma^\mu Q_{j} +  \frac{\hbar v_\theta^2 \lambda_{11}^2}{144 m_\theta^4}\Big[9 \left( \lvert y^d_{32} \rvert^2 + \lvert y^d_{33} \rvert^2 \right) \delta_{3i} \delta_{3j} \left(5+ 2\mathbb{L}\right) 
    \\
    \nonumber
   & + 9 \left( y^u y^{u\dagger} \right)_{ij} \left(5+ 2\mathbb{L}\right)  - g^2 \delta_{ij} \left(17 + 6 \mathbb{L}\right)\Big] \left(H^\dagger \DarrI H\right) \overline{Q}_i \tau^I \gamma^\mu Q_{j} \\
\nonumber
    &+ \frac{\hbar \lambda_{11}^2 v_\theta^2 }{216 m_\theta^4}\Big[ g_Y^2 \delta_{ij}\left(17+ 6 \mathbb{L}\right)  - 27 \left(y^{d\dagger}y^d\right)_{ij} \left(5 + 2\mathbb{L}\right)   \Big] \left(H^\dagger \Darr H\right) \overline{d}_{i} \gamma^\mu d_j \\
    \nonumber
    &- \left\{ \frac{ \hbar \lambda_{11}^2 v_\theta^2 }{4 m_\theta^4}\Big[\left(y^{u\dagger}y^d\right)_{ij}\left(5+ 2\mathbb{L}\right) \Big] \left(\tilde{H}^\dagger i D_\mu H\right) \overline{u}_{i} \gamma^\mu d_j + \text{h.c.} \right\}\\
    \nonumber
    &- \frac{\hbar \lambda_{11}^2 v_\theta^2}{216 m_\theta^4} \left[ 2 g_Y^2 \delta_{ij} (17 + 6 \mathbb{L} ) - 27 ({y^u}^{\dagger} y^u )_{ij} (5 + 2 \mathbb{L})\right] ( H\hc \Darr H ) ( \overline{u}_i \gamma^\mu u_j)  \\
    \nonumber
     &- \frac{ \hbar \lambda_{11}^2 v_\theta^2}{72 m_\theta^4} \left[ 9 (y^{e \dagger} y^e )_{ij} (5 + 2 \mathbb{L} ) - g_Y^2 \delta_{ij} (17 + 6 \mathbb{L} )  \right] ( H\hc \Darr H ) (\overline{e}_i \gamma^\mu e_j ) \\
     \nonumber
      &+ \frac{\hbar \lambda_{11}^2 v_\theta^2}{144 m_\theta^4} \left[ 9 (y^e y^{e\dagger} )_{ij} (5 + 2 \mathbb{L} ) + g_Y^2 \delta_{ij}(17 + 6 \mathbb{L} ) \right] ( H\hc \Darr H ) ( \overline{L}_i \gamma^\mu L_j) \\
      &+ \frac{\hbar \lambda_{11}^2 v_\theta^2}{144 m_\theta^4} \left[ 9 (y^e y^{e\dagger} )_{ij} (5 + 2 \mathbb{L} ) - g^2 \delta_{ij}(17 + 6 \mathbb{L}) \right] ( H\hc  \DarrI H ) ( \overline{L}_i \tau^I \gamma^\mu L_j) \,,
     \label{eq:dim4loopmatchHiggsKinetic}
\end{align}
where the product of down-type Yukawa matrices can be written in matrix form as 
\begin{equation}
\label{eq:toyModelYDaggerY}
	\left( y^{d\dagger} y^d \right)_{ij} = 
	\begin{pmatrix} 
		0 & 0 & 0 \\
		0 & \lvert y^d_{32} \rvert^2 & y^{d*}_{32} y^d_{33}  \\
		0 & y^d_{32} y^{d*}_{33} & \lvert y^{d}_{33} \rvert^2 \\
	\end{pmatrix} 
\end{equation}
for the toy model in question.
To help with intuitive understanding, Fig.~\ref{fig:dim4Loops}\footnote{All the Feynman diagrams in this paper have been drawn using \pkg{TikZ-Feynman} \cite{Ellis:2016jkw}.} presents some Feynman diagrams that contribute to the Wilson coefficient of the operator $ ( H\hc \Darr H ) \left( \overline{d}_i \gamma^\mu d_j  \right)$ upon integrating out the flavon.

\begin{figure}[th!]
	\usetikzlibrary {arrows.meta} 
	\centering
    \begin{subfigure}[t]{\textwidth}
        \centering
	\raisebox{-0.45\height}{
	\begin{tikzpicture}
		\begin{feynman}
			\vertex  (a);
			\vertex [left=1.5cm of a] (i1) {$ H\hc  $};
			\vertex [right=2cm of a] (b);
			\vertex [right=1.5cm of b] (f1) {$ d_j $};
			\vertex [below=2cm of a] (c);
			\vertex [below=2cm of b] (d);
			\vertex [left=1.5cm of c] (i2) {$ H $};
			\vertex [right=1.5cm of d] (f2) {$ \overline{d}_i $};

			\diagram* {
				(i1) -- [anti charged scalar] (a) -- [anti charged scalar, edge label = $ H\hc $] (b) -- [fermion] (f1) [particle = $ d_i $],
				(a) -- [dashed, double, edge label = $  \vartheta$] (c),
				(i2) -- [charged scalar] (c) -- [charged scalar, edge label = $ H $] (d) -- [anti fermion] (f2),
				(b) -- [anti fermion, edge label = $ Q_k $] (d),
			};
		\end{feynman}
	\end{tikzpicture}
	}
	\qquad
	\qquad
	\raisebox{-0.59\height}{
	\begin{tikzpicture}
		\begin{feynman}
			\vertex  (a);
			\vertex [left=1.5cm of a] (i1) {$ H\hc  $};
			\vertex [right=2cm of a] (b);
			\vertex [right=1.5cm of b] (f1) {$ d_j $};
			\vertex [below=2cm of a] (c);
			\vertex [below=2cm of b] (d);
            \vertex [left=1cm of d] (e);
            \vertex [below right=1cm of e] (j3) {$B_\mu$};
			\vertex [left=1.5cm of c] (i2) {$ H $};
			\vertex [right=1.5cm of d] (f2) {$ \overline{d}_i $};

			\diagram* {
				(i1) -- [anti charged scalar] (a) -- [anti charged scalar, edge label = $ H\hc $] (b) -- [fermion] (f1) [particle = $ d_i $],
				(a) -- [dashed, double, edge label = $  \vartheta$] (c),
				(i2) -- [charged scalar] (c) -- [charged scalar, edge label = $ H $] (e) -- [charged scalar, edge label = $H$] (d) -- [anti fermion] (f2),
				(b) -- [anti fermion] (d),
			    (e) -- [photon] (j3),   
			};
		\end{feynman}
	\end{tikzpicture}
	}
	\caption{These two diagrams give a contribution to the Wilson coefficient  $ C_{Hd} $ proportional to $ \lambda_{11}^2 v_\theta^2 ({y^d}\hc y^d)_{ij} / m_\theta^4  $. 
	}
	\end{subfigure}
  \begin{subfigure}[t]{\textwidth}
  \centering
  	\begin{equation*}
	\begin{tikzpicture}[baseline=-1.1cm]
		\begin{feynman}
			\vertex (i1) {$ H $};
			\vertex [below=2cm of i1] (i2) {$ H\hc $};
			\vertex [right=1.8cm of i1] (a);
			\vertex [right=1.8cm of i2] (b);
			\vertex at ($(a) + (-30:2cm) $) (c);
			\vertex [right=1.4cm of c] (f1) {$ B^\mu $};

		\diagram* {
			(i1) -- [charged scalar] (a) -- [charged scalar] (c) -- [charged scalar] (b) -- [charged scalar] (i2),
			(a) -- [dashed, double, edge label' = $ \vartheta $] (b),
			(c) -- [boson] (f1),
		};
		\end{feynman}
		  \draw[->]        (6,-1)   -- (7.3,-1);
	\end{tikzpicture}
 \quad
	\left( H\hc \Darr H \right) \p^\nu B_{\nu \mu} 
	\stackrel{\text{EoM}}{=}
	q_\psi g_Y \left( H\hc \Darr H \right)  \left( \overline{d}_i \gamma_\mu d_i \right) 
	\end{equation*}
	\caption{A diagram contributing to the term proportional to $ g_Y^2 $ in the Wilson coefficient $C_{Hd}$.}
 \end{subfigure}
 \caption{Some diagrams contributing to the Wilson coefficient $C_{Hd}$ of the Higgs-kinetic operator $( H\hc \Darr H ) \left( \overline{d}_i \gamma^\mu d_j  \right)$ coming from the dimension-4 UV Lagrangian at one-loop.
 \label{fig:dim4Loops} 
 }
\end{figure}
	
One also finds a number of \emph{four-fermion} ($\psi^4$) operator classes:
\begin{align}
\nonumber
\mathcal{L}_{\text{SMEFT}} &\supset \left\{ \frac{\hbar \lambda_{11}^2 v_\theta^2 y^{e}_{ij} y^{d\dagger}_{kl}  }{6 m_\theta^4}\left(\overline{L}^\alpha_i e_j\right) \left( \overline{d}_{k} Q_l^\alpha\right) 
+ \frac{\hbar \lambda_{11}^2 v_\theta^2 y^e_{ij} y^u_{kl}  }{6 m_\theta^4}\, \epsilon_{\alpha \beta}\left(\overline{L}^{\alpha}_i e_j\right) \left(\overline{Q}^{\beta}_k u_l\right) \right. \\
\nonumber
& \left. + \frac{\hbar \lambda_{11}^2 v_\theta^2 y^u_{ij} y^d_{kl}  }{6 m_\theta^4} \epsilon_{\alpha \beta}\left(\overline{Q}^{\alpha}_{i} u_j\right) \left(\overline{Q}^{\beta}_{k} d_l\right) + \text{h.c.} \right\}
- \frac{\hbar \lambda_{11}^2 v_\theta^2 y^{e}_{il}y^{e\dagger}_{kj}  }{12 m_\theta^4}(\overline{L}_i \gamma_\mu L_j)  (\overline{e}_{k} \gamma^\mu e_l) \\
\label{eq:dim4loopmatchFourFermi}  
&- \frac{\hbar \lambda_{11}^2 v_\theta^2 y^d_{il} y^{d\dagger}_{kj}   }{12 m_\theta^4}(\overline{Q}_i \gamma_\mu Q_j) (\overline{d}_{k} \gamma^\mu d_l)
- \frac{\hbar \lambda_{11}^2 v_\theta^2 y^u_{il} y^{u\dagger}_{kj} }{12 m_\theta^4}(\overline{Q}_i \gamma_\mu Q_j)  (\overline{u}_{k} \gamma^\mu u_l),
\end{align}
where $\epsilon_{\alpha\beta}$ is the antisymmetric SU(2)$_\text{L}$ tensor. 

The one-loop matching coefficients presented in this section already exist in the literature in the context of the Standard Model singlet extension, obtained using functional methods \cite{Jiang:2018pbd,Jiang:2018pbd-err} and diagrammatic techniques \cite{Haisch:2020ahr}.
Our results from \pkg{Matchete} are in full agreement with those in the references once we account for the fact that we have restricted the form of the down-type Yukawa matrices in a flavour non-universal way.
The restriction of these Yukawas to only the (33) and (32) elements leads to patterns in family space best illustrated by equations~\eqref{eq:dim4loopmatchYukawa} and \eqref{eq:dim4loopmatchHiggsKinetic}--\eqref{eq:dim4loopmatchFourFermi}.
We see for instance that the Wilson coefficient $(C_{Hd})_{ij}$ of the Higgs-kinetic operator with down-type quarks, understood as a matrix in generation space, contains a fully populated $2 \times 2$ block shown partly in \eqref{eq:toyModelYDaggerY} and a (11) element proportional to $g_Y^2$.
The other elements in the first row and columnn are zero.
This flavour pattern, determined by the charge assignments, can be recognised as a prediction of the toy model.

In conclusion, we find that, before even considering the $\theta$-dependent operators of principal interest to the FN mechanism (those populating the Yukawa matrix in \eqref{eq:toymD}), a wide array of SMEFT operator classes appear already from $\DFN=4$ UV contributions.
Many of the Wilson coefficients that appear contain patterns in generation space dictated by the renormalizable Yukawa elements.
In fact, we will find that tree- and loop-level corrections sourced from $\DFN=5$ operators will only introduce a few  more operator classes beyond those found in this subsection, although the associated Wilson coefficients of most of the operators appearing between \eqref{eq:dim4loopmatchYukawa}--\eqref{eq:dim4loopmatchFourFermi} will also be augmented by higher-order corrections.

\subsubsection{Dimension Five ($\normalfont{\DFN=5}$)}
\label{sec:dim5matching}
At $\DFN=5$ one only has the Weinberg operator \cite{Weinberg:1979sa} and the effective Yukawa operator implied in \eqref{eq:toymD},
\begin{equation}
\label{eq:toyLdim5}
\mathcal{L}^5_{\text{FN}} = \frac{c_{ij}^W}{\Lambda_{\text{UV}}} \left(H \overline{L}^c_i\right) \left(L_j H\right)+ \frac{c^d_{31}}{\Lambda_{\text{UV}}}\, \overline{Q}_3 H d_1 \theta + \text{h.c.},
\end{equation}
assuming that the (unspecified) \SFN{}-charge of the SU(2)$_\text{L}$ lepton doublet $L$ respects $q_{L_i} + q_{L_j}\overset{!}{=}-2q_H$ and that the \SFN{} charges of the up quarks and/or charged leptons do not allow for a leading $\DFN=5$ contribution to the Yukawa matrix.\footnote{
We will discuss FN charge generalizations to the lepton sector in Section \ref{sec:GENERALIZE} below.} 
Substituting the expanded flavon field into this Lagrangian and ignoring operators with no $\vartheta$ dependence results in
\begin{equation} 
\label{eq:L5FNexpanded}
	 \mathcal{L}_{\slashed{\text{FN}}}^5 \supset
	c^d_{31} \,\lambda\,\overline{Q}_3 H d_1 +  \frac{c^d_{31}}{\sqrt{2} \Lambda_\text{UV}} \overline{Q}_3 H d_1 \vartheta + \text{h.c.}
\end{equation}
 The first term is of course the operator driving the core FN mechanism, in that it populates the Yukawa couplings with a non-zero (31) matrix element, suppressed by one power of $\lambda~=~v_\theta / (\sqrt{2} \LUV)$, while the second term allows for a propagating $\vartheta$ that again needs to be integrated out in order to make contact with the SMEFT.

\subsubsection*{Tree-Level Matching}

Following exactly the same procedure as between \eqref{eq:ThetaEOMDim4}--\eqref{eq:ThetaIterativeDim4} above, one quickly arrives at the tree-level $\DFN=5$ contribution to the $\DSM=6$ SMEFT,
\begin{equation} 
\label{eq:dim5treematchfinal}
	\mathcal{L}_{\text{SMEFT}} \supset  
	-   \frac{\lambda_{11} \lambda \, c^d_{31}}{m_\theta^2} \left(H\hc H \right) \overline{Q}_3 H d_1  + \text{h.c.} ,
\end{equation}
which is of course understood to be in \emph{addition} to those operators found in Section \ref{sec:dim4matching}.
We see that the enhanced Yukawa-sector operator  $ \mathcal{O}_{d H} $ of the Warsaw basis is turned on, with a distinctive Wilson coefficient suppressed both by $\lambda$ and two powers of the flavon mass $m_\theta^2$.  Additionally, the presence of $\lambda_{11}$ and $c^d_{31}$ indicates that this term is sourced by an interplay between scalar potential (Higgs-flavon mixing) and Yukawa potential EFT operators. 
As Fig.~\ref{fig:dim5-tree-level-enhanced-yuk} shows, this is easy to understand in terms of diagrammatic matching.
The contribution arises from connecting a $\hdh \vartheta$ vertex of \eqref{eq:L4thetaexpanded} to the second vertex of \eqref{eq:dim5treematchfinal} via an internal flavon propagator, yielding the local $d_\text{SM} = 5$ term in \eqref{eq:dim5treematchfinal}.
 We have again checked \eqref{eq:dim5treematchfinal} with {\pkg{Matchete}}, finding perfect agreement.

\begin{figure}[htpb]
    \centering
	\feynmandiagram[horizontal = c to d, baseline=-1.1cm ]{
		a [particle = $H$] --[charged scalar] c --[charged scalar] b [particle = $H^\dagger$],
		c --[double, dashed, edge label = $\vartheta$] d,
		e [particle = $ \overline{Q}_3 $] --[fermion] d --[charged scalar] f [particle  = $ H $],
		d   --[fermion] g [particle = $ d_1 $],
	};
	\begin{tikzpicture} [baseline=1.15cm]
		  \draw[->]        (0,1.2)   -- (1,1.2);
	\end{tikzpicture}
	$
	\quad
C_{dH}^{31}\left( H^{\dagger}H \right) \overline{Q}_3 H d_1  
	$
    \caption{A tree-level flavon exchange, leading to an enhanced Yukawa operator in the Warsaw basis. 
    Note that above the electroweak scale, the Higgs field ($H$) carries hypercharge $Y=1/2$, which is denoted by the arrows in the external Higgs legs.
    }
    \label{fig:dim5-tree-level-enhanced-yuk}
\end{figure}

As a final comment for intuitive readers, one also obtains Higgs-enhanced four-fermions ($H^2 \psi^4$) from $\DFN=5$ vertices at tree-level (via a single internal $\vartheta$ propagator), but these represent $\DSM=8$ SMEFT operators (which we briefly discuss in our Summary and Outlook). 
At low energies, upon the Higgs acquiring a vev, these will turn into Low-Energy EFT (LEFT) \cite{Jenkins:2017jig,Dekens:2019ept} operators $\propto v_H^2 \psi^4$.

\subsubsection*{One-Loop Matching} 
We now move to our final loop-level analysis by enumerating the contributions to IR SMEFT operators coming, in the diagrammatic interpretation, from loops that can be drawn with both renormalizable and $\DFN=5$ vertices.  We again use {\pkg{Matchete}} to perform this analysis using the functional matching method, and later do analytic computations of individual diagrams to crosscheck the results.

Unsurprisingly, we find that \eqref{eq:toyLdim5} does not source any novel contributions to the purely bosonic sector of the SMEFT beyond those already presented in Section \ref{sec:dim4matching}.  We do of course find additional contributions to the enhanced Yukawa sector given by 
\begin{align} \label{eq:dim5loopmatchYukawa}
\nonumber
&\mathcal{L}_{\text{SMEFT}} \supset -   \frac{\hbar  c_{31}^d\lambda \lambda_{11} }{4 m_\theta^2}\biggl[ 6 \left( \lvert y^{d}_{32} \rvert^2 + \lvert y^d_{33} \rvert^2 \right) ( 1 + \mathbb{L} ) - 6 \frac{m_\theta^2}{v_\theta^2} (1 - \mathbb{L} ) -  \frac{\lambda_{11}^2 v_\theta^2}{m_\theta^2} (15 + 4 \mathbb{L} ) \\
& + 2 \lambda_{11} (4 - 6 \mathbb{L}) + \lambda_{20}' (7 + 6 \mathbb{L})  \biggr] \left(H^\dagger H\right) \overline{Q}_3 H d_1 + \frac{\hbar y^d_{kl}}{144  m_\theta^2 } \biggl[ 18  \frac{\lambda^2 m_\theta^2 \lvert c_{31}^d \rvert^2}{v_\theta^2}  (1 + 2 \mathbb{L})   \\
\nonumber
& + \frac{ 4 \lambda_{11}^2 v_\theta^2}{m_\theta^2} \left( 9  \left( \lvert y^{d}_{32} \rvert^2 + \lvert y^d_{33} \rvert^2 \right) (5 + 6 \mathbb{L}) -  g^2 (31 + 30\mathbb{L}) + 6 \lambda_{20}' (29 + 6 \mathbb{L})  \right) \biggr] \left( \hdh \right) \overline{Q}_k H d_l  + \text{h.c.},
\end{align}
where $ (k,l )  \neq (3,1) $, as well as contributions to Higgs-kinetic operators
\begin{align} \label{eq:dim5loopmatchHiggsDerivative}
\nonumber
&\mathcal{L}_{\text{SMEFT}} \supset \frac{\hbar}{8 m_\theta^2} \left[ 4 \lambda \lambda_{11} \left( {c^d}\hc y^d + {y^d}^{\dagger} c^d \right)_{ij} + \frac{2 \lambda ^2 m_\theta^2 \lvert c_{31}^d \rvert^2}{v_\theta^2}  \delta_{i1} \delta_{j1} (1 + 2 \mathbb{L}) \right]   \left(H^\dagger \Darr H\right)\left(\overline{d}_{i} \gamma^\mu d_j\right) \\
\nonumber
&+ \left\{ \frac{\hbar \lambda \lambda_{11}}{2 m_\theta^2} \left( {y^u}^{\dagger} c^d \right)_{ij}  ( \tilde{H}\hc iD_\mu H  ) ( \overline{u}_i \gamma^\mu d_j ) + \text{h.c.} \right\} - \frac{\hbar \lambda^2 \lvert c_{31}^d \rvert^2}{16 v_\theta^2}  \delta_{i3} \delta_{j3} (1 + 2 \mathbb{L}) \left(H^\dagger \Darr H\right)\left(\overline{Q}_i \gamma^\mu Q_j\right) \\
& - \frac{\hbar \lambda^2 \lvert c_{31}^d \rvert^2}{16 v_\theta^2} \delta_{i3} \delta_{j3} (1 + 2 \mathbb{L})  \left(H^\dagger  \DarrI H\right) \left(\overline{Q}_i \tau^I \gamma^\mu Q_j\right).
\end{align}
It is again instructive to show the emergent flavour structure for the toy model defined in Table~\ref{tab:DownToy}.
We may write the product of flavour space matrices on the first line of \eqref{eq:dim5loopmatchHiggsDerivative} as 
\begin{equation}
    \left( {c^d}\hc y^d + {y^d}^{\dagger} c^d \right)_{ij} 
    = 
    \begin{pmatrix} 
		0 &  c_{31}^{d*} y^d_{32} &  c_{31}^{d*} y^d_{33}\\
		c_{31}^d y^{d*}_{32} & 0 & 0  \\
		c_{31}^d y^{d*}_{33} & 0 & 0 \\
	\end{pmatrix},
\end{equation}
whereas ${y^u}^{\dagger} c^d$ from the line that follows is represented by
\begin{equation}
    \left( y^{u \dagger} c^d \right)_{ij}
    =
    \begin{pmatrix} 
		c_{31}^d y^{u*}_{31} &  0 &  0\\
		c_{31}^d y^{u*}_{32} & 0 & 0  \\
		c_{31}^d y^{u*}_{33} & 0 & 0 \\
	\end{pmatrix}.
\end{equation}
Note again that we have not made any assumptions about the up-type Yukawa matrix elements $y^u_{ij}$ of the renormalizable Lagrangian.
Finally there are also four-fermion operators,
\begin{align}
\nonumber
\mathcal{L}_{\text{SMEFT}} &\supset \left\{ -\frac{\hbar \lambda \lambda_{11} y^e_{ij}c^{d*}_{13}}{2 m_\theta^2}\left(\overline{L}^\alpha_i e_{j}\right) \left( \overline{d}_{1} Q_3^\alpha\right)  
 -\frac{\hbar \lambda \lambda_{11} y^u_{ij}c^d_{31}}{2 m_\theta^2} \epsilon_{\alpha \beta}\left(\overline{Q}^{\alpha}_{i} u_j\right) \left(\overline{Q}^{\beta}_{3} d_1\right) + \text{h.c.} \right\} \\
  \label{eq:dim5loopmatchFourFermion}
&- \frac{\hbar \lambda^2 \lvert c^d_{31} \rvert^2 }{2 v_\theta^2} \left( 1 + \mathbb{L} \right) \left( \overline{Q}_3 \gamma_\mu Q_3 \right) \left( \overline{d}_1 \gamma^\mu d_1 \right) \,.
\end{align}
Hence \eqref{eq:dim5loopmatchYukawa}--\eqref{eq:dim5loopmatchFourFermion} complete the modifications to SMEFT Wilson coefficients sourced from ${\DFN=5}$ loop corrections. 
As in Section \ref{sec:dim4matching}, in Fig.~\ref{fig:dim5loops} we have sketched a number of diagrams that allow us to physically intuit the {\tt{Matchete}} output in \eqref{eq:dim5loopmatchYukawa}--\eqref{eq:dim5loopmatchFourFermion}, and we have explicitly calculated a subset of those diagrams, confirming the associated Wilson coefficients that appear above.

It is worth pausing at this stage to study the results.
Firstly, we see that the inclusion of a single FN term in the dimension-5 Lagrangian generates substantial flavour structure in the Higgs-enhanced Yukawa, Higgs-kinetic and four-fermion operators of the SMEFT output.
For the Wilson coefficient $ (C_{Hd})_{ij} $ associated with the Higgs-kinetic operator with down-type quarks $  (H^\dagger \Darr H)(\overline{d}_{i} \gamma^\mu d_j)$ in \eqref{eq:dim5loopmatchHiggsDerivative}, a $ 2 \times 2 $ block of zero elements in flavour space is predicted, whilst the first column and row are populated.
These findings are hardly a surprise, given that the renormalizable terms of \eqref{eq:toyLdim4} and the dimension-5 term in \eqref{eq:toyLdim5} single out flavoured pairs of quarks which, after the flavon is integrated out, are mapped to SMEFT operators with flavour-non-universal Wilson coefficients.
Nonetheless, the results demonstrate that at a fixed operator dimension in the SMEFT, the terms associated with FN dynamics are capable of generating 
distinguishable flavour patterns at the one-loop level.

To gain a better understanding of the patterns in the Wilson coefficients, we combine the results obtained above with those coming out of the ${\DFN = 4} $ part of the Lagrangian in Section~\ref{sec:dim4matching}.
We may illustrate this by again considering the Wilson coefficients $ (C_{Hd})_{ij} $ of the Higgs-kinetic operator with a down-current.
Adding the Wilson coefficients from the two stages of matching gives 

{{\scriptsize
	\begin{align} 
		\nonumber
	&\left( C_{Hd} \right) = \dfrac{\hbar}{2 m_\theta^2} \times\\
	& 
	\begin{pmatrix}
		\dfrac{54 \lambda^2 \left( 1 + 2 \mathbb{L} \right) m_\theta^4 \left| c^d_{31} \right| ^2 + g_Y^2 \lambda_{11}^2 v_\theta^4 \left( 17 + 6\mathbb{L} \right)  }{108 m_\theta^2 v_\theta^2}&  \lambda \lambda_{11} c^{d*}_{31} {y^d_{32}} & \lambda \lambda_{11} c^{d*}_{31} {y^d_{33}}\\
		\lambda \lambda_{11} c^d_{31} y^{d*}_{32} &  \dfrac{\lambda_{11}^2 v_\theta^2 \left( g_Y^2 \left( 17 + 6 \mathbb{L} \right) - 27 \left( 5 + 2 \mathbb{L} \right) \left| y_{32}^d \right| ^2 \right)  }{108 m_\theta^2} &  - \dfrac{\lambda_{11}^2 v_\theta^2 \left( 5 + 2 \mathbb{L} \right) y_{32}^{d*} y_{33}^{d}  }{4 m_\theta^4}\\
		\lambda \lambda_{11} c^d_{31} y^{d*}_{33}& - \dfrac{\lambda_{11}^2 v_\theta^2 \left( 5 + 2 \mathbb{L} \right) y_{32}^{d} y_{33}^{d*}  }{4 m_\theta^4} & \dfrac{\lambda_{11}^2 v_\theta^2 \left( g_Y^2 \left( 17 + 6 \mathbb{L} \right) - 27 \left( 5 + 2 \mathbb{L} \right) \left| y_{33}^d \right| ^2 \right)  }{108 m_\theta^2}  \\
	\end{pmatrix}
\end{align}
}}
We may assume $ c_{31} \sim y_{ij}^d \sim 1 $, as required by our toy model, and $ v_\theta \sim m_\theta $ which stems from the expectation that the flavon vev and mass originate from the same physical mechanism.
The matrix $ (C_{Hd})_{ij} $ then becomes
\begin{equation} \label{eq:CHd_scaling}
	 C_{Hd} \sim 
	\dfrac{\hbar}{2 m_\theta^2} 
	 \begin{pmatrix} 
		 \dfrac{54 \lambda^2 \left( 1 + 2 \mathbb{L} \right)  + \lambda_{11}^2 \left( 17 + 6 \mathbb{L} \right) }{108} & \lambda \lambda_{11} & \lambda \lambda_{11} \\
		 \lambda \lambda_{11} & - \dfrac{\lambda_{11}^2 \left( 59 + 24 \mathbb{L} \right) }{54} & - \dfrac{\lambda_{11}^2 \left( 5 + 2 \mathbb{L} \right) }{4} \\ 
		 \lambda \lambda_{11} & - \dfrac{\lambda_{11}^2 \left( 5 + 2 \mathbb{L} \right) }{4} & - \dfrac{\lambda_{11}^2 \left( 59 + 24 \mathbb{L} \right) }{54} \\
	 \end{pmatrix},
\end{equation}
under the assumption that $g_Y \sim 1$ near the matching scale.  We see that the elements of the matrix come with three different parametric dependencies. 
The bottom $ 2 \times 2 $ block is proportional to $ \lambda_{11}^2 $, the off-diagonals on the first row and column scale as $ \lambda \lambda_{11},  $ whereas the $ (11) $ element has two competing terms controlled by the squares of each dimensionless parameter.
If one makes the further assumption that $ \lambda_{11} \sim 1 $ as befits an otherwise undetermined dimensionless parameter, recalls that $ \lambda \sim 0.1 $ and evaluates the Wilson coefficient at the matching scale, one arrives at the prediction $ \left( C_{Hd} \right)_{1i} \sim \left( C_{Hd} \right)_{i1} \sim 0.1 \left( C_{Hd} \right)_{jk} $, where $ i \in \{1,2,3\} $ and $ j,k \in  \{2,3\} $.
This relation can be understood using spurion analysis.
For a SMEFT operator with a flavourful Wilson coefficient $ (C_{\mathcal{O}})_{ij} $, one generically expects the coefficients to be proportional to a power of $ \lambda $ equal to the number of flavon insertions required to restore the broken \SFN{} symmetry. 
The choice of \SFN{} charges $ q_{d_1} = 5, \, q_{d_2} = 3, \, q_{d_3} = 3$ in Table~\ref{tab:DownToy} implies a leading-order $ \lambda^0 $-scaling for the (11) element and the $ 2  \times 2$ block of the heaviest two generations, whereas $ (C_{Hd})_{1i} \sim (C_{Hd})_{i 1} \sim \lambda $ for $ i \in \{1,2 \} $.
This explains the leading-order pattern seen in \eqref{eq:CHd_scaling}.
The (11) entry also contains a sub-leading correction $\propto \lambda^2$ thanks to the order at which we truncate the UV FN EFT. 
Allowing for higher operator dimensions in the FN EFT when calculating our loop-level SMEFT matching would lead to further corrections to the leading-order results, suppressed by more powers of $ \lambda $.

Despite its simplicity, this analysis shows that it is possible to predict the relative sizes of flavourful Wilson coefficients in our model whilst making minimal assumptions about the UV details of the FN model in question. 
Given two different UV complete models, where the hitherto unknown model parameters are fixed, the precise form of the predicted flavour structure could be used to discriminate between the two. 

In the same vein, there are also hierarchies between different classes of SMEFT operators.
Unsurprisingly, the relative sizes are dictated by the powers of $ \lambda $ at which  the operators appear, and whether they appear at tree- or loop-level.
There is for instance a hierarchy between the Wilson coefficients of the operator $ (H\hc H ) \overline{Q}_3 H d_1 $ and that of $ \left(H^\dagger \Darr H\right)\left(\overline{d}_{i} \gamma^\mu d_j\right) $.
In this case, the distinction between the two is caused by the former receiving contributions from tree-level matching and the latter only appearing at the one-loop level, leading to the $ 16 \pi^2 $ suppression factor. 

\begin{figure}[h!]
	\usetikzlibrary {arrows.meta} 
	\centering
    \begin{subfigure}[t]{\textwidth}
        \centering
        \feynmandiagram [medium, horizontal = b to c,baseline = 1.1cm] {
            a [particle =\( d_1\) ]-- [fermion] b -- [fermion] d [particle = \(\overline{Q}_3 \)],
            b -- [charged scalar, half left, looseness = 1.5, edge label = \(H\)] c --[dashed, double, half left, looseness = 1.5 ,edge label = \( \vartheta\)] b,
            e [particle  = \( \overline{Q}_3 \)] -- [fermion] c -- [fermion] f [particle =\(d_1\) ],
        };
        \begin{tikzpicture} [baseline=1.1cm]
              \draw[->]        (0,1.2)   -- (1,1.2);
        \end{tikzpicture}
        $
        \
        - 2 C_{qd}^{3311}\left( \overline{Q}_3 d_1 \right) \left( \overline{d}_1 Q_3 \right) 
        \
        \stackrel{\mathclap{\normalfont\mbox{\scriptsize {Fierz}}}}{=}
        \
        C_{qd}^{3311} \left( \overline{Q}_3 \gamma^\mu Q_3 \right) \left( \overline{d}_1 \gamma_\mu d_1 \right) 
        $
        \caption{A four-fermion operator at dimension-6 sourced by the FN model vertices of \eqref{eq:L5FNexpanded}. 
The diagram illustrates why the resulting Wilson coefficient, shown in the last term of \eqref{eq:dim5loopmatchFourFermion}, is proportional to $ \lvert c^d_{31} \rvert^2 $.}
    \end{subfigure}
	\begin{subfigure}[t]{\textwidth}
		\centering
		\feynmandiagram [medium, horizontal = b to c,baseline = 1.1cm] {
			a [particle =\( H \) ]-- [charged scalar] b -- [anti fermion] d [particle = \( d_1\)],
			b -- [fermion, half left, looseness = 1.5, edge label = \(Q_3\)] c --[dashed, double, half left, looseness = 1.5 ,edge label = \( \vartheta\)] b,
			e [particle  = \( d_1 \)] -- [anti fermion] c -- [charged scalar] f [particle =\(H\) ],
		};
     \begin{tikzpicture} [baseline=1.1cm]
		  \draw[->]        (0,1.2)   -- (1,1.2);
	\end{tikzpicture}
	$
	\quad
	 ( H\hc \Darr H ) \left( \overline{d}_1 \gamma^\mu d_1  \right)
	$
	\caption{A contribution to the Higgs-derivative operator $ ( H\hc \Darr H ) \left( \overline{d}_1 \gamma^\mu d_1  \right) $ built out of dimension-5 FN vertices and thus proportional to $ \lvert c^d_{31} \rvert^2 $. 
    \label{fig:higgs-derivative}}
	\end{subfigure}
\begin{subfigure}[t]{\textwidth}
	\centering
	\begin{tikzpicture}[baseline = 0cm]
	\begin{feynman}[small]
		\vertex (b);
		\vertex [above left=of b] (a) {$ \overline{Q}_3 $};
		\vertex [below left=of b] (d) {$ d_1 $};
		\vertex [above right=of b] (g);
		\vertex [below right=of g] (c);
		\vertex [above right=of c] (e) {$ H $};
		\vertex [below right=of c] (f) {$ \langle \theta \rangle $};
		\vertex [above right=of g] (h) {$ H $};
        \vertex [above left=of g] (i) {$ H $};
		\diagram* {
			(a) -- [fermion] (b) -- [fermion] (d),
			(b) -- [charged scalar, quarter left, edge label = $ H $] (g),
			(g) -- [charged scalar, quarter left, edge label = $ H $] (c),
			(c) -- [double, dashed, half left ,edge label = $ \vartheta $] (b),
			(c) -- [charged scalar] (e),
			(c) -- [scalar] (f),
			(g) -- [charged scalar] (h),
            (g) -- [anti charged scalar] (i), 
		};
	\end{feynman}
\end{tikzpicture}
	\begin{tikzpicture} [baseline=1.1cm]
		  \draw[->]        (0,1.2)   -- (1,1.2);
	\end{tikzpicture}
	$
	\quad
	 (H\hc H) \overline{Q}_3 H d_1 
	$
\caption{A contribution to the Yukawa-like operator $(H\hc H) (\overline{Q}_3 H d_1)$ proportional to $c^d_{31}$.
\label{fig:Higgs-enhanced-yukawa}
}
\end{subfigure}
	\caption{
 Some Feynman diagrams contributing to the one-loop matching results coming out of a dimension-5 UV Lagrangian. 
	We have verified the Wilson coefficients arising from the above diagrams manually and found agreement with \pkg{Matchete}.
	}
	\label{fig:dim5loops}
\end{figure}


\subsubsection{Dimension Six ($\normalfont{\DFN=6}$)}
\label{sec:dim6matching}
Many more operators appear at mass-dimension 6.  For example, the $\theta$-dependent scalar potential and \SFN{}-type operators driving our UV flavour mechanism give 11 additional interactions on their own,
\begin{align}
\label{eq:L6FNpartial}
	\mathcal{L}^6_{\text{FN}} &\supset  \frac{1}{\LUV^2} \left[
 \lambda_{03} \left(\theta^\star \theta \right)^3 + \lambda_{12} \left(H^\dagger H\right)\left(\theta^\star \theta\right)^2 + \lambda_{21} \left(H^\dagger H\right)^2\left(\theta^\star \theta\right) \right] \\
 \nonumber
 &+ \frac{1}{\LUV^2} \left[c^{d_6}_{22} \,\overline{Q}_{2} H d_2\, \theta^2 + c^{d_6}_{23} \,\overline{Q}_{2} H d_3\, \theta^2 + c^{d_6}_{32}\,\overline{Q}_{3} H d_2\, \left(\theta^\star \theta \right) + c^{d_6}_{33} \,\overline{Q}_{3} H d_3\, \left(\theta^\star \theta \right) + \text{h.c.} \right]\,,
\end{align}
although this is but a subset of the full $\DFN=6$ Lagrangian, and would be yet further augmented upon considering $\left(\theta^\star \theta\right)$-enhanced renormalizable Yukawa terms for up quarks and leptons.

 One option is to simplify the $\DFN = 6$ analysis by only considering (e.g.) these operators in our matching since they are expected to be the most relevant for the flavour dynamics.  
 However, doing so potentially misses important physics effects that might be sourced from energies around $\sim \LUV$, where we remain purposefully agnostic about hypothetical UV completions to the FN class of EFTs.  
 This might of course occur directly, upon integrating out some new vector-like fermion or novel (gauge) boson, or it could occur via Renormalization-Group-Evolution (RGE) effects sourced from running between the scales $\LUV$ and $\LFN$, where we recall that $\LFN$ sits at or around the flavon mass.  
 As is well known \cite{Jenkins:2013zja,Jenkins:2013wua,Alonso:2013hga} in the case of (e.g.) the SMEFT, RGE mixes EFT operators via anomalous dimensions such that Wilson coefficients in the IR are functions of \emph{multiple} Wilson coefficients (associated to different EFT operators) that may have existed in the UV.  In other words, in the absence of an explicit UV completion, it is likely important to incorporate the complete basis of EFT operators in the IR (here meaning  physics propagating around $\sim\LFN$), such that said effects can be parameterized in the full set of otherwise unconstrained Wilson coefficients. 
\subsubsection*{A Complete and Minimal Basis}
 The construction of a complete, minimal, on-shell operator basis in EFTs is, a priori, tricky.
 However, assuming the charge assignments of Table~\ref{tab:DownToy} and that all other SM multiplets have trivial \SFN{} charges, one can readily calculate the Hilbert Series $\mathbb{H}$ associated to the UV Lagrangian of our theory at successively higher orders.
 This then serves as a guide for the construction of a sensible basis before pursuing matching to the SMEFT, either at tree- or loop-level.  
 
 We have calculated the relevant Hilbert Series using {\pkg{ECO}} \cite{Marinissen:2020jmb,Calo:2022jqv}, an automated {\pkg{form}}-based package with a number of functionalities that builds on prior Hilbert Series results \cite{Henning:2015daa,Henning:2015alf,Henning:2017fpj}.   {\pkg{ECO}} includes the ability to add BSM field content as well as U(1) (global or gauged) BSM symmetries to the computation, and is furthermore able to apply integration-by-parts and EOM identities to reach a minimal on-shell basis.  
 This is all we need in the simplified FN class of models we are currently considering.  We collect our results for the Hilbert Series in terms of the number of insertions of $\theta^{(\star)}$:\footnote{Recall that the RHS of these equations are not to be interpreted as operators.} 
\begin{align}
\label{eq:HSdim5}
&\mathbb{H}(\lbrace \Phi \rbrace_{\text{SM}},\theta) \vert_{d=5} = \theta \left(H \overline{Q}_3 d_1\right) + \theta^\star \left(H^\dagger Q_3 \overline{d}_1\right)\,, \\
\nonumber
&\mathbb{H}(\lbrace \Phi \rbrace_{\text{SM}},\theta) \vert_{d=6} = \theta^2 H \overline{Q}_2\left(d_3 + d_2 \right) + \theta^{\star 2}H^\dagger Q_2\left(\overline{d}_3+\overline{d}_2\right) \\
\label{eq:HSdim6}
&+ \theta^\star \theta \left(2 G^2 +2 W^2 + 2 B^2 + H^2 H^{\dagger 2} + H\overline{Q}_3 (d_2+d_3) + H^\dagger Q_3(\overline{d}_2+\overline{d}_3)\right) \\
\nonumber
&+\theta^\star \theta \,p\,\left(\sum_{e,u,d,L,Q} \psi_i \overline{\psi_i} + \sum_{e,u,L} \psi_i \overline{\psi}_j + d_3\overline{d}_2 + d_2\overline{d}_3+ 2 p H H ^\dagger \right) + \left(\theta^\star \theta\right)^2\left(p^2 + H H^\dagger\right) + \left(\theta^\star \theta\right)^3\,,
\end{align}
where a summation over latin (flavour) indices and fermions $\psi \in \lbrace e,u,d,...\rbrace$ is implied in the third row of \eqref{eq:HSdim6},
where $i \neq j \in \lbrace 1,2,3 \rbrace$, where $ p $ stands for a (covariant) derivative, and where there are not yet terms (e.g.) $\propto \theta^\star \theta H L e$ or similarly for the up quarks (we have effectively set up and lepton \SFN{} charges to zero for the above sample HS calculation, which also explains the off-diagonal flavour summation in the first bracketed term in the last line of \eqref{eq:HSdim6}).
Note that we have neglected terms with no $\theta^{(\star)}$-dependence.

We can now look at \eqref{eq:HSdim5} and readily confirm that no other terms besides the FN operator in \eqref{eq:toyLdim5} and its Hermitian conjugate appear at $\DFN=5$, while from \eqref{eq:HSdim6} we see the three $\theta$-dependent terms expected in the $\DFN=6$ scalar potential,
\begin{align}
 \label{eq:dim6Scalarops}
    \mathcal{O}_\theta &= \left(\theta^\star \theta\right)^3,\,\,\,\,\,\,\,\,\,\,\, \mathcal{O}_{H\theta} = \left(H^\dagger H\right)^2\left(\theta^\star \theta\right)\,,\,\,\,\,\,\,\,\,\,\,\,\mathcal{O}_{\theta H} = \left(H^\dagger H\right)\left(\theta^\star \theta\right)^2,
\end{align}
as well as three $\theta$-dependent kinetic operators
\begin{align}
 \label{eq:dim6Kineticops}
\mathcal{O}_{\theta \Box \theta} = \left(\theta^\star \theta\right)\Box\left(\theta^\star \theta\right)\, ,\,\,\,\,\,\,\,\,\,\,\mathcal{O}_{H\Box\theta} = \left(H^\dagger H\right)\Box\left(\theta^\star \theta \right)\,,\,\,\,\,\,\,\,\,\,\,\mathcal{O}_{\theta \Box H} = (D^\mu H \hc D_\mu H) \theta^*\theta \, .
\end{align}
Additionally, we can draw one-to-one correspondences between many other terms in \eqref{eq:HSdim6} and $\DSM=6$ operators of the SMEFT Warsaw basis \cite{Grzadkowski:2010es} (also see \cite{Gripaios:2016xuo}), albeit with $H^{(\dagger)} \rightarrow \theta^{(\star)}$.  Critically, the $\theta$-enhanced Yukawa operators driving the FN mechanism of principal interest to our study are obvious in \eqref{eq:HSdim6},
\begin{align}
 \label{eq:dim6Yukawaops}
    \mathcal{O}_{\overline{Q}_3d_i\theta}&= \left(\theta^\star\theta\right) \left(\overline{Q}_3 H d_2 + \overline{Q}_3 H d_3\right) + \text{h.c.}\,,\,\,\,\,\,\,\,\,\,\,\,&&\mathcal{O}_{\overline{Q}_2d_i\theta} = \theta^2\left( \overline{Q}_2 H d_3 + \overline{Q}_2 H d_2\right) + \text{h.c.}\,, 
\end{align}
and we also acknowledge that, for charged lepton and up quark flavours whose Yukawa elements are realized at the renormalizable level, additional operators will appear:
\begin{equation}
\label{eq:dim6lep&upYukawaops}
\mathcal{O}_{\overline{L}_i e_j\theta} = \left(\theta^\star \theta\right)\, \overline{L}_i H e_j + \text{h.c.}\,, \,\,\,\,\,\,\,\,\,\,\,\mathcal{O}_{\overline{Q}_i u_j\theta} = \left(\theta^\star \theta\right)\, \overline{Q}_i \tilde{H} u_j + \text{h.c.}
\end{equation}
More care must of course be taken if the leading Yukawa structure in these family sectors appears at $\DFN=6$, as in the (22) and (23) matrix elements of the toy model in \eqref{eq:toymD}, or if the Weinberg operator in \eqref{eq:toyLdim5} can only appear at $\DFN=6$ via a $\theta^{(\star)}$ insertion (cf. Section \ref{sec:GENERALIZE} for more commentary on this point).

Six scalar-field strength interaction operators, the last three of which are CP violating, are also obvious in \eqref{eq:HSdim6}:
\begin{align}
\nonumber
    \mathcal{O}_{\theta G} &= \left(\theta^\star\theta\right)\,  G_{\mu\nu}^A G^{A\mu\nu},\,\,\,\,\,\,\,\,\,\,\mathcal{O}_{\theta W}= \left(\theta^\star\theta\right)\,  W_{\mu\nu}^I W^{I\mu\nu},\,\,\,\,\,\,\,\,\,\,\mathcal{O}_{\theta B}= \left(\theta^\star\theta\right)\,  B_{\mu\nu} B^{\mu\nu}, \\
     \label{eq:dim6FieldStrengthops}
    \mathcal{O}_{\theta \tilde{G}} &=  \left(\theta^\star\theta\right)\,  \widetilde{G}_{\mu\nu}^A G^{A\mu\nu},\,\,\,\,\,\,\,\,\,\,\mathcal{O}_{\theta \tilde{W}}= \left(\theta^\star\theta\right)\,  \widetilde{W}_{\mu\nu}^I W^{I\mu\nu},\,\,\,\,\,\,\,\,\,\,\mathcal{O}_{\theta \tilde{B}}= \left(\theta^\star\theta\right)\,  \widetilde{B}_{\mu\nu} B^{\mu\nu}.
\end{align}
Note that, contrary to field strength operators constructed using the SM Higgs doublet, no cross term of the type $\theta^* \theta W B$ appears because such an operator would not be $\text{SU(2)}_\text{L}$ invariant.
 
We also have contact interactions between a flavon current and a fermion current,
\begin{align}
\label{eq:dim6GaugeFlavonFermionops}
\mathcal{O}_{\partial\overline{\psi}_i\psi_j} \in  \theta^\star \parr \theta \, \left(\sum_{e,u,d,L,Q} \overline{\psi}_i \gamma^\mu \psi_i + \sum_{e,u,L}\overline{\psi}_i \gamma^\mu \psi_j + \overline{d}_{2} \gamma^\mu d_3 + \overline{d}_{3} \gamma^\mu d_2\right),
\end{align}
where a summation over latin (flavour) indices and fermions is again implied, and where the number and flavour structure of off-diagonal elements in the latter three terms is subject to modification depending on the complete set of unspecified \SFN{} charges for the charged leptons and up-type quarks.
In total, \eqref{eq:dim6Scalarops}--\eqref{eq:dim6GaugeFlavonFermionops} represent the 
$\theta$-dependent operators of the non-redundant, flavour-dependent $\DFN=6$ UV FN Lagrangian.\footnote{See Section \ref{sec:GENERALIZE} for a discussion of other operators that can appear when we generalize the \SFN{} charges.}  

There are of course additional contributions from all \SFN{}-symmetric SMEFT operators at dimension six \cite{Grzadkowski:2010es},  $\mathcal{L}^6_{\text{SMEFT}^\prime}$, which have no $\theta$-dependence.   While in the absence of a complete set of \SFN{} charges for all SM multiplets it is again impossible to fully enumerate $\mathcal{L}^6_{\text{SMEFT}^\prime}$ (or analogous bases at higher mass dimensions for that matter), they will generically include a host of flavour-symmetric four-fermion ($\psi^4$), Higgs-enhanced Yukawa ($\psi^2 H^3$),\footnote{After \SFN{} symmetry breaking, these types of operators lead to $\mathcal{O}(\lambda^2)$ corrections to the (23) and (33) entries of $M^d$, for instance. Higher-order corrections of this type appear generically in FN-type models and have implications for flavoured phenomenology, despite often being neglected.} dipole-type ($\psi^2 X H$), and Higgs-kinetic ($\psi^2 H^2 D$) operators, in addition to purely bosonic structures. However, we will only attempt tree-level matching at $\DFN=6$, and hence we only need \eqref{eq:dim6Scalarops}--\eqref{eq:dim6GaugeFlavonFermionops} for our calculations.  

Furthermore, given that we are matching in the \emph{broken} \SFN{} phase, it turns out that we will only need a subset of the operators enumerated between \eqref{eq:dim6Scalarops}--\eqref{eq:dim6GaugeFlavonFermionops}.  Indeed, when $\theta$ can effectively be treated as a real scalar, the operators of \eqref{eq:dim6GaugeFlavonFermionops} go to zero due to their left-right derivative structure, as do the first ($\theta$-dependence only) and third (mixed $\theta$-$H$ contribution) kinetic terms in \eqref{eq:dim6Kineticops}. 
We have performed an additional Hilbert Series calculation confirming the results in \cite{Gripaios:2016xuo}, which found a minimal EFT basis of 18 operators at $d=6$ for a real, gauge-singlet scalar enhancement of the SM, since our results in \eqref{eq:dim6Scalarops}--\eqref{eq:dim6GaugeFlavonFermionops} map to \cite{Gripaios:2016xuo} up to flavour-symmetry assumptions.
The \SFN{}-breaking expansion of $\theta$ about the true vacuum transforms dimension-6 operators of the type $ (\theta^* \theta) \times \mathcal{O}^{d=4} $, which constitute the majority of the basis in \eqref{eq:dim6Scalarops}--\eqref{eq:dim6GaugeFlavonFermionops}, into dimension-5 operators of the type $ v_\theta \vartheta \times \mathcal{O}^{d=4}$. 
It is this set of dimension-5 operators that is responsible for generating EFT operators when matching at tree-level to the SMEFT: diagrammatically, one connects the flavon $\vartheta$ of the dimension-5 vertex to the three-point vertex $-\lambda_{11} v_\theta H\hc H \vartheta$ in a fashion illustrated in Fig.~\ref{fig:dim-6-matching-illustration} where we have drawn a matching contribution to the SMEFT operator $(H\hc H) \overline{Q}_2 H d_2$.

\begin{figure}
	\centering
	\begin{tikzpicture}[baseline = 0cm]
	\begin{feynman}
		\vertex (c);
		\vertex [above left= of c] (ia) {$ H $};
		\vertex [below left= of c] (ib) {$H^\dagger$}; 
		\vertex [right= of c] (d);
		\path (d) ++(+60:1.8cm) node (fa) {$ \overline{Q}_2 $};
		\path (d) ++(+0:1.8cm) node (fb) {$ H $};
		\path (d) ++(-60:1.8cm) node (fc) {$ d_2 $};

		\diagram* {
			(ia) -- [charged scalar] (c) -- [charged scalar] (ib),
			(c) -- [dashed, double, edge label = $\vartheta$] (d),
			(fa) -- [fermion] (d) -- [fermion] (fc),
			(d) -- [charged scalar] (fb),
		};
	\end{feynman}
		\newcommand\Xcoord{3.7}
		\newcommand\bracY{1.7}
		  \draw[->]        (\Xcoord,0)   -- (\Xcoord+1.3,0);
		  \draw (\Xcoord+3,0) node {$ C_{dH} (H^\dagger H) \overline{Q}_2 H d_2 $};
		\draw [decorate, align=center, decoration={brace, amplitude=1.1ex, raise=1ex}]
		  (-1.8, \bracY) -- (0.71, \bracY) node[pos=.5, above=2ex] {Dim-3 \\ $ v_\theta \vartheta (H^\dagger H) $};
		\draw [decorate, align=center, decoration={brace, amplitude=1.1ex, raise=1ex}]
		  (0.79, \bracY) -- (3.5, \bracY) node[pos=.5, above=2ex] {Dim-5 \\ $ v_\theta \vartheta (\overline{Q}_2 H d_3 ) $};
	\end{tikzpicture}
	\caption{A diagrammatic illustration of how tree-level matching of the FN model yields the SMEFT operator $(H^\dagger H) \overline{Q}_2 H d_2$.
  A flavon line is used to glue together a dimension-3 vertex and a dimension-5 vertex both arising from the \SFN{} symmetry breaking; the dimension-5 vertex in the figure is sourced from a $\DFN=6$ operator $(\overline{Q}_2 H d_2) \theta^2 $ in the unbroken phase.
  The full Wilson coefficient $C_{dH}$, which obtains a second contribution from a diagram not drawn here, is written explicitly in \eqref{eq:SMEFT66Yukawas}.  \label{fig:dim-6-matching-illustration}}
\end{figure}

In summary, we find that there is a subset of 18 operators in \eqref{eq:dim6Scalarops}--\eqref{eq:dim6GaugeFlavonFermionops} that are required when matching the $\DFN=6$ toy FN model to the SMEFT, not including the unspecified number of $\left(\theta^\star \theta\right)\left[\overline{Q}_i \tilde{H} u_j + \overline{L}_i H e_j \right] + \text{h.c.}$ operators allowed.  We will now discuss the IR SMEFT operators that appear as each class of UV operators in said subset turns on.
\subsubsection*{Matching Scalar Potential and Derivative Operators}
As is well known for the SMEFT in the broken EW phase, the presence of higher-order scalar potential operators shifts the minimum of the classical vacuum.  This is also true for the FN EFT we are considering here, with $\mathcal{O}_\theta$ responsible for the shift in the flavon vev.
Following the analogous discussion in \cite{Alonso:2013hga} concerning the Higgs, the $\theta$ vev from \eqref{eq:thetavev} is modified to
\begin{equation}
\label{eq:newvac}
\langle \theta^\star \theta \rangle = \frac{v_\theta^2}{2}\left(1 + \frac{3\, \lambda_{03}\, v_\theta^4}{4\, \mu_\theta^2\, \LUV^2}\right) \equiv \frac{\overline{v}_\theta^2}{2}
\end{equation}
at leading order in the implied expansion parameters on the right-hand side of the equation. 
We can then re-expand the complete scalar potential from \eqref{eq:Ltheta}, now including all of the operators in \eqref{eq:dim6Scalarops}, about its new vacuum,
\begin{equation}
H \rightarrow H,\,\,\,\,\,\,\,\,\,\,\theta \rightarrow \left(\overline{v}_\theta + \vartheta\right)/\sqrt{2}\,,
\end{equation}
to obtain the potential in the broken \SFN{} phase analogous to \eqref{eq:L4thetaexpanded},
\begin{align}
  \nonumber
  \mathcal{L}_{\slashed{\text{FN}}}^{6} &\supset - \frac{\overline{m}_\theta^2}{8 \overline{v}_\theta^2} \vartheta^4-\frac{\overline{m}_\theta^2}{2 \,\overline{v}_\theta}\vartheta^3 - \left(\frac{\lambda_{11}}{2} H^\dagger H +\frac{\overline{m}_\theta^2}{2}\right) \vartheta^2 - \overline{v}_\theta \lambda_{11}  H^\dagger H \,\vartheta - \frac{\overline{v}_\theta^2 \lambda_{11}}{2} H^\dagger H +\frac{\overline{v}_\theta^2 \lambda_{21}}{2 \LUV^2} \left(H^\dagger H\right)^2 \\
  \nonumber
  &+  \frac{1}{\LUV^2} \biggl\{ \frac{\lambda_{03}}{8}\vartheta^6 + \frac{3\overline{v}_\theta\lambda_{03}}{4}\vartheta^5 + \left(\frac{\lambda_{12}}{4} H^\dagger H +\frac{3\,\overline{v}_\theta^2 \,\lambda_{03}}{2} \right)\vartheta^4
  + \left(\overline{v}_\theta\, \lambda_{12}\,H^\dagger H + \overline{v}_\theta^3 \,\lambda_{03} \right)\vartheta^3\\
 \label{eq:brokendim6scalarpotential}
    & + \left[\frac{3 \overline{v}_\theta^2 \lambda_{12}}{2} H^\dagger H + \frac{\lambda_{21}}{2} (H^\dagger H)^2\right]\vartheta^2 +  \left[\overline{v}_\theta^3 \lambda_{12} H^\dagger H + \overline{v}_\theta \,\lambda_{21}(H^\dagger H)^2 \right]\vartheta + \frac{1}{4}v_\theta^4 \lambda_{12} H^\dagger H \biggl\} \,,
\end{align}
where we have ignored static terms, have only kept contributions up to linear order in the dimension-6 Wilson coefficients of \eqref{eq:dim6Scalarops},
and have defined the modified flavon mass as 
\begin{equation}
\overline{m}^2_\theta \equiv 2\, \mu_\theta^2 - \frac{3\, v_\theta^4 \,\lambda_{03}}{2\,\LUV^2}.
\end{equation}
Upon implementing \eqref{eq:brokendim6scalarpotential} and $\mathcal{O}_{H \Box \theta}$ from \eqref{eq:dim6Kineticops} in {\tt{Matchete}}, we then match to $d_{\text{SM}}=6$ at tree-level, obtaining  
\begin{align}
\label{eq:SMEFT66Scalar}
	\nonumber
	&\mathcal{L}_{\text{SMEFT}} \supset + \frac{ \overline{\lambda}^2 \lambda_{12} \overline{v}_\theta^2}{2} H\hc H
	+ \frac{1}{ \overline{m}_\theta^2} 
 \left[\overline{\lambda}^2 \left(\lambda_{21} m_\theta^2 - 2 \lambda_{11} \lambda_{12}  \overline{v}_\theta^2 \right)
 + 2 \overline{\lambda}^4 \lambda_{12}^2 \overline{v}_\theta^2  \right] \left( H\hc H \right)^2 \\
	\nonumber
	& +\left[  \frac{2 c^d_{31} \overline{\lambda}^3 \lambda_{12} }{\overline{m}_\theta^2} \left( H\hc H \right) \overline{Q}_3 H d_1  + \text{h.c.} \right] - \frac{2}{m_\theta^6} \left[ \lambda_{11} \overline{\lambda}^2 \left(  \lambda_{03} \lambda_{11}^2 \overline{v}_\theta^4 - \lambda_{11} \lambda_{12} m_\theta^2 \overline{v}_\theta^2 +  \lambda_{21} m_\theta^4 \right) \right. \\
\nonumber
	& \left. - 2 \overline{\lambda}^4 \left( 3  \lambda_{03} \lambda_{11}^2 \lambda_{12} \overline{v}_\theta^4 - 2 \lambda_{11} \lambda_{12}^2 \overline{m}_\theta^2 \overline{v}_\theta^2 + \lambda_{12} \lambda_{21} \overline{m}_\theta^4 \right) + \mathcal{O}\Big( \overline{\lambda}^6 \Big)   \right]  \left(H^\dagger H\right)^3 \\
	& + \frac{2}{\overline{m}_\theta^4} \left[ \lambda_{11} \overline{\lambda}^2 \left( \lambda_{12} \overline{v}_\theta^2 - c_{H\Box\theta} \overline{m}_\theta^2 \right)  - \lambda_{12} \overline{\lambda}^4 \left(  \lambda_{12} \overline{v}_\theta^2 - 2 c_{H \Box \theta} m_\theta^2\right)  \right] \left( H\hc H \right) \Box \left( H\hc H \right) \,,  
\end{align}
where we have defined the shifted FN expansion parameter $\overline{\lambda}\equiv \overline{v}_\theta/(\sqrt{2}\LUV)$ and expanded the results in powers of this order $\mathcal{O}(0.1)$ parameter.
The expansions are shown up to and including order $\overline{\lambda}^4$ terms, but we note that contributions from dimension-8 operators in the scalar potential will also enter at this order and are expected to augment the results.
Note in \eqref{eq:SMEFT66Scalar} the renormalizations of the Higgs potential and kinetic terms, in addition to the novel Higgs-flavon mixing term impacting the (31) Yukawa coupling that first gets turned on at $\DFN = 5$. 
We emphasize that the shifted vacuum in \eqref{eq:newvac} will also impact the lower-order matching presented in above sections, but do not show these results explicitly, for brevity.
Furthermore, the shifts in $ \overline{m}_\theta^2 $ and $ \overline{v}_\theta $ are formally of order $ (v_\theta / \LUV)^2 $ and so enter at the same order as dimension-8 SMEFT operators.
\subsubsection*{Matching Yukawa-Like FN Operators}
Continuing with the Yukawa-like operators in \eqref{eq:dim6Yukawaops}, our tree-level {\tt{Matchete}} matching analysis gives 
\begin{multline}
    \label{eq:SMEFT66Yukawas}
    \mathcal{L}_{\text{SMEFT}} \supset \Big[\overline{Q}_2 H\left(c^{d_6}_{22}\,  d_2 +  c^{d_6}_{23}\, d_3\right) + \overline{Q}_3 H\left(c^{d_6}_{32}\, d_2+ c^{d_6}_{33}\,  d_3 \right)\Big] \\
    \times \left[\overline{\lambda}^2 + \frac{2\overline{\lambda}^2\left(2\overline{\lambda}^2\,\lambda_{12} -\lambda_{11}\right)}{\overline{m}_\theta^2} \left(H^\dagger H\right)\right] +\text{h.c.} 
\end{multline} 
As in \eqref{eq:L5FNexpanded}, in the first bracketed term we have kept the explicit FN contributions to the down-type Yukawas, responsible for the texturing of mass and mixing in the family.  
The second bracketed term represents the novel contributions to the Higgs-enhanced Yukawa operators of the $\DSM = 6$ SMEFT sourced by the $\DFN = 6$ FN theory.
Comparing the dimension-6 Higgs-enhanced Yukawa terms for the (31) and (22) flavour index pairs in \eqref{eq:dim5treematchfinal} and \eqref{eq:SMEFT66Yukawas}, respectively, we see that the latter Wilson coefficient is suppressed by one more power of $\lambda$.
This finding generalizes to matching to the SMEFT at higher $\DFN$; as higher-dimension FN terms, of type $\overline{\psi}_i H \psi_j \theta^n $, are added to the UV Lagrangian and matching is done at tree-level, the Wilson coefficients of the resulting Higgs-enhanced Yukawa operators show the same hierarchies in $\overline{\lambda}$ as the Yukawa matrices themselves. 
That the two flavour patterns are the same is fully consistent with a spurion analysis.
The $ S_\text{FN} $ violation structure of the Higgs-enhanced Yukawa terms is identical to that of the SM Yukawas, implying the same hierarchies in powers of the spurion $ \lambda $.
The toy model thus predicts the following hierarchies in the down-type Higgs-enhanced Yukawas:
\begin{equation}
    (C_{dH})_{ij} \sim 
    \begin{pmatrix} 
		\overline{\lambda}^4 &  \overline{\lambda}^3 &  \overline{\lambda}^3\\
		\overline{\lambda}^3 & \overline{\lambda}^2 & \overline{\lambda}^2  \\
		\overline{\lambda} & \overline{\lambda}^0 & \overline{\lambda}^0 \\
	\end{pmatrix},
\end{equation}
where the $\overline{\lambda}^0 $ terms come from $\mathcal{L}_{\text{SMEFT}^{'}}$, the part of the \SFN{}-symmetric UV Lagrangian without factors of $\theta$, which contains $(C_{dH})_{33}$ and $(C_{dH})_{32}$ with order one coefficients.  

Finally, for the flavour scenario where charged-lepton Yukawa couplings appear at the renormalizable level, the tree-level matching effects sourced from the $\DFN=6$ operators in \eqref{eq:dim6lep&upYukawaops} are given by
\begin{equation}
\mathcal{L}_{\text{SMEFT}} \supset  c^{e_6}_{ij} \overline{L}_i \,H\,e_j\left[\overline{\lambda}^2 + \frac{2\overline{\lambda}^2\left(2\overline{\lambda}^2\,\lambda_{12} -\lambda_{11}\right)}{\overline{m}_\theta^2} \left(H^\dagger H\right)\right] +\text{h.c.} 
\end{equation}
where we see the same structure as above for the (flavour-specific) down-quark sector.  We expect that the same structure appears for the up-quarks.

\subsubsection*{Matching Field-Strength Operators}
Finally, upon including \eqref{eq:dim6FieldStrengthops} into the matching analysis, one finds the following contributions to the $\DSM=6$ SMEFT:
\begin{align}
\label{eq:SMEFT66fieldstrengths}
    \mathcal{L}_{\text{SMEFT}} &
    \supset \frac{2\overline{\lambda}^2\left(2\overline{\lambda}^2\,\lambda_{12} -\lambda_{11}\right)}{\overline{m}_\theta^2} \left(H^\dagger H\right) 
    \left( 
    \ptwiddle{C}_{\theta B}\,\ptwiddle{B}_{\mu\nu}B^{\mu\nu} +
    \ptwiddle{C}_{\theta W}\,\ptwiddle{W}_{\mu\nu}W^{\mu\nu} 
    + \ptwiddle{C}_{\theta G}\,\ptwiddle{G}_{\mu\nu}G^{\mu\nu} 
    \right)\,,
\end{align}
with the $\left(\sim\right)$ notation implying that both the $H^2X^2$ and $H^2 \tilde{X} X$ field-strength operators are sourced with the same Wilson coefficients.  
The terms in \eqref{eq:SMEFT66fieldstrengths} are again implied in addition to the operators presented above. 
We see contributions to the Higgs-enhanced gluonic field strength tensors of the Warsaw basis for the first time, while the contributions to the electroweak analogues are in addition to those already found in Section~\ref{sec:dim5matching}, sourced at one-loop from $\DFN=4$ operators.
For those choices of parameters where the loop-matched Higgs-gauge Wilson coefficients of \eqref{eq:dim4loopmatchHiggsGauge} are larger than those shown above, the model then predicts a relative size difference in the Higgs-gauge operators of 
\begin{equation}
    \frac{C_{HW}}{C_{HG}}  
    =  
    \frac{g_Y^2 \lambda_{11} \overline{v}_\theta^2}{384 \pi^2 \overline{m}_\theta^2 \overline{\lambda}^2} \sim \frac{g_Y^2 \lambda_{11}}{40} \,,
    \quad
     \frac{C_{HB}}{C_{HG}} 
     =
     \frac{g^2 \lambda_{11} \overline{v}_\theta^2}{384 \pi^2 \overline{m}_\theta^2 \overline{\lambda}^2}
     \sim 
     \frac{g^2 \lambda_{11}}{40} \,,
\end{equation}
where we have, once again, assumed $\overline{m}_\theta \sim \overline{v}_\theta $ and $\overline{\lambda} \sim 0.1$.
However, given that one typically takes $\lambda_{11} \lesssim 1 $, it becomes clear that it is more sensible to assume that the tree-level result of \eqref{eq:SMEFT66fieldstrengths} dominates over the loop-level contribution and one therefore predicts $C_{HW} \sim C_{HB} \sim C_{HG}$.

\subsubsection{Matching Summary and Phenomenological Outlook}
\label{sec:phenooutlook}

The results above conclude our matching of the toy FN model implied in \eqref{eq:toymD} to the SMEFT up to $\DSM=6$.  In summary, we have considered the one-loop (tree-level) matching contributions to the SMEFT sourced by operators up to $\DFN = 5$ (6) in the FN theory. The results in \eqref{eq:dim4treematchfinal}, \eqref{eq:dim4loopmatchYukawa}--\eqref{eq:dim4loopmatchFourFermi}, \eqref{eq:dim5treematchfinal}--\eqref{eq:dim5loopmatchHiggsDerivative}, \eqref{eq:dim5loopmatchFourFermion}, \eqref{eq:SMEFT66Scalar}--\eqref{eq:SMEFT66Yukawas} and \eqref{eq:SMEFT66fieldstrengths} represent the core findings of our study, and reveal that Higgs-scalar ($H^6$), Higgs-derivative ($H^4 D^2$), Higgs-kinetic ($H^2 D \psi^2$), Higgs-enhanced Yukawa ($H^3 \psi^2$), Higgs-gauge ($H^2 X^2$) and four-fermion ($\psi^4$) operators are all sourced with non-zero Wilson coefficients in the IR SMEFT.
Their specific forms are also summarized in Tables~\ref{tab:treeleveldim4and5}--\ref{tab:4fermions}.  

While a dedicated phenomenological study of these results is beyond our scope in this paper, we have already attempted to highlight some of the most relevant signatures in the text above, as results have appeared. Here we simply summarize possible directions to falsify the FN mechanism with the SMEFT, recalling that falsification includes not only the FN mechanism itself, where \SFN{}-symmetry breaking occurs via a BSM flavon, but also the ability to distinguish one FN-type flavour model from another.  

Concerning the former, the first and perhaps most obvious point to make regards which operators do \emph{not} appear in Sections \ref{sec:dim4matching}--\ref{sec:dim6matching}.  Indeed, at least up to the orders we have considered (in the UV and IR), we do not see dipole-type  ($\psi^2 X H$) nor triple-field-strength ($X^3$) operators appearing in our matching analysis, although we do of course recall that such operators can appear in $\mathcal{L}_{\text{SMEFT}^\prime}$ and will filter down to the IR theory if so.  This will be ubiquitous for the triple-field-strength operators which have no flavour content, but only flavour-specific combinations of the dipole SMEFT$^\prime$ operators will appear (the same flavour combinations as allowed at the renormalizable level in \eqref{eq:toymD}).  Hence the presence of sizable non-zero Wilson coefficients in dipole operators outside of these limited combinations would signal a deviation and/or extension from the FN-type dynamics we have studied here. 

Similarly, the generic FN mechanism relies on the presence of a BSM flavon and, as we have seen in Sections \ref{sec:dim4matching}--\ref{sec:dim6matching}, its mixing with the SM Higgs boson, either via the renormalizable Higgs portal $\propto \lambda_{11}$ in \eqref{eq:toyLdim4} or its $\DFN=6$ portal analogues $\propto \lambda_{12,21}$ in \eqref{eq:dim6Scalarops}, plays a major role in our matching analysis.  In fact, we already pointed out that a non-zero $C_{H\Box}$ ($\propto \lambda_{11}^2$) is generated from renormalizable operators at tree level in \eqref{eq:dim4treematchfinal}, and leads to a straightforward flavon mass bound of $m_\theta \gtrsim 1$ TeV (also see \cite{Ellis:2020unq}). 
Better bounds on $C_{H \Box}$ from (e.g.) Higgs coupling analyses at the LHC will immediately translate to stricter bounds on the two-dimensional $\lambda_{11}$-$m_\theta$ parameter space. 
We also notice that at $\DFN=6$ all of the newly obtained Wilson coefficients contain terms proportional to (powers of) $\overline{\lambda}^2 \lambda_{12}$ whose origin is the $\DFN=6$ Higgs-flavon scalar potential portal term $(H\hc H) (\theta^* \theta)^2$ which, after \SFN{} symmetry breaking, takes the form
\begin{equation}
    \frac{\lambda_{12}}{\LUV^2} (H\hc H) (\theta^* \theta)^2 \rightarrow 
    2 \lambda_{12} \overline{\lambda}^2 v_\theta \vartheta H\hc H + \ldots 
\end{equation}
This $\lambda^2$-suppressed three-point vertex can be used to build Feynman diagrams just like the familiar $\lambda_{11} v_\theta \vartheta H\hc H$ term from \eqref{eq:L4thetaexpanded}, explaining its frequent appearance in the Wilson coefficients of this section. 
These terms thus help generate non-trivial dimension-6 SMEFT operators even when the renormalizable Higgs-flavon mixing parameter $ \lambda_{11} $ is small. 
In general though, it is clear that in the absence of any sizable Higgs-flavon mixing term in the IR, regardless of its UV operator source, the number of phenomenologically relevant operators appearing in Sections \ref{sec:dim4matching}--\ref{sec:dim6matching} is greatly reduced, as expected.

On the other hand, phenomenologically relevant BSM effects can also be sourced from four-fermion interactions, and we observe from \eqref{eq:dim4loopmatchFourFermi} and \eqref{eq:dim5loopmatchFourFermion} that these are turned on at one-loop level in our FN setup by both $\DFN=4$ and $\DFN=5$ operators. 
Such four-fermion terms can contribute to flavour-changing neutral-current (FCNC) observables, such as neutral meson mixing, and often produce the most stringent constraints on flavourful BSM models.
For the FN model in question the four-fermions may thus be expected to provide the leading lower bound on $m_\theta$. 
However, as evident in \eqref{eq:dim4loopmatchFourFermi} and \eqref{eq:dim5loopmatchFourFermion}, the four-fermion operators are suppressed not only by powers of $m_\theta$ but also $16 \pi^2$ and (powers of) $\lambda$. 
Furthermore, by construction, the FN charge assignments ensure that four-fermion operators involving the lightest quarks, which contribute to increased neutral meson mixing, will be more suppressed than those containing the heavy quarks. 
One caveat of this analysis is that to achieve a realistic comparison with the FCNC bounds, the fermions would have to be rotated from the interaction eigenbasis to the mass eigenbasis. 
Setting the $c^{d_i}_{ij}$ coefficients to unity in the down-quark Yukawa matrix of \eqref{eq:toymD} yields a rank-1 matrix with a degenerate eigenspace, and so the quark rotation matrices cannot be uniquely determined without making more elaborate assumptions about the Wilson coefficients.
We do, of course, expect nearly diagonal quark rotation matrices, with the off-diagonal elements controlled by increasing powers of $\lambda$.
All in all, the suppression structures discussed in this section suggest that the commonly-held assumption that the presence of four-fermion operators will push $m_\theta$ to PeV scales may require further thought in FN-type setups. Exploratory studies implementing a set of four-fermion, Higgs-kinetic and Higgs-enhanced Yukawa operators involving down-type quarks, with coefficients corresponding to our matching results, in {\tt{smelli}} \cite{Aebischer:2018iyb}\footnote{{\tt{smelli}} is a global likelihood optimization tool which automatically runs down and matches SMEFT Wilson coefficients to the LEFT and subsequently compares them against electroweak scale and low-energy flavour data.} indicate $m_\theta \gtrsim \mathcal{O}\left(5\right)$ TeV.
In contrast, including a full set of four-fermion operators of the type $\mathcal{O}_{dd}$, $\mathcal{O}_{qd}^{(1) }$ and $\mathcal{O}_{qd}^{(8)}$ with no suppression other than the power of $ \lambda $ required by spurion analysis lifts the bounds to the multi-PeV scale. 
The majority of simple UV completions will of course not give rise to such an anarchic scenario and we leave for future work the classification of possible UV completions that are able to circumvent the subset of four-fermion operators that are the most constrained.

The above discussion highlights the importance of studying SMEFT operators besides the four-fermion ones, given that other Wilson coefficients may exhibit significantly less suppression after the FN EFT is matched to the SMEFT at a given order. For example, measuring Higgs observables that acquire corrections from the Higgs-enhanced Yukawa terms found above (or the Higgs-kinetic $\psi^2 H^2 D $ operators), could yield constraints on $m_\theta$ and other FN parameters that are competitive with FCNC bounds.  
Furthermore, the form of the Higgs-enhanced Yukawas following from our matching shows that their Wilson coefficients exhibit the same flavour hierarchies, organised in powers of $\lambda$, as the Yukawa elements themselves (see \cite{Asadi:2023ucx} for a discussion on dynamics and/or symmetries in the UV theory which could lead to deviations from this pattern).  
This indicates a distinct phenomenological SMEFT signature in the FN setup.
More broadly, our results also indicate the importance of matching UV theories beyond tree level, and also potentially motivate studying the SMEFT output to higher orders than $\DSM=6$ --- one may find that less parametrically suppressed higher-dimension operators contribute as much to the IR physics as more parametrically suppressed lower-dimension ones.

With respect to distinguishing different FN models in the Abelian \ZFN{}-type constructions presented above, it is (unsurprisingly) clear that precision flavour constraints are required.  After all, \SFN{} is a UV flavour symmetry! 
As already discussed in Section \ref{sec:dim5matching}, the differences in the parametric dependence of the flavourful Wilson coefficients appearing in Sections \ref{sec:dim4matching}--\ref{sec:dim6matching} suggest that their relative sizes may be used to distinguish independent FN models --- one UV charge scenario will generate different hierarchies to another.  Consider (e.g.) the final four-fermion operator $\propto (\overline{Q}_3 \gamma_\mu Q_3)(\overline{d}_1 \gamma^\mu d_1)$ in \eqref{eq:dim5loopmatchFourFermion}.  That this particular flavour ensemble appears with no $\lambda_{11}$-dependence is evident from Table~\ref{tab:4fermions} and the loop diagram in Fig.~\ref{fig:dim5loops}(a).  Were the charges of a particular UV model aligned such that different quark flavours $\lbrace i, j \rbrace$ leave the combination $\overline{Q}_i H \theta d_j$ \SFN{}-invariant, then those flavours would be the ones apparent in the four-fermion operator. Similar suppression patterns can be found throughout all of the flavoured SMEFT operators that appear.

We also find it important to recall the obvious; the main goal of FN-type constructions as introduced in Section \ref{sec:DOWNTOY} is to dynamically explain observed patterns of fermionic mass, mixing, and CP violation.  Hence data on extracted quark mass eigenvalues and CKM mixing angles and phases from (e.g.)\ the PDG \cite{ParticleDataGroup:2022pth} serves as the first point of phenomenological comparison between different FN models. 
However, care must be taken in the context of the SMEFT, given that the Higgs-enhanced Yukawa operators that appear throughout our matching (cf. e.g. \eqref{eq:dim4loopmatchYukawa} or \eqref{eq:dim5treematchfinal}) will, after EWSB, effectively renormalize the Yukawa couplings and therefore mass and mixing outputs.  Indeed, in \cite{Talbert:2021iqn} the definitions of mass eigenvalues and CKM parameters are given \emph{exactly} in terms of Yukawa couplings, and these relations hold regardless of how many $(H^\dagger H)$ insertions serve to enhance these parameters' formal theory definitions.  Furthermore, \cite{Descotes-Genon:2018foz} correctly notes that, upon extracting a mass or mixing parameter by comparing theory (the SMEFT, at a given mass dimension) to data, there are indirect  SMEFT constraints sourced in all processes that then depend on those extracted parameters, e.g. heavy flavour or W-boson decays.  

We are of course aware that these speculations do not amount to a robust phenomenological study of the matching analysis in Sections \ref{sec:dim4matching}--\ref{sec:dim6matching}.  Such a study would ideally incorporate both a likelihood analysis with respect to existing global $\DSM=6$ SMEFT fits (see e.g. \cite{Ellis:2020unq,Ethier:2021bye,Iranipour:2022iak,Brivio:2022hrb}), as well as more tailored studies with {\tt{smelli}}.  We will briefly reemphasize this point in our Summary and Outlook in Section \ref{sec:CONCLUDE}.

\section{Generalizations and Extensions Beyond the Toy Model}
\label{sec:GENERALIZE}
In the analysis of Section \ref{sec:MODEL} we have assumed FN dynamics only in the down-quark sector and in a particular charge setup (cf.\ \eqref{eq:toyLdim4}).
We will now discuss generalizations and extensions to this toy model analysis that more completely characterize the matching of generic FN models to the SMEFT.  Note that our goal here is not to explore every scenario in full generality as above (for example, loop-level matching will be largely unconsidered in what follows) but rather to highlight the most relevant and/or obvious points regarding realistic FN theories beyond Table~\ref{tab:DownToy}, many of which will evolve into future study.

\subsection{Charge Generalizations and Extensions to Different Families}
\label{sec:GENERALCHARGE}
The model encapsulated in \eqref{eq:toymD} is but one of many possible choices in an effort to resolve down quark mass hierarchies.  Furthermore, we have left the \ZFN{} charges for the up quarks and charged leptons unspecified in our matching procedures above, having simply allowed for the presence of renormalizable couplings in these sectors without the assumption that FN dynamics successfully structure their Yukawas. 
In this section we discuss the generalization of {\bf{(1)}} the assignment of \ZFN{} charges within our toy model and  the extension of structured \ZFN{} charges to {\bf{(2)}} the up quark and {\bf{(3)}} lepton families.

\subsubsection*{Generalization of Down Quark Charges}
\begin{table}[htpb]
	\centering
 \begingroup
\addtolength{\jot}{-0.25em}
    \begin{minipage}[t]{.49\linewidth}
	\tabletitle{Tree-level matching at $\DFN=4$}
	\begin{alignat*}{2}
		\\[-1.8cm]
		\cline{1-4}\\[-1.cm]
	&\text{Operator} &\qquad&   \text{Wilson coefficient} \\[-.5cm]
	\cline{1-4}\\[-.9cm]
	 &H^\dagger H  &&-\frac{\lambda_{11} v_\theta^2}{2} \\
	 &(H^\dagger H)^2 &&\quad \, \frac{\lambda_{11}^2 v_\theta^2}{2 m_\theta^2} \\ 
	 &\left(H^\dagger H\right)\Box\left(H^\dagger H\right) &&-\frac{\lambda_{11}^2 v_\theta^2}{2 m_\theta^4} \\[-0.2cm]
	\cline{1-4}
\end{alignat*}
    \end{minipage}
\begin{minipage}[t]{.49\textwidth}
\tabletitle{Tree-level matching at $\DFN=5$}
	\begin{alignat*}{2}
		\\[-1.8cm]
		\cline{1-4}\\[-1.cm]
	&\text{Operator} &\qquad&   \text{Wilson coefficient} \\[-.5cm]
	\cline{1-4}\\[-.9cm]
	 &\overline{Q}_i H d_j  &&\quad \, c_{ij}^d\, \lambda \hspace{.3cm}\\
	 &(H^\dagger H) \overline{Q}_i H d_j &&-\frac{\lambda_{11}\, \lambda\, c_{ij}^d}{m_\theta^2} \\[-0.2cm]
	\cline{1-4}
\end{alignat*}
\end{minipage}
\\[-0.5cm]
\caption{The (SM)EFT operators and associated Wilson coefficients appearing from a tree-level matching of the $\DFN=4 (5)$ FN Lagrangian in \eqref{eq:toyLdim4} and \eqref{eq:toyLdim5}, respectively. 
\label{tab:treeleveldim4and5}}
	\begin{tabularx}{0.9\textwidth}{>{\centering\arraybackslash}X}
		Tree-level matching at $\DFN=6$\\[-0.25cm]
	\end{tabularx}
\begin{align*}
	\hline
	 & \text{Operator} && \text{Wilson coefficient} \\
	 \hline \\[-.6cm]
	 &H^\dagger H  && \frac{ \,\overline{\lambda}^2 \lambda_{12} \overline{v}_\theta^2}{2} \hspace{7.9cm}\\
	 &\left(H^\dagger H\right)^2 && \frac{1}{ \overline{m}_\theta^2} 
     \left[\overline{\lambda}^2 \left(\lambda_{21} m_\theta^2 - 2 \lambda_{11} \lambda_{12}  \overline{v}_\theta^2 \right)
    + 2 \overline{\lambda}^4 \lambda_{12}^2 \overline{v}_\theta^2  \right] \\
	 &\left(H^\dagger H\right)^3 && - \frac{2}{m_\theta^6} \left[ \lambda_{11} \overline{\lambda}^2 \left(  \lambda_{03} \lambda_{11}^2 \overline{v}_\theta^4 - \lambda_{11} \lambda_{12} m_\theta^2 \overline{v}_\theta^2 +  \lambda_{21} m_\theta^4 \right) \right. \\
	&&& \left. - 2 \overline{\lambda}^4 \left( 3  \lambda_{03} \lambda_{11}^2 \lambda_{12} \overline{v}_\theta^4 - 2 \lambda_{11} \lambda_{12}^2 \overline{m}_\theta^2 \overline{v}_\theta^2 + \lambda_{12} \lambda_{21} \overline{m}_\theta^4 \right) + \mathcal{O}\Big( \overline{\lambda}^6 \Big)   \right] \\
     &\left(H^\dagger H\right)\Box \left(H^\dagger H\right) && \frac{2}{\overline{m}_\theta^4} \left[ \lambda_{11} \overline{\lambda}^2 \left( \lambda_{12} \overline{v}_\theta^2 - c_{H\Box\theta} \overline{m}_\theta^2 \right)  - \lambda_{12} \overline{\lambda}^4 \left(  \lambda_{12} \overline{v}_\theta^2 - 2 c_{H \Box \theta} m_\theta^2\right)  \right] \\ 
     &\overline{Q}_{i} H d_j &&  c^{d_6}_{ij} \,\overline{\lambda}^2\\
     &\left(H^\dagger H\right)\overline{Q}_{i} H d_j && \frac{2\,c_{ij}^d \,\overline{\lambda}^3\,\lambda_{12}}{\overline{m}_\theta^2} + \,\frac{2 c^{d_6}_{ij} \overline{\lambda}^2\big(2\overline{\lambda}^2\,\lambda_{12}-\lambda_{11}\big)}{\overline{m}_\theta^2} \\
      & \left(H^\dagger H\right)  
    \ptwiddle{X}_{\mu\nu}X^{\mu\nu} && \ptwiddle{C}_{\theta X}\,\frac{2\overline{\lambda}^2\left(2\overline{\lambda}^2\,\lambda_{12} -\lambda_{11}\right)}{\overline{m}_\theta^2} \\[0.1cm]
	\hline
\end{align*}
\\[-0.6cm]
\caption{The SMEFT operators appearing from a tree-level matching of the $\DFN=6$ FN Lagrangian. Recall the barred notation accounts for corrections due to expanding about the new minimum of the associated scalar potential. In the sixth and seventh lines, the notation in the RH column indicates that the index pattern on $c^d_{ij}$ and $c^{d_6}_{ij}$ follows that of the indices on $\overline{Q}_i$ and $d_j$ in the left-hand column. 
Finally, $X \in \lbrace B, W, G \rbrace$ in the final line, with tilded notation following that in the text around \eqref{eq:SMEFT66fieldstrengths}.
\label{tab:treeleveldim6}
}
\endgroup
\end{table}

\begin{table}[htpb]
	\centering
	\begin{tabularx}{0.9\textwidth}{>{\centering\arraybackslash}X}
		One-loop matching at $\DFN{} = 5$ \, --- \, $\psi^2 H $ (Yukawas)\\[-0.15cm]
	\end{tabularx}
\begin{align*}
	\hline
	 & \text{Operator} && \text{Wilson coefficient} \\
	 \hline \\[-.7cm]
	 &\overline{L}_i H e_j  && \frac{\hbar y^e_{ij} \lambda_{11}^2 v_\theta^2}{12 m_\theta^4} \left[ 3 m_\theta^2 - m_H^2 \left( 17  + 6 \mathbb{L} \right)  \right] \hspace{7.9cm}\\
	 &\overline{Q}_i H d_j && \frac{\hbar}{12 m_\theta^4 v_\theta^2} \biggl[ 6  c_{ij}^d \lambda m_\theta^2 \bigl(3 m_\theta^4 (1 + \mathbb{L} ) - 2 \lambda_{11} v_\theta^2 m_\theta^2 (1 + \mathbb{L} ) + \lambda_{11} m_H^2 v_\theta^2 (3 + 2 \mathbb{L} )  \bigr) \\ 
	 &&&+y^d_{ij} \lambda_{11}^2 v_\theta^4  \bigl(3 m_\theta^2 - m_H^2 (17 + 6 \mathbb{L} ) \bigr)\biggr] \\
	 &\overline{Q}_i \tilde{H} u_j && \frac{\hbar y^u_{ij}  \lambda_{11}^2 v_\theta^2}{12 m_\theta^4} \bigl[ 3 m_\theta^2 - m_H^2 (17 + 6 \mathbb{L} ) \bigr] \\[0.1cm]
	\hline
\end{align*}
\\[-0.5cm]
\caption{The modified renormalizable Yukawa terms at one-loop. 
Whilst not SMEFT operators, we have presented them here to illustrate the one-loop corrections to the Yukawa matrices.
Though not shown here for brevity, one also finds corrections to the coefficients of the renormalizable $ H^2 $ and $ H^4 $ operators.
}
\label{tab:renormalisableTerms}
\end{table}

\begin{table}[htpb]
	\centering
	\begin{tabularx}{0.9\textwidth}{>{\centering\arraybackslash}X}
		One-loop matching at $\DFN{} = 5$ \, --- \,  $H^6$ and $ H^4 D^2 $ \\[-0.2cm]
	\end{tabularx}
\begin{align*}
	\hline
	 & \text{Operator} && \text{Wilson coefficient} \\
	\hline\\[-0.75cm]
	 &\left( \hdh  \right)^3 && \frac{\hbar \lambda_{11}^2}{9 m_\theta^8} \biggl[ 6 \lambda_{11} m_\theta^6 -6 \lambda_{11}^4 v_\theta^6   + 18\lambda_{11}^2 v_\theta^4 m_\theta^2   ( \lambda_{11} - 6 \lambda_{20}' )  \\
	 &&&  + v_\theta^2 m_\theta^4 \Bigl( - 18 \lambda_{11}^2 + 108 \lambda_{20}' \lambda_{11} +  \lambda_{20}' \bigl( 12 \lambda_{20}' (41 + 30 \mathbb{L} ) - g^2 (31 + 30 \mathbb{L} )   \bigr) \Bigr) \biggr] \\
	 & \left( \hdh \right) \Box \left( \hdh \right) &&   - \frac{\lambda_{11}^2 v_\theta^2}{2 m_\theta^4} + \frac{\hbar \lambda_{11}^2}{12 m_\theta^2} \bigg[ 16 - 18 \mathbb{L}  - \frac{2 \lambda_{11} v_\theta^2 }{ m_\theta^2} \left( 7  - 12 \mathbb{L}  \right) + \frac{\lambda_{11}^2 v_\theta^4 }{m_\theta^4} \left( 19 + 12 \mathbb{L}  \right) \\
		 	&&&  - \frac{g_Y^2 v_\theta^2}{6 m_\theta^2} \left( 31 + 30 \mathbb{L}  \right)  - \frac{g^2 v_\theta^2}{2 m_\theta^2} \left( 31 + 30 \mathbb{L} \right)  \bigg]\\
	&( H\hc D^\mu H )\hc ( H\hc D_\mu H ) && - \frac{\hbar g_Y^2 \lambda_{11}^2 v_\theta^2 }{18 m_\theta^4 } \left[ 31 + 30 \mathbb{L}  \right]   \\[0.1cm]
	\hline
\end{align*}

 \begin{tabularx}{0.9\textwidth}{>{\centering\arraybackslash}X}
		One-loop matching at $\DFN{} = 5$ \, --- \,  $\psi^2 H^3 $\\[-0.2cm]
	\end{tabularx}
\begin{align*}
	\hline
	 & \text{Operator} && \text{Wilson coefficient} \\
	\hline\\[-0.75cm]
	 &( \hdh  ) \overline{L}_i He_j   && \frac{\hbar \lambda_{11}^2 v_\theta^2}{36 m_\theta^4} \Bigl[ 9 (y^e y^{e\dagger} y^e)_{ij}  (5 + 6 \mathbb{L}) + 6 y^e_{ij} \lambda_{20}' (29 + 6 \mathbb{L}) - y^e_{ij} g^2  (31 + 30 \mathbb{L}) \Bigr]   \\
	 &( \hdh  )  \overline{Q}_i H d_j   && \frac{- \lambda_{11} \lambda c_{ij}^d}{ m_\theta^2} + \frac{\hbar}{144 v_\theta^2 m_\theta^6} \Bigl\{ 18 v_\theta^2 m_\theta^4 \Bigl[ \frac{ m_\theta^2}{v_\theta^2} \lambda^2 (y^d {c^d}^{\dagger} c^d + c^d {c^d}^{\dagger} y^d )_{ij} (1 + 2 \mathbb{L}) \\
	 &&&+ 4 \lambda_{11} \lambda  (y^d {c^d}^\dagger y^d )_{ij}  \Bigr] + 36 v_\theta^2 m_\theta^2 \Bigl[ \frac{4 \lambda^2 m_\theta^4}{v_\theta^2} (c^d {y^d}^{\dagger} c^d )_{ij} (1 + \mathbb{L}) \\
	 &&& - 6 \lambda_{11}\lambda m_\theta^2  \big(c^d {y^d}^{\dagger} y^d +y^d {y^d}^{\dagger} c^d \big)_{ij}  (1 + \mathbb{L} ) + \lambda_{11}^2 v_\theta^2 (y^d {y^d}^{\dagger} y^d )_{ij} (5 + 6 \mathbb{L})  \Bigr]\\
	 &&&   - 2 \lambda_{11}  \Bigl[ 2 \lambda_{11} m_\theta^2 v_\theta^4 y^d_{ij} \Bigl( -6  \lambda_{20}' (29 + 6 \mathbb{L} ) + g^2 ( 31 + 30 \mathbb{L}) \Bigr)\\
	 &&&  + 18 \lambda m_\theta^2 c^d_{ij} \Bigl( 6 m_\theta^4 (-1 + \mathbb{L} ) - \lambda_{11}^2 v_\theta^4 (15 + 4 \mathbb{L} ) \\
	 &&& + 2 v_\theta^2 m_\theta^2 \big(\lambda_{11} (4 - 6 \mathbb{L} ) + 2 \lambda_{20}' (7 + 6 \mathbb{L} ) \big) \Bigr)  \Bigr] \Bigr\}\\
	 &( \hdh  )   \overline{Q}_i \tilde{H}u_j  && \frac{\hbar \lambda_{11}^2 v_\theta^2 }{36 m_\theta^4} \left[ 9 (y^u {y^u}^{\dagger} y^u)_{ij}   (5 + 6 \mathbb{L} ) + 6 y^u_{ij} \lambda_{20}' (29 + 6 \mathbb{L} ) - y^u_{ij} g^2  (31 + 30 \mathbb{L} ) \right] \\[0.05cm]
	\hline
\end{align*}
\\[-0.5cm]
\caption{One-loop matching contributions from the dimension-5 FN EFT to the Higgs and Higgs-enhanced Yukawa operators in the Warsaw basis. 
The down-quark charges under \SFN{} are taken to be arbitrary. 
\label{tab:HiggsAndHiggsEnhancedYuk}}
\end{table}

\begin{table}[htpb]
	\centering
	\tabletitle{One-loop matching at $\DFN{} = 5$ \, --- \,  $ X^2 H^2 $}
	\begin{alignat*}{2}
		\\[-1.5cm]
		\cline{1-4}\\[-1.2cm]
	&\text{Operator} &\qquad \qquad&   \text{Wilson coefficient} \\[-.6cm]
	\cline{1-4}\\[-1.2cm]
	&( \hdh ) B_{\mu \nu} B^{\mu \nu} && \frac{\hbar g_Y^2 \lambda_{11}^2 v_\theta^2 }{12 m_\theta^4}  \\
			&( H\hc \tau^I H ) W_{\mu \nu}^I B^{\mu \nu} && \frac{\hbar g g_Y \lambda_{11}^2 v_\theta^2 }{6 m_\theta^4}  \\
			&( \hdh ) W_{\mu \nu}^I W^{I, \mu \nu} && \frac{\hbar g^2 \lambda_{11}^2 v_\theta^2 }{12 m_\theta^4} \\[-.3cm]
			\cline{1-4}
\end{alignat*}
\\[-0.5cm]
\caption{One-loop matching contributions from the dimension-5 FN EFT to the Higgs-field strength operators in the Warsaw basis.
    These operators receive contributions only from Higgs-flavon mixing.
    \label{tab:fieldStrengths}}
\end{table}

\begin{table}[htpb]
	\centering
	\tabletitle{One-loop matching at $\DFN{} = 5$ \, --- \, $ \psi^2 H^2 D $}
	\begin{align*}
		\hline
		 & \text{Operator} && \text{Wilson coefficient} \\
		 \hline \\[-0.75cm]
		 &  ( H\hc \Darr H ) (\overline{d}_i \gamma^\mu d_j ) &&  \frac{\hbar}{432 m_\theta^4} \Bigl[ 108 \lambda_{11} \lambda  m_\theta^2 \big( {c^d}^\dagger y^d + {y^d}^{\dagger} c^d \big)_{ij} + \frac{54 \lambda^2 m_\theta^4}{v_\theta^2} ({c^d}\hc c^d )_{ij} (1 + 2\mathbb{L} ) \hspace{1.7cm}\\
		 &&& + 2 \lambda_{11}^2 v_\theta^2 \bigl(  g_Y^2  \delta_{ij} (17 + 6 \mathbb{L} )  - 27   ({y^d}^{\dagger} y^d )_{ij} (5 + 2 \mathbb{L} )  \bigr)\Bigr]\\
		 &  i( \tilde{H}\hc D_\mu H  ) ( \overline{u}_i \gamma^\mu d_j )    && \frac{\hbar \lambda_{11} }{4 m_\theta^4 } \Bigl[ 2 \lambda m_\theta^2 ( {y^u}^\dagger c^d )_{ij} - \lambda_{11} v_\theta^2 ({y^u}^{\dagger}y^d )_{ij} (5 + 2 \mathbb{L} ) \Bigr] \\
		 &	( H\hc \Darr H ) (\overline{e}_i \gamma^\mu e_j ) &&  \frac{ - \hbar \lambda_{11}^2 v_\theta^2}{72 m_\theta^4} \left[ 9 ({y^e}^{\dagger} y^e )_{ij} (5 + 2 \mathbb{L} ) - g_Y^2 \delta_{ij} (17 + 6 \mathbb{L} )  \right]   \\
		 &	( H\hc \Darr H ) ( \overline{L}_i \gamma^\mu L_j)&& \frac{\hbar \lambda_{11}^2 v_\theta^2}{144 m_\theta^4} \left[ 9 (y^e {y^e}^{\dagger} )_{ij} (5 + 2 \mathbb{L} ) + g_Y^2 \delta_{ij}(17 + 6 \mathbb{L} ) \right] \\
		 &	( H\hc  \DarrI H ) ( \overline{L}_i \tau^I \gamma^\mu L_j)&& \frac{\hbar \lambda_{11}^2 v_\theta^2}{144 m_\theta^4} \left[ 9 (y^e {y^e}^{\dagger} )_{ij} (5 + 2 \mathbb{L} ) - g^2 \delta_{ij}(17 + 6 \mathbb{L}) \right] \\
		 &	( H\hc \Darr H ) ( \overline{Q}_i \gamma^\mu Q_j)&&  \frac{ - \hbar}{288 m_\theta^4 } \Bigl[ 36 \lambda_{11} \lambda m_\theta^2  \big( y^d {c^d}\hc + c^d {y^d}^{\dagger} \big)_{ij} + \frac{18 m_\theta^4 \lambda^2}{v_\theta^2} (c^d {c^d}\hc )_{ij}  (1 + 2\mathbb{L}) \\
		 &&& + 2 \lambda_{11}^2 v_\theta^2 \Bigl( \frac{g_Y^2}{3} \delta_{ij} (17 + 6 \mathbb{L} ) - 9    (y^d {y^d}^{\dagger} - y^u {y^u}^{\dagger} )_{ij} (5 + 2\mathbb{L} )     \Bigr) \Bigr] \\
		 &	( H\hc  \DarrI H ) ( \overline{Q}_i \tau^I \gamma^\mu Q_j)&&  \frac{ - \hbar}{288 m_\theta^4} \Bigl[ 36  \lambda_{11} \lambda m_\theta^2   \big( y^d {c^d}\hc + c^d {y^d}^{\dagger} \big)_{ij} + \frac{18 \lambda^2 m_\theta^4}{v_\theta^2} (c^d {c^d}\hc )_{ij}  (1 + 2\mathbb{L}) \\
		&&& + 2 \lambda_{11}^2 v_\theta^2 \Bigl( g^2    \delta_{ij} (17 + 6 \mathbb{L} ) - 9   (y^d {y^d}^\dagger + y^u {y^u}^{\dagger} )_{ij} (5 + 2\mathbb{L} )    \Bigr) \Bigr] \\
		&	( H\hc \Darr H ) ( \overline{u}_i \gamma^\mu u_j)&& \frac{\hbar \lambda_{11}^2 v_\theta^2}{216 m_\theta^4} \left[ 27 ({y^u}^{\dagger} y^u )_{ij} (5 + 2 \mathbb{L}) - 2 g_Y^2 \delta_{ij} (17 + 6 \mathbb{L} ) \right] \\
		\hline
	\end{align*}
 \\[-0.5cm]
	\caption{One-loop matching contributions from the dimension-5 FN EFT to the Higgs kinetic operators in the Warsaw basis.
		The down-quark charges under \SFN{} are taken to be arbitrary.
    \label{tab:HiggsKinetic}}
\end{table}

\begin{table}[htpb]
	\centering
	\tabletitle{One-loop matching at $\DFN{} = 5$ \, --- \,$ \psi^4: \; (\overline{L} R ) (\overline{R} L )$ and $ ( \overline{L} R) ( \overline{L} R) $}
	\begin{alignat*}{4}
		\\[-1.4cm]
		\cline{1-8}\\[-1.2cm]
		 &\qquad && \text{Operator} &\qquad & \text{Wilson coefficient} &\qquad &  \\[-0.6cm]
		 \cline{1-8}\\[-1.2cm]
		 &&&  (\overline{L}^\alpha_i e_j) ( \overline{d}_k Q_l^\alpha) && \frac{\hbar \lambda_{11} y^e_{ij} }{12 m_\theta^4} \left[ 2 \lambda_{11} ({y^d}^{\dagger})_{kl} v_\theta^2 - 6 \lambda  ({c^d}\hc)_{kl} m_\theta^2 \right] && \\
		 &&& (\overline{L}_i^\alpha e_j) \epsilon_{\alpha\beta} (\overline{Q}_k^\beta u_l ) && \frac{\hbar \lambda_{11}^2 v_\theta^2 y^e_{ij} y^u_{kl} }{6 m_\theta^4} && \\
		 &&& (\overline{Q}_i^\alpha u_j ) \epsilon_{\alpha\beta} (\overline{Q}^\beta_k d_l )&& \frac{\hbar \lambda_{11} y^u_{ij} }{12 m_\theta^4 } \left[ 2 \lambda_{11} v_\theta^2 y^d_{kl} - 6 c^d_{kl} \lambda m_\theta^2 \right] && \\[-.3cm]
		 \cline{1-8}
	\end{alignat*}
    \\[-0.3cm]
	\tabletitle{One-loop matching at $\DFN{} = 5$ \, --- \, $ \psi^4: \; (\overline{L} L ) (\overline{R} R ) $}
	\begin{alignat*}{4}
		\\[-1.4cm]
		\cline{1-8}\\[-1.2cm]
		 &\quad&& \text{Operator} &\qquad& \text{Wilson coefficient} &\quad& \\[-0.6cm]
		 \cline{1-8}\\[-1.2cm]
		 &&&  (\overline{Q}_i \gamma_\mu Q_j) (\overline{d}_k \gamma^\mu d_l ) && - \frac{\hbar}{12} \left[ \frac{\lambda_{11}^2 v_\theta^2 y^d_{il} ({y^d}^{\dagger})_{kj} }{m_\theta^4}  +\frac{6 \lambda^2 c^d_{il} ({c^d}\hc)_{kj}}{v_\theta^2} (1 + \mathbb{L} ) \right] && \\
		 &&& (\overline{L}_i \gamma_\mu L_j)  (\overline{e}_k \gamma^\mu e_l) && - \frac{\hbar \lambda_{11}^2 v_\theta^2 y^e_{il} ({y^e}^{\dagger} )_{kj}}{12 m_\theta^4} && \\
		 &&& (\overline{Q}_i \gamma_\mu Q_j)  (\overline{u}_k \gamma^\mu u_l)&& - \frac{\hbar \lambda_{11}^2 v_\theta^2 y^u_{il} ({y^u}^{\dagger})_{kj} }{12 m_\theta^4} && \\[-0.3cm]
		 \cline{1-8}
	\end{alignat*}
 \\[-0.5cm]
	\caption{One-loop matching contributions from the dimension-5 FN EFT to the four-fermion operators in the Warsaw basis.
    The down-quark charges under \SFN{} are taken to be arbitrary.
    \label{tab:4fermions}
 }
\end{table}

The simplest generalization to Section \ref{sec:MODEL} is to still consider only down-type FN dynamics, but to allow for charge assignments beyond those in Table~\ref{tab:DownToy}. It is clear that, in such a scenario, the results of Section \ref{sec:DOWNTOY} will hold, up to the explicit flavour labels that appear, which will of course be altered to reflect the new model's charges. 

For example, everywhere one sees terms $\propto y_{32,33}^d$ in (e.g.)\ \eqref{eq:dim4loopmatchYukawa} or \eqref{eq:dim4loopmatchHiggsDerivative}, one would instead expect to see terms $\propto y^d_{ij}$ in the new model, where $y^d_{ij}$ are any renormalizable Yukawa matrix elements $\propto \lambda^0$ appearing in the model --- cf.~\eqref{eq:toymD}. One would clearly also expect straightforward modifications to the Wilson coefficients $c_{ij}^d$ that appear at different mass dimensions in \eqref{eq:toymD}, and therefore contribute with different relative magnitudes to the Wilson coefficients of the SMEFT.  We have been careful when presenting our results in Section \ref{sec:MODEL} to point out those features which are readily generalizable to different charge configurations, and in fact Tables~\ref{tab:renormalisableTerms}--\ref{tab:4fermions} are already presented in a fully generic flavour notation.

One might suspect that an interesting extension of the results in Section \ref{sec:MODEL} could appear if the charges populating the $\DFN=5$ Yukawa sector were modified such that two or more Wilson coefficients akin to $c_{31}^d$ appear in the UV Lagrangian.  However, we have checked this possibility by imagining the (23) matrix element is $\lambda$ (as opposed to $\lambda^2$) suppressed.  
We find that the only additional SMEFT operators that appear are given by
\begin{equation}
\mathcal{L}_{\text{SMEFT}} \supset -  \frac{\lambda_{11} \lambda \, c^d_{23}}{m_\theta^2} \left(H\hc H \right)  \overline{Q}_2 H d_3  + \text{h.c.},
\end{equation}
when matching to the $\DSM=6$ SMEFT at tree level. These conclusions are subject to modification when matching at $\DSM \ge 6$ and/or when considering loop-level matching effects.

\subsubsection*{Extension to Up Quarks}
A complete description of quark flavour physics requires \SFN{} dynamics in both the up and down quark sector.  
On the one hand, the up quark family exhibits the largest mass hierarchy, given a top mass on the order of the Higgs vev.  
On the other hand, the CKM matrix is constructed out of a product of two unitary matrices which transform the up- and down-type left-handed fermions from the gauge eigenbasis to the mass eigenbasis, $V_\text{CKM} = V_{u_L}\hc V_{d_L}$. 
Each element of the matrix is measurable in charged-current interactions and ties together the up- and down-quark Yukawa matrices.

The simplest way to study the up quarks is to assign them a set of charges in complete analogy to Table~\ref{tab:DownToy}, accounting, of course, for the fact that the Higgs boson enters the Yukawa elements in question as $\tilde{H} $, which has charge +3 under \SFN{}.
For instance, in order to achieve the same hierarchies in the up and down quark Yukawa matrices, the right-handed up-type quarks could be assigned charges $q_{u_1} = -1$, $q_{u_2} = -3$ and $q_{u_3} = -3$ under the Abelian symmetry.
We stay rather agnostic about the up-sector charges in this section and simply assume, in analogy to the down-sector, that some up-type Yukawa elements are allowed at the tree-level, denoted by $y^u_{ij}$, that others enter at dimension-5 level, with coefficients $c^u_{ij} / \LUV$ and so on.

We first approach the matching to the SMEFT by turning off the down-type FN sector of the model and keeping only those FN terms which are responsible for up-type quark masses.
In this instance we find all the same SMEFT operator structures as presented in Tables~\ref{tab:treeleveldim4and5}--\ref{tab:4fermions}, but with straightforward ``$d \rightarrow u$'' replacements. 
For instance, when matching the $\DFN = 5$ UV Lagrangian to the SMEFT at tree-level up to dimension-6, the Higgs-enhanced Yukawa term changes as
\begin{equation}  
	-  \frac{\lambda_{11} \lambda \, c^d_{ij}}{m_\theta^2} \left(H\hc H \right) \left( \overline{Q}_i H d_j  + \text{h.c.} \right) 
 \longrightarrow 
 -  \frac{\lambda_{11} \lambda \, c^u_{ij}}{m_\theta^2} \left(H\hc H \right) \left( \overline{Q}_i \tilde{H} u_j  + \text{h.c.} \right)
\end{equation}
as intuitively expected. 
At one-loop, the story remains largely the same.
In Table~\ref{tab:renormalisableTerms}, the Wilson coefficients of $ \overline{Q}_i H d_j $ and $ \overline{Q} \tilde{H} u_j $ are exchanged together with the swap $ y^d_{ij} \leftrightarrow y^u_{ij}$ and replacement $ c^d_{ij} \rightarrow c^u_{ij} $. 
The same applies to the Higgs-enhanced Yukawa terms of Table~\ref{tab:HiggsAndHiggsEnhancedYuk}, whereas the purely bosonic operators from Tables~\ref{tab:HiggsAndHiggsEnhancedYuk} ($H^6$) and \ref{tab:fieldStrengths} ($X^2 H^2$) remain unchanged.
When it comes to the Higgs-kinetic terms in Table~\ref{tab:HiggsKinetic}, we can once again make the obvious substitutions and exchanges as explained above.
The coefficients of $( H\hc \Darr H ) (\overline{d}_i \gamma^\mu d_j )$ and $( H\hc \Darr H ) ( \overline{u}_i \gamma^\mu u_j) $ are traded, and those of the operators $( H\hc  \DarrI H ) ( \overline{Q}_i \tau^I \gamma^\mu Q_j)$ and $( H\hc \Darr H ) ( \overline{Q}_i \gamma^\mu Q_j)$ receive a ``$d \rightarrow u$'' relabelling.
In the same vein, the four-fermion operators of Table~\ref{tab:4fermions} change in the obvious way.

We next consider the matching analysis for the combined toy model of \eqref{eq:toymD} and its up quark analogue, with both sectors turned on at once.  
There is hardly any mixing between the two sectors of the model.
For nearly all operators one may collect the Wilson coefficients arising from the down-type FN sector, shown in the tables above, and add to these coefficients the terms which appear for the first time when turning on only the up-type sector in the UV.

The exception to this rule is the contributions to the Higgs kinetic operator $i( \tilde{H}\hc D_\mu H  ) ( \overline{u}_i \gamma^\mu d_j )$.
When the up-sector is turned on in addition to the down-sector, the Wilson coefficient of this operator contains  terms which cross the dimension-5 coefficients $c^d_{ij}$ and $c^u_{ij}$:
\begin{align}
        \nonumber
          \frac{\hbar}{4 m_\theta^4 } \Bigl[ 2 \lambda_{11} \lambda m_\theta^2 ( {y^u}^\dagger c^d )_{ij} - \lambda_{11}^2 v_\theta^2 ({y^u}^{\dagger}y^d )_{ij} (5 + 2 \mathbb{L} ) & \Bigr] i( \tilde{H}\hc D_\mu H  ) ( \overline{u}_i \gamma^\mu d_j ) + \text{h.c.} \\
          \longrightarrow 
          \frac{\hbar }{4 m_\theta^4 } 
          \Bigl[ 2 \lambda_{11} \lambda m_\theta^2 ( {y^u}^\dagger c^d )_{ij} 
          - \lambda_{11}^2 v_\theta^2 ({y^u}^{\dagger}y^d )_{ij}  (5 &+ 2  \mathbb{L} ) + 2 \lambda_{11} \lambda m_\theta^2 ({c^u}^\dagger y^d)_{ij} \\
          \nonumber
          +  \lambda^2 & ({c^u}^\dagger c^d)_{ij} \frac{m_\theta^4}{v_\theta^2} (1 + 2 \mathbb{L})\Bigr]  i( \tilde{H}\hc D_\mu H  ) ( \overline{u}_i \gamma^\mu d_j ) + \text{h.c.}
\end{align}
The first two terms carry over from the down-quark-only scenario and the third term follows from the ``$d \rightarrow u$'' recipe outlined above, but the final term, proportional to $({c^{u\dagger}} c^d)_{ij}$, is new.

\subsubsection*{Extension to Leptons}
As with the up quarks, in the toy model analysis presented above we have not specified the \SFN{} charges of $L_i$ and/or $e_i$ which, assuming a non-trivially charged Higgs, must respect 
\begin{equation}
\label{eq:CLchargecondition}
q_{\overline{L}_i}+q_{e_j} \overset{!}{=} -(q_H + x_{ij}\, q_\theta)\,
\end{equation}
to allow for generically flavoured charged-lepton mass terms a la \eqref{eq:basicidea}.  
Given the parity between the charged lepton and quark sectors of the (SM)EFT, we do not expect any substantive deviations in the matching of FN models where \eqref{eq:CLchargecondition} is respected and those already presented in Section \ref{sec:MODEL} and generalized in \ref{sec:GENERALCHARGE}.
In other words, if/when FN dynamics controls $y^e_{ij}$ along the lines of \eqref{eq:toymD}, we expect the flavour structure of all derived Wilson coefficients dependent on $L_i,e_i$ to be analogous to that found for the down quarks in Section \ref{sec:MODEL}.  
As a final note, we again caution that, when constructing gauged FN models with the aim of explaining all fermion masses and mixings, gauge anomalies provide an additional constraint on the charge assignment, and this is especially true when incorporating \SFN{}-charged leptons into the spectrum.

On the other hand, non-zero neutrino masses require a generation mechanism that is fundamentally BSM.  One approach is to introduce SM-gauge-singlet, right-handed neutrinos analogous to $\lbrace u,d,e \rbrace$ into the dynamical particle spectrum, which immediately leads to renormalizable Dirac neutrino masses in analogy to the charged leptons and quarks (not to mention a lepton-number-violating Majorana mass for the gauge-singlets).  Doing so is beyond our scope in this paper and, in any event, if said gauge singlets are sufficiently heavy, integrating them out of the Lagrangian leads to Weinberg operators \cite{Weinberg:1979sa} via the see-saw mechanism \cite{Minkowski:1977sc}, and these already generically appear in our construction (cf.~\eqref{eq:toyLdim5}). 
Indeed, it is well known that a Weinberg operator generating non-zero Majorana neutrino masses appears at $\DSM=5$, and this is also possible in a FN construction which generalizes the Weinberg operator class and its charge constraints as follows:
\begin{equation}
    \label{eq:WeinbergDFN6}
     \mathcal{L}^{5+x_{ij}+y_{ij}}_{\text{FN}} 
     \supset \frac{c_{ij}^{W_{xy}}}{\LUV^{1+x_{ij}+y_{ij}}} \left(H \overline{L}^c_i\right) \left(L_j H\right)\theta^{x^{ij}}\theta^{\star^{y_{ij}}} +\text{h.c.}
     \quad \text{iff} \quad
     \left(x_{ij}-y_{ij}\right) \, q_{\theta} 
     = -(2q_H + q_{L_i}+q_{L_j})\,.
\end{equation}
In this scenario FN dynamics also control hierarchies of neutrino masses, which is generally appealing for a complete flavour model.  

However, implementing a flavour-generic, dimension-6 version of \eqref{eq:WeinbergDFN6} into {\tt{Matchete}}, and subsequently asking the program to match to the $\DSM = 6$ SMEFT, only generates a Weinberg operator with the characteristic Wilson coefficient: 
\begin{align}
\label{eq:SMEFT66weinberg}
    \mathcal{L}_{\text{SMEFT}} &
    \supset c_{ij}^{W_{10}}\frac{\overline{\lambda}}{\LUV}\left(H \overline{L}^c_i\right) \left(L_j H\right) + \text{h.c.}
\end{align} 
 It is expected that, if matching to higher-orders in the SMEFT, more elaborate results will appear. 
 Regardless, it is fun to observe that a Majorana neutrino mass term suppressed by $\overline{\lambda}/\LUV$ is naturally generated in this naive FN setup.
 The suppression of the Weinberg operator by some power of $\overline{\lambda}$ will modify the typical lower bounds on the see-saw scale derived from neutrinoless double-beta decay searches.

A final comment regards the overlap of neutrino and charged lepton masses, whose diagonalization to their respective mass-eigenstate bases results in the Pontecorvo-Maki-Nakagawa-Sakata (PMNS) mixing matrix $U^{\text{PMNS}}$ in the SMEFT. 
PMNS mixings are now quite well-constrained experimentally (cf.~\cite{Esteban:2020cvm}) and, unlike the CKM quark mixing matrix, are generically large and non-hierarchical ($\vert U_{ij}^{\text{PMNS}} \vert \sim \mathcal{O}(10^{-1})-\mathcal{O}(1)$).  Hence a successful FN model of lepton flavour must simultaneously explain strong charged-lepton mass hierarchies alongside of weak PMNS hierarchies, both as a function of $\lambda$.  While not impossible, this challenge (along with the desire to limit the number of free UV model parameters) has often led model-builders to consider non-Abelian theories of flavour --- cf. Section \ref{sec:BEYONDABELIAN} below.   

\subsubsection*{Flavour Patterns for Arbitrary Charges}
Before concluding, let us see how the leading-order FN scalings exposed above can be understood using spurion methods.
As we have already seen in Sections \ref{sec:dim5matching} and \ref{sec:dim6matching}, by treating the FN EFT power-counting parameter $ \lambda $ as the spurion of the spontaneously broken $ \SFN{} $ symmetry one may intuit the flavour-dependent suppression patterns within each fermionic SMEFT operator (see \cite{Feldmann:2006jk} and \cite{Bordone:2019uzc} for a comparable scenario where the FN charge assignment determines the flavour structure in next-to-minimally-flavour-violating\footnote{The concept of minimal flavour violation is introduced in \cite{DAmbrosio:2002vsn}.} spurions).
Under the assumption that the flavon vev controls all violations of the \SFN{} symmetry, the leading-order $\lambda$-dependence of both tree- and loop-level generated fermionic Wilson coefficients is fully determined by the FN charge assignments.  Indeed, using spurion analysis one can easily derive the leading-order $\lambda$-scaling, $\lambda^x$, of a generic SMEFT operator containing Higgs bosons and fermions using
\begin{equation}
\label{eq:spurionequation}
x = \frac{\left| \sum_{\overline{\psi}} N_{\overline{\psi}} \, q_\psi - \sum_\psi  N_\psi \, q_{\psi} +  N_{H\hc , \tilde{H}} \, q_H -  N_{H, \tilde{H}\hc} \, q_H \right| }{\left| q_\theta \right| }\,,
\end{equation}
where $x$ is the number of spurion $\lambda$ insertions required to make the IR SMEFT operator invariant under \SFN, and the RHS of \eqref{eq:spurionequation} contains a sum over all species of Weyl fermions and Higgs conjugation structures ($H$,  $\tilde{H}$, $H^\dagger$, $ \tilde{H}^\dagger $) with $ N_i $ denoting the number of times each field appears in the operator. 
Using \eqref{eq:spurionequation} one finds the following scalings for a number of the flavourful building blocks appearing in the SMEFT operators above:
\begin{equation}
\label{eq:samplescalings}
\overline{d}_i \gamma^\mu d_j \sim \lambda^{\left|\frac{q_{d_i} - q_{d_j}}{q_\theta}\right|}\,, \,\,\,\,\,\,\,\,\,\,\overline{Q}_i \tilde{H} u_j \sim \lambda^{\left|\frac{q_{Q_i} +q_H - q_{u_j}}{q_\theta}\right|}\,,\,\,\,\,\,\,\,\,\,\, \left( \overline{Q}_i^\alpha  u_j \right) \epsilon_{\alpha \beta} \left( \overline{Q}_k^\beta  d_l \right) \sim \lambda^{\left|\frac{q_{Q_i} - q_{u_j} + q_{Q_k} - q_{d_l}}{q_\theta}\right|}\,.
\end{equation}
The generalisation to other operators, including purely leptonic SMEFT operators, follows trivially.

As discussed in detail in \cite{Feldmann:2006jk}, a generic FN model allows for more flexibility in generating spurions of the approximate $ \text{SU(3)}^5 $ symmetry of the SM fermion sector than the minimal flavour violation scenario, leading to flavour transitions with smaller $ \lambda $-suppression compared to expectations based on the latter, especially for FCNC transitions containing right-handed quarks.
In the absence of any assumptions about the UV completion of the FN EFT, all spurionic representations of the approximate global symmetry could appear in the SMEFT as long as operators of a sufficiently high operator dimension are considered.


\subsection{Non-Abelian and/or Continuous Flavour Models}
\label{sec:BEYONDABELIAN}
We now consider how our matching analysis might change in flavour models that are based on broken non-Abelian symmetries, and/or models based on continuous flavour symmetries, as opposed to the global, discrete, Abelian considerations of Section \ref{sec:MODEL}. 
\subsubsection*{Non-Abelian Flavour Models}
Non-Abelian flavour models typically assign multiple generations of a SM fermion family to a single BSM flavour-symmetry multiplet, which is not possible in Abelian frameworks. These symmetries can be continuous  (e.g. U(2), SO(3) or SU(3)) or discrete  (e.g. A4, S4, or $\Delta(27)$), and the literature concerning both is vast.  A common feature of these constructions is the reduction of UV parameters compared to Abelian models, which can be understood thanks to their multi-dimensional irreducible representations --- contracting $N$-plets of a given flavour symmetry populates \emph{multiple} effective Yukawa couplings with a single free parameter (Wilson coefficient), and hence typically leads to increased predictivity. A trade-off is that multiple new flavon fields are often introduced to allow for distinct contractions, which typically have vacua that are especially aligned in flavour space to realize characteristic Yukawa textures.  

As an example of a relatively recent and successful discrete model, we consider the Universal Texture Zero (UTZ)  \cite{deMedeirosVarzielas:2017sdv,Bernigaud:2022sgk} introduced by one of the authors and collaborators, based on a $\Delta(27) \times \mathbb{Z}_N$ flavour symmetry and consistent with an underlying $\text{SO}(10) \rightarrow \text{SU}(4) \times \text{SU}(2)_\text{R} \times \text{SU}(2)_\text{L}$ Grand Unified Theory (GUT).  The UTZ is capable of controlling the Dirac masses\footnote{Majorana neutrino masses (and associated PMNS mixings) are also controlled with a similar equation.  Note that the UTZ also assumes supersymmetry breaking, whose complications we will ignore in this discussion.} of both the quark and lepton sectors in terms of three flavons $\lbrace \theta_{3,23,123}\rbrace$ and two additional scalars $\lbrace S, \Sigma \rbrace$, with $\Sigma$ related to the GUT symmetry breaking and $S$ designed, along with the additional $\mathbb{Z}_N$ symmetry, to `shape' the allowed interactions between $\theta_i$:
\begin{equation}
\label{eq:UTZLag}
\mathcal{L}_{\text{UTZ}}  \supset \psi_i\left[\frac{c^{(6)}_3}{ \Lambda_{3,f}^2}\theta_3^i  \theta_3^j  + \frac{c^{(7)}_{23}}{ \Lambda_{23,f}^3}\theta_{23}^i\theta_{23}^j\Sigma+\frac{c^{(7)}_{123}}{\Lambda_{123,f}^3}(\theta_{123}^i\theta_{23}^j+\theta_{23}^i\theta_{123}^j)  S    \right]\psi_j^{c}H + \mathcal{O}(1/\Lambda_i^4) + \text{...}
\end{equation}
Here $\psi$ is a generic SM fermion with flavour $\lbrace i,j \rbrace$ of family $f$, $f \in \lbrace u, d, e, \nu \rbrace$, assigned to a  $\Delta \left(27\right)$ triplet.  As indicated, these flavour multiplets contract with the (anti-triplet) flavon fields, while $\lbrace H, S, \Sigma \rbrace$ are all trivial flavour singlets.  Thanks to its elaborate $\Delta \left(27\right)$-invariant scalar potential --- cf. \cite{deMedeirosVarzielas:2017sdv} for details --- the UTZ flavons develop the following vevs in flavour space ($\alpha$ and $\beta$ are free complex phases):
\begin{equation}
\label{eq:UTZvacua}
\left\langle {{\theta_{3}}} \right\rangle  = {{\rm{v}}_{\theta 3}}\left( {\begin{array}{*{20}{c}}
0\\
0\\
1
\end{array}} \right),\quad \left\langle {{\theta _{123}}} \right\rangle  = \frac{{{\rm{v}}_{123}}}{{\sqrt 3 }}\left( {\begin{array}{*{20}{c}}
e^{i\beta}\\
e^{i\alpha}\\
{ - 1}
\end{array}} \right), \quad
 \left\langle {{\theta _{23}}} \right\rangle  = \frac{{{\rm{v}}_{23}}}{{\sqrt 2 }}\left( {\begin{array}{*{20}{c}}
0\\
e^{i\alpha}\\
1
\end{array}} \right),
\end{equation}
from which it is easy to intuit from \eqref{eq:UTZLag} that the UTZ predicts very specific Yukawa/mass matrix textures:
\begin{equation}
\label{eq:UTZMDIR}
M^{f}_{\text{Dirac}} \propto
\left(
\begin{array}{ccc}
0 & a\,e^{i(\alpha+\beta+\gamma)}  & a\,e^{i(\beta+\gamma)} \\
a\,e^{i(\alpha+\beta+\gamma)}  & (b\,e^{-i\gamma}+2a\,e^{-i\delta})\,e^{i(2\alpha+\gamma + \delta)} & b\,e^{i(\alpha + \delta)}\\
a\,e^{i(\beta + \gamma)} & b\,e^{i(\alpha + \delta)} & 1 - 2 a\,e^{i\gamma} + b\,e^{i\delta}
\end{array}
\right) \,,
\end{equation}
which we have written after EW symmetry breaking, and where $\lbrace a, b, \gamma, \delta \rbrace$ are all rescaled functions of the (complex) coefficients and parameters given in \eqref{eq:UTZLag}--\eqref{eq:UTZvacua}.  One immediately sees in \eqref{eq:UTZMDIR} a concrete manifestation of the non-Abelian mechanism mentioned above in that, for a given family sector, only six \emph{real} parameters structure Dirac mass and mixing up to EFT order implied in \eqref{eq:UTZLag}. This is to be compared to a \emph{complex} parameter for each matrix element in the simple  $\mathbb{Z}_N$ construction of Section \ref{sec:MODEL}.

From \eqref{eq:UTZLag}--\eqref{eq:UTZMDIR} we can already anticipate a number of complications, but also results, when  matching the UTZ to the SMEFT, in comparison to the FN setup of Section \ref{sec:MODEL}.  
First and foremost, in \emph{each} family sector $f$ there are potentially three independent UV scales $\Lambda_{3,23,123}$ identified in \eqref{eq:UTZLag}, associated with the first appearance of each of the three flavons $\theta_{3,23,123}$. This allows for hierarchies amongst the flavon and/or UV mediator masses, which further implies a potentially complex sequential matching to intermediate EFTs. This would generate multiple expansion parameters that ultimately filter down to characterize the SMEFT Wilson coefficients (analogous to $\lambda$ above).  A naive option is to assume no hierarchy amongst the UV scales, $\Lambda_3 \sim \Lambda_{23} \sim \Lambda_{123}$, although as discussed in \cite{deMedeirosVarzielas:2017sdv,Bernigaud:2022sgk} phenomenological considerations and the desire to avoid fine-tuned Wilson coefficients quickly demands $\Lambda_{3,u}/\Lambda_{23,u} < \Lambda_{3,d}/\Lambda_{23,d}$, while GUT considerations lead to an implied equivalence in the Yukawa expansion parameters between down quark and charged lepton sectors.  Finally, it is straightforward to deduce from \eqref{eq:UTZLag} that at least $\langle \theta_3 \rangle / \Lambda_{3,u}$ should be relatively large to account for the top mass (cf. \cite{deMedeirosVarzielas:2005ax}). 
The takeaway is that the presence of multiple UV scales, associated to multiple flavons, might rapidly complicate the matching we have performed in Section \ref{sec:MODEL} and that, in the absence of a fully generic / sequential matching to intermediate EFTs, one risks a heavily model-dependent analysis. 

At the same time, \eqref{eq:UTZMDIR} reveals that 
$M^d_{12,13,23}=M^d_{21,31,32}$, since these matrix elements are sourced by the same UTZ operators.  This indicates that, upon integrating out the relevant flavon(s), each term in \eqref{eq:UTZLag} will induce multiple SMEFT operators (corresponding to distinct IR flavour structures), each with the same Wilson coefficient.  As an obvious example, we would expect an equivalent contribution to Higgs-enhanced Yukawa operators akin to \eqref{eq:dim6Yukawaops} to be induced at the first tree-level matching order relevant to the UTZ, for each matrix element associated to the UTZ operator.  This represents a distinct signature with respect to the Abelian FN construction, and it also holds up to the most important next-to-leading-order EFT operators in the UTZ (see \cite{deMedeirosVarzielas:2017sdv,Bernigaud:2022sgk} for higher-order contributions to \eqref{eq:UTZMDIR}).  Additionally, we see that there are no renormalizable Yukawa couplings present in the UTZ, and so the analogous matching to Section \ref{sec:dim4matching} will be sourced fully from the scalar potential. Hence we would naively expect all one-loop matching terms proportional to $y_{32,33}^d$ to go to zero.  It is also worth noting that while fewer terms in \eqref{eq:UTZLag} populate $M^d$ than in \eqref{eq:toymD}, and that the maximum mass dimension of the contributing operators is also less than in \eqref{eq:toymD}, the first order at which interesting Yukawa structure appears is \emph{higher} than in \eqref{eq:toymD} (mass-dimension 6, not 4).  

As a final preliminary consideration, we recall that a rather particular extended scalar potential is required to dynamically drive the minimization of the vevs in \eqref{eq:UTZvacua}.  This is true already at the renormalizable level, while the model-building impacts of effective scalar potential operators akin to \eqref{eq:dim6Scalarops} is to our knowledge somewhat lacking in the literature (see \cite{deMedeirosVarzielas:2012ylr} for a related counterexample, albeit in a different class of model), not to mention the role of effective operators in the UTZ theory outside of the scalar or Yukawa sectors (akin to the $\DFN=6$ basis we built in Section \ref{sec:dim6matching}).  As seen at virtually all stages of the matching in Section \ref{sec:MODEL}, the impact of Higgs-flavon mixing is extremely important in understanding the structure of the IR SMEFT Wilson coefficients.  This will become even more intricate in non-Abelian, UTZ-type models.  Besides the presence of more scalar operators, the special vev alignment in \eqref{eq:UTZvacua} requires hierarchies amongst their coefficients, which will have immediate implications for the SMEFT Wilson coefficients they populate as well.  These must be understood in order to fully appreciate the IR imprint of the UTZ.   

While the UTZ of \eqref{eq:UTZLag} represents but one of many non-Abelian flavour models on the market, it shares a number of salient features with other constructions.  For example, the renowned A4 models of Altarelli-Feruglio \cite{Altarelli:2005yp,Altarelli:2005yx} or Babu-Ma-Valle \cite{Babu:2002dz}, the SU(3) models of Ross-King \cite{King:2001uz,King:2003rf}, or even the more recent gauged SO(3) model of Reig-Valle-Wilczek \cite{Reig:2018ocz} all depend on  flavon(s) with particular vev alignments in flavour space, which, as mentioned above, will likely generate the dominant source of complexity when matching non-Abelian flavour models to the SMEFT.   Distilling their most generic features beyond Points (1)--(3) above represents future work.
\subsubsection*{Goldstone Modes and Gauged Flavour Models}
As mentioned below \eqref{eq:thetavevexpand}, the toy FN model we consider in Section \ref{sec:MODEL} is effectively controlled by a discrete, Abelian FN symmetry ($\mathbb{Z}_N$), since we do not consider the implications of the (pseudo-)Goldstone boson associated to the breaking of a continuous \FN{}, regardless of whether it is global or gauged.  

In the instance where we generalize to a \emph{global} \FN{}, the $\pi_\theta$ in \eqref{eq:thetavevexpand} are massless Goldstone particles in the strict symmetry limit, or light (with respect to the flavon $\vartheta$) pseudo-Goldstone particles when \FN{} is only approximate.   
In the absence of hypothetical model building that forces $\pi_\theta$ to be very heavy, we are left with an additional, dynamical BSM degree of freedom in either case, and so formally one should not match directly to the SMEFT as we have above, but instead to the IR EFT that also incorporates $\pi_\theta$ (which we will refer to as the $\pi$SMEFT). Furthermore, even if $\pi_\theta$ is heavy with respect to the SM, one would have to purposely integrate it out alongside of or sequential to $\vartheta$, which would ultimately amount to matching a (flavour-specific) UV model with two real SM gauge singlets, since we work in the broken \FN{} phase. 

While beyond our scope to explore these situations in detail, we do note that, up to $d\le7$, a non-redundant basis for the proposed $\pi$SMEFT is effectively given in \cite{Gripaios:2016xuo} up to flavour-symmetry assumptions.  One can anticipate that, since the $\pi$SMEFT $\supset$ SMEFT, the contributions outlined in Section \ref{sec:MODEL} will still hold.  
Also, up to $\DFN=6$, the only Lagrangian interactions allowed between $\vartheta$ and $\pi_\theta$ in the UV \FN{} theory are sourced by the $\theta$-current-fermion-current interaction in \eqref{eq:dim6GaugeFlavonFermionops} and $\mathcal{O}_{\overline{Q}_2 d_i \theta}$ in \eqref{eq:dim6Yukawaops} (cross terms proportional to $\pi_\theta\, \vartheta$ cancel in $\theta^\star \theta$ operator pairs). 
Hence some contributions to matched IR Wilson coefficients in the $\pi$SMEFT would simply follow trivially from $\vartheta$-independent UV interactions (e.g. the $\DFN=5$ operator $\propto \bar{\psi}_L H \psi_R \pi_\theta$). 
On the other hand, flavon-Goldstone mixing will complicate the results when matching up to higher operator dimensions in $\pi$SMEFT and especially in a loop-level analysis.
Furthermore, even upon successfully matching to $\pi$SMEFT, utilizing the hypothetical $\pi$SMEFT-matched FN theory for precision phenomenology would require knowledge of the complete set of RGE equations for generic $d=6$ $\pi$SMEFT Wilson coefficients, analogous to \cite{Alonso:2013hga,Jenkins:2013wua,Jenkins:2013zja} for the SMEFT, to run down to scales relevant for experimental comparison or subsequent EFT matching(s). 
We note that some relatively recent work has been done on SMEFT interference effects when an additional, light, axion-like particle (akin to $\pi_\theta$) is present \cite{Galda:2021hbr}. 

If instead the \FN{} is \emph{gauged}, we must incorporate an additional $Z^\prime$ vector in the UV spectrum. 
In the unitary gauge, it will eat the Goldstone to acquire a mass proportional to the flavon vev and will therefore be integrated out of the spectrum alongside the scalar flavon as above. 
There are a host of papers discussing the matching of a $Z^\prime$ model to the SMEFT, but generic tree-level results are available from (e.g.) \cite{deBlas:2017xtg}. 
There it is clear that a number of dimension-6 SMEFT operators are turned on as a result of integrating out the $Z^\prime$.  
Owing to its nature as a gauge boson, the $Z'$ may only couple to the SM fermions via the covariant derivative $\overline{\psi}_i i \slashed{D} \psi_i$ at the renormalizable level.
If $g_\text{F} $ denotes the gauge coupling of \SFN{} and $q_{\psi_i}$ the charge of the fermion under the Abelian symmetry, we arrive at the interaction vertex $- g_\text{F} \,q_{\psi_i} Z'_\mu \overline{\psi}_i \gamma^\mu \psi_i$ where one may assume without loss of generality that the interaction is diagonal in flavour space. 
For the down quark toy model of Table~\ref{tab:DownToy}, the interaction may be expressed as
\begin{equation} \label{eq:zprimeFermionInt}
    \mathcal{L}_\text{FN} \supset - g_\text{F} Z_\mu' \left( \Omega_{ij} \overline{d}_i \gamma^\mu d_j + \xi_{ij} \overline{Q}_i \gamma^\mu Q_j  \right)\,,
\end{equation}
where the two flavour matrices are 
\begin{equation}
    \Omega = 
    \begin{pmatrix} 
    	5 & 0 & 0 \\ 
	0 & 3 & 0 \\
	0 & 0 & 3 \\
    \end{pmatrix}
    \quad \text{and} \quad
    \xi = -
    \begin{pmatrix} 
    	6 & 0 & 0 \\ 
	0 & 4 & 0 \\
	0 & 0 & 0 \\
    \end{pmatrix} .
\end{equation}

When integrating out the $ Z' $ diagrammatically, \eqref{eq:zprimeFermionInt} leads to four-fermion interactions via a $ Z' $ exchange:
\begin{multline}
	\mathcal{L}_\text{SMEFT}^{d=6}
	\supset
	- \frac{g_\text{F}^2 \, \Omega_{i j} \Omega_{k l}}{2 M_{Z'}^2} \left( \overline{d}_i \gamma^\mu d_j \right) \left( \overline{d}_k \gamma_\mu d_l \right) 
	- \frac{g_\text{F}^2 \, \xi_{i j} \xi_{k l}}{2 M_{Z'}^2} \left( \overline{Q}_i \gamma^\mu Q_j \right) \left( \overline{Q}_k \gamma_\mu Q_l \right) \\
	- \frac{g_\text{F}^2 \, \Omega_{i j} \xi_{k l}}{ M_{Z'}^2} \left( \overline{d}_i \gamma^\mu d_j \right) \left( \overline{Q}_k \gamma_\mu Q_l \right).
\end{multline}
This of course generalizes to the remaining fermion species which would generate similar four-fermion terms $ \mathcal{O}_{\psi \psi} $ for $ \left( \psi \psi \right) \in \{L L, L Q, e e, u u, e d, e u, u d, L e, L d, L u, Q e, Q u \} $.
Here, too, the results are presented in the interaction eigenbasis and a rotation to the mass basis is required in order to get to serious phenomenological predictions.

Provided that the Higgs is charged under \SFN{}, the Higgs kinetic term $D_\mu H\hc D^\mu H$ gives more renormalizable interactions between the $Z'$ and the SM fields, one linear and one quadratic in the $ Z' $.
When it comes to dimension-6 results from tree-level matching, it is only the term linear in the novel gauge boson, $ \mathcal{L}_\text{FN} \supset - i g_\text{F} q_H Z'_\mu H\hc D^\mu H + \text{h.c.} $, that contributes to the Wilson coefficients.
In our toy model, where $ q_H  = -3$, connecting two such vertices together via a $ Z' $ propagator gives rise to
\begin{equation}
	\mathcal{L}_\text{SMEFT}^{d=6} \supset  
	- \frac{9 g_\text{F}^2 }{2 M_{Z'}^2} \left( H\hc H \right)  \Box \left( H\hc H \right) 
	- \frac{18 g_\text{F}^2 }{ M_{Z'}^2} \left( H\hc D^\mu H \right)^*  \left( H\hc D_\mu H \right) \,,
\end{equation}
whereas connecting a fermion current from \eqref{eq:zprimeFermionInt} with a $ Z'_\mu H\hc D^\mu H $ vertex contributes to Higgs kinetic operators
 \begin{equation}
	\mathcal{L}_\text{SMEFT}^{d=6} \supset 
	- \frac{3 g_\text{F}^2 \Omega_{i j}}{M_{Z'}^2}( H\hc \Darr H ) ( \overline{d}_i \gamma^\mu d_j)
	- \frac{3 g_\text{F}^2 \xi_{i j}}{M_{Z'}^2}( H\hc \Darr H ) ( \overline{Q}_i \gamma^\mu Q_j).
\end{equation}
Again, a similar set of terms is obtained for the other fermion species, too.

One may also expect matching contributions that hinge on vertices that depend on both the flavon $ \vartheta $ and the $ Z' $, originating from the flavon kinetic term $ D_\mu \theta^* D^\mu \theta $.
These vertices, however, will only yield SMEFT operators with $\DSM > 6$ when matching at tree-level, and hence will not be addressed here.
Similarly, when going beyond the renormalizable level in the FN EFT, it is possible to write down more complex interactions between the $ Z' $, the flavon, and the SM fields, but these vertices will not play a role in generating dimension-6 SMEFT operators at tree-level, and can introduce further complications regarding flavour-violating basis changes as well. 

From the above analysis we conclude that, at tree-level, integrating out a  $Z^\prime$ does not  induce $\DSM=6$ dipole-type ($\psi^2 X H$) or triple-field-strength ($X^3$) operators in the SMEFT.  
As mentioned above, the $X^3$ operators will still be allowed in the IR with arbitrary Wilson coefficients due to their presence in $\mathcal{L}_{\text{SMEFT}^\prime}$, but IR SMEFT dipole operators will only appear with limited flavour combinations from $\mathcal{L}_{\text{SMEFT}^\prime}$.  Hence the presence of other dipole signals (via sizable non-zero Wilson coefficients) in global SMEFT fits might serve as a powerful falsification tool for FN-type models, regardless of whether they are gauged or global.  
Simultaneously, signals for sizable four-fermion or Higgs-kinetic ($H^2 D \psi^2)$ Wilson coefficients could, in principle, be used to distinguish the gauged \FN{} vs. global scenario, given that we found these operators only at loop-level in Section \ref{sec:MODEL}.
What is more, the hierarchies of such operators in flavour space are different for the contributions driven by the flavon and controlled by powers of $\lambda$, versus those coming from the $ Z' $ and controlled by the fermion charges under \SFN{}.
For the toy model we find, for instance, that the Wilson coefficients of the four-fermion operators grow as we move to lower generations, a feature that is capable of producing stringent lower bounds on the $ M_{Z'} $ mass and thus the \SFN{} breaking scale in the gauged scenario.

Finally, all of the above discussion can be further generalized to the case of continuous, \emph{non-Abelian} flavour symmetries, as have been explored in the models of (e.g.) \cite{King:2001uz,King:2003rf} and, more recently, \cite{Reig:2018ocz}. 
Here the presence of multiple (pseudo-)Goldstone bosons will source the obvious complication(s) beyond those already mentioned in the subsection above.  
As with a \FN, breaking a global non-Abelian flavour symmetry will necessitate matching to a larger EFT than the SMEFT or even $\pi$SMEFT (incorporating \emph{all} of the light Goldstones), while breaking a gauged non-Abelian flavour symmetry will require understanding the IR impacts coming from subsequently (or simultaneously) integrating out multiple heavy BSM vectors.

\section{Summary and Outlook}
\label{sec:CONCLUDE}

We have systematically matched the Froggatt-Nielsen (FN) \cite{Froggatt:1978nt} flavour generation mechanism to the mass-dimension 6 $(\DSM=6)$ SMEFT, considering tree- and one-loop level contributions sourced from UV operators in a FN effective theory up to mass-dimension 6 and 5 ($\DFN=6,5$), respectively.  
Beginning from a simplified \ZFN{}-symmetric construction, in Section \ref{sec:MODEL} we studied an Abelian toy model where FN dynamics resolve fermionic mass and mixing relations derived from the down-quark Yukawa matrix in \eqref{eq:toymD}.  This included enumerating a minimal, complete basis of EFT operators at $\DFN=6$ in the FN theory, to be fully generic in our analysis. 
Upon executing (and checking!)\ the functional matching procedure embedded in {\tt{Matchete}} \cite{Fuentes-Martin:2022jrf} we saw that, besides renormalizations of SM terms, a host of SMEFT operators also appear with characteristic Wilson coefficients.  
These include Higgs-scalar ($H^6$) and -derivative ($H^4 D^2$), Higgs-gauge ($H^2 X^2$), Higgs-kinetic ($H^2 D \psi^2$), Higgs-enhanced Yukawa ($H^3 \psi^2$), and four-fermion ($\psi^4$) operators. 
Many of these IR signatures are effectively flavour independent, appearing simply as a result of integrating out a heavy BSM scalar (the FN flavon, in this instance), while others are flavour-specific, reflecting the particular FN charge assignment (and therefore patterns of flavour suppression) of our UV toy model. Our final results are collected in Tables~\ref{tab:treeleveldim4and5}--\ref{tab:4fermions}. 

Our theoretical analysis can be further improved in a number of ways.  
In Section \ref{sec:GENERALIZE} we already discussed scenarios that obviously generalize our analysis, including the straightforward extension(s) of the FN dynamics we studied in the down quark sector to different fermion families. We also anticipated the implications of considering continuous and/or non-Abelian flavour models, either gauged or global, where we saw that a number of complications can arise due to the presence of additional IR (e.g. Goldstone boson) or UV (e.g. flavon and/or vector gauge boson) physics scales, depending on the particular flavour symmetry realization.  We pointed out that, in the presence of additional light degrees of freedom, a better understanding of the matching to an intermediate EFT, with BSM propagating degrees of freedom, was necessary.  We also commented on the possibility of distinguishing gauged vs.\ global UV flavour scenarios via the relative magnitudes of SMEFT Wilson coefficients that appear in each respective scenario.  Indeed, all of the discussions in Section \ref{sec:GENERALIZE} stand to be refined with more dedicated theory analyses in the future, perhaps even by matching UV-complete (fully renormalizable) model(s) of FN flavour to the SMEFT, although this would be much less model-independent than the EFT approach taken in this paper.  

On the other hand, in Section \ref{sec:GENERALIZE} we did not consider operators with $d>6$, either in the IR or UV. From the UV perspective, the mass matrix in \eqref{eq:toymD} has matrix elements that first appear at $\DFN \ge 7$.  While highly suppressed and very much model dependent, understanding such higher-order contributions in the context of the SMEFT may be interesting, if technically challenging.  From the IR perspective, matching the SMEFT at $\DSM>6$ would also yield interesting results.  Preliminary studies indicate a plethora of new, dimension-8 operators can appear, some of which may be phenomenologically relevant.  
One may for instance join together two dimension-5 FN vertices of the toy model from \eqref{eq:L5FNexpanded} in a way illustrated in Fig.~\ref{fig:dim-8-operator} to obtain the Higgs-enhanced four-fermion terms
\begin{align}
    \nonumber
    \mathcal{L}_\text{SMEFT}^{d=8} \supset &\left\{\frac{(c^d_{31})^2 \tilde{\lambda}^2}{2 \tilde{m}_\theta^2 \tilde{v}_\theta^2} (\overline{Q}_3 H d_1) (\overline{Q}_3 H d_1) + \text{h.c.} \right\}
    + \frac{\lvert c^d_{31}\rvert^2 \tilde{\lambda}^2}{4 \tilde{m}_\theta^2 \tilde{v}_\theta^2} \left( H\hc H \right) \left(\overline{Q}_3 \gamma^\mu Q_3 \right) 
    \left(\overline{d}_1 \gamma_\mu d_1 \right)\\ 
    &+ \frac{\lvert c^d_{31}\rvert^2 \tilde{\lambda}^2}{4 \tilde{m}_\theta^2 \tilde{v}_\theta^2} \left( H\hc \tau^I H \right) \left(\overline{Q}_3 \gamma^\mu \tau^I Q_3 \right) 
    \left(\overline{d}_1 \gamma_\mu d_1 \right).
\end{align}
The tildes above the parameters remind us that they will be modified by contributions coming from the dimension-8 scalar potential in the broken \SFN{} phase.
To obtain a full set of $\DSM=8$ results, it would of course be important to adopt a non-redundant, complete operator basis, for instance that of \cite{Murphy:2020rsh} or \cite{Li:2020gnx}.

\begin{figure}[htbp]
	\centering
	\begin{tikzpicture}[baseline = 0cm]
	\begin{feynman}
		\vertex (c);
		\path (c) ++(+120:1.8cm) node (ia) {$ \overline{Q}_3 $};
		\path (c) ++(-180:1.8cm) node (ib) {$ H $};
		\path (c) ++(-120:1.8cm) node (ic) {$ d_1 $}; 
		\vertex [right= of c] (d);
		\path (d) ++(+60:1.8cm) node (fa) {$ \overline{Q}_3 $};
		\path (d) ++(+0:1.8cm) node (fb) {$ H $};
		\path (d) ++(-60:1.8cm) node (fc) {$ d_1 $};

		\diagram* {
			(ia) -- [fermion] (c) -- [fermion] (ic),
            (c) -- [anti charged scalar] (ib),
			(c) -- [dashed, double, edge label = $\vartheta$] (d),
			(fa) -- [fermion] (d) -- [fermion] (fc),
			(d) -- [charged scalar] (fb),
		};
	\end{feynman}
		\newcommand\Xcoord{3.7}
		\newcommand\bracY{1.7}
		  \draw[->]        (\Xcoord,0)   -- (\Xcoord+1.3,0);
		  \draw (\Xcoord+4.8,0) node {$ C^{(1)}_{H^2 Q^2 d^2} \Big[ \left(H\hc H \right) \left(\overline{Q}_3 \gamma^\mu Q_3 \right) \left( \overline{d}_1 \gamma_\mu d_1 \right)$};
        \draw (\Xcoord+6.8,-1) node {$ + \left(H\hc \tau^I H \right) \left(\overline{Q}_3 \gamma^\mu \tau^I Q_3 \right) \left( \overline{d}_1 \gamma_\mu d_1 \right) \Big]$};
	\end{tikzpicture}
	\caption{A diagrammatic illustration of how tree-level matching of the toy FN model yields the dimension-8 SMEFT operators $  \left(H\hc H \right) \left(\overline{Q}_3 \gamma^\mu Q_3 \right) \left( \overline{d}_1 \gamma_\mu d_1 \right)$ and $  \left(H\hc \tau^I H \right) \left(\overline{Q}_3 \gamma^\mu \tau^I Q_3 \right) \left( \overline{d}_1 \gamma_\mu d_1 \right)$.
  To get to the final operators, we use a Fierz identity and the Pauli matrix completeness relation.  
  The Wilson coefficient follows the notation of \cite{Murphy:2020rsh}. \label{fig:dim-8-operator}}
\end{figure}

Finally, we have not considered matching FN theories in the \emph{unbroken} flavour-symmetry phase, nor have we considered matching to so-called geometric EFTs, where all-$v_H/\Lambda$-orders generalizations of (flavoured) Wilson coefficients, with implied geometries in BSM field space, appear (see e.g. \cite{Helset:2020yio,Finn:2020nvn,Talbert:2021iqn,Talbert:2022unj,Gattus:2023gep,Assi:2023zid} for recent work incorporating fermions in these formalisms).  All of these theoretical avenues can and should be explored.

Besides theory improvements, a major long-term goal of the SMEFT community is the robust, global fit of $\DSM=6$ Warsaw basis operators to experimental data across a range of energy scales, and the subsequent projection of those operator bounds to concrete BSM physics conclusions.  As mentioned in the Introduction, a number of studies have already made significant progress on the global fitting front, including in the flavour sector \cite{Falkowski:2017pss,Falkowski:2019hvp,Cirigliano:2019vfc,Aoude:2020dwv,Bissmann:2020mfi,Cirigliano:2021img,Bellafronte:2023amz,Grunwald:2023nli,Allwicher:2022gkm,Fajfer:2023gie,Allwicher:2023shc}. Hence an obvious next step in our study of FN-type effects on the SMEFT is to pursue precision phenomenology given the IR matching signatures we have outlined in Section \ref{sec:MODEL}. There we briefly mentioned bounds on flavon mass scales ($m_\theta \gtrsim \mathcal{O}(1)$ TeV) sourced only from $\mathcal{O}_{H\Box}$ in \eqref{eq:dim4treematchfinal}, as well as from certain fermionic operators ($m_\theta \gtrsim \mathcal{O}(5)$ TeV) that appeared in the Section.  However, a more dedicated and systematic phenomenological study could be approached with the automated matching, running, and optimization tools embedded in (e.g.) {\tt{smelli}}. Indeed, data procured at energy scales spanning many orders of magnitude will likely be required to falsify FN-type models in general, and/or to distinguish one FN-type model from another.

In summary, by matching the FN mechanism to the SMEFT, we have presented an important first step in the effort to utilize the SMEFT and all of its associated experimental and theoretical technologies to efficiently gain concrete, falsifiable insight on one of the SM's oldest and most perplexing problems:  the flavour puzzle.  We are hopeful that our work will help motivate a more EFT-centered approach to BSM flavour physics, and therefore both experimental and theoretical advances to that end.

\section*{Acknowledgements}
We thank Tyler Corbett for his help understanding certain SMEFT Feynman Rules in the unbroken electroweak phase, Rodrigo Alonso for reminding us about how to treat Goldstone bosons in non-renormalizable operators, the Cambridge Pheno Working Group for helpful discussions while completing this work, and Ben Allanach for useful comments in the finishing stages of the paper.
The work of EL is supported by STFC consolidated grants ST/T000694/1 and ST/X000664/1.
 JT gratefully acknowledges prior funding from the European Union's Horizon 2020 research and innovation programme under the Marie Sk\l{}odowska-Curie grant agreement No. 101022203, and guest support from the T-2 Group at LANL.  JT also thanks the participants of HEFT2023 for valuable insight, Steve King for inspiring early discussions that slowly evolved into this work, and Robert Szafron for clarifying remarks regarding matching formalisms. 
 
\bibliographystyle{unsrt}
\bibliography{biblioFNxSMEFT}
\end{document}